\documentclass[a4paper,aps,preprintnumbers,showpacs,twocolumn,superscriptaddress,nofootinbib]{revtex4-1}

\usepackage{amsmath}
\usepackage{amssymb}
\usepackage{graphicx}
\usepackage{epsfig}
\usepackage{color}
\usepackage{url}
\usepackage{times}
\usepackage{bm}
\usepackage{mathrsfs}
\usepackage[utf8]{inputenc}
\usepackage{hyperref}
\usepackage{enumerate}
\usepackage{amsthm}

\newcommand{\beq}{\begin{equation}}
\newcommand{\eeq}{\end{equation}}
\newcommand{\bea}{\begin{eqnarray}}
\newcommand{\eea}{\end{eqnarray}}
\newcommand{\bit}{\begin{itemize}}
\newcommand{\eit}{\end{itemize}}
\newcommand{\ben}{\begin{enumerate}}
\newcommand{\een}{\end{enumerate}}
\newcommand{\nn}{\nonumber}

\def\scri{\mathscr{I}^+}

\renewcommand{\d}[1]{\,\textnormal{d}#1}

\newtheorem{conjecture}{Conjecture}

\newcommand{\blue}[1]{\textcolor{blue} {#1}}

\begin{document}
\title{
Hyperboloidal slicing approach to quasi-normal mode expansions: the Reissner-Nordstr\"om case
}
\author{Rodrigo Panosso Macedo} 
\affiliation{School of Mathematical Sciences, Queen Mary, University of
  London, \\ Mile End Road, London E1 4NS, United Kingdom}

\author{Jos\'e Luis Jaramillo} 
\affiliation{Institut de Math\'ematiques de Bourgogne (IMB), UMR 5584, CNRS, Universit\'e de
Bourgogne Franche-Comt\'e, F-21000 Dijon, France}

\author{Marcus Ansorg} 
\affiliation{Theoretisch-Physikalisches Institut,
    Friedrich-Schiller-Universit\"at Jena,\\ Max-Wien-Platz 1,
          D-07743 Jena, Germany}

\date{\today}

\begin{abstract}
  We study quasi-normal modes of black holes, with a focus on resonant (or quasi-normal mode) expansions,
    in a geometric frame
  based on the use of conformal compactifications together with
  hyperboloidal foliations of spacetime. Specifically, this work extends
  the previous study of Schwarzschild in this geometric approach to
  spherically symmetric asymptotically flat black hole spacetimes,
  in particular   Reissner-Nordstr\"om. The discussion involves, first,
  the non-trivial technical developments needed to address
  the choice of appropriate hyperboloidal slices in the extended setting as well as the generalization of the
  algorithm determining the  coefficients in the expansion of the solution
  in terms of the quasi-normal modes. In a second stage, we discuss how the
  adopted framework provides a geometric insight into the origin of regularization
  factors needed in Leaver's Cauchy based foliations, as well as into the discussion
  of quasi-normal modes in the extremal black hole limit.
 \end{abstract}

\maketitle
\section{Introduction}\label{e:Intro}

Black-hole perturbation theory represents one of the cornerstones in the development of General Relativity. Initially developed in the
context of astrophysical problems, its applications nowadays are spread among many areas in gravitational physics.
Focusing on the decaying properties of propagating fields,
the time evolution of perturbative fields on a  background containing a black hole presents (after an initial
transitory) a characteristic behaviour at intermediate
 time scales: a decay in terms of a discrete set of exponentially damped oscillations, the so-called quasi-normal modes (QNMs).
Most importantly, the decay and oscillation time scales are signatures of the background spacetime.
QNMs have found a wide range of application in the settings of gravitational wave detection, mathematical relativity,
the gauge/gravity duality, string theory,
brane-world models and quantum gravity --- see e.g. \cite{Chandrasekhar83} for a classical reference and
\cite{Kokkotas99a,Nollert99,Berti:2009kk,Konoplya:2011qq}
for a revision along the past few decades.
For spacetimes whose curvature (more precisely the effective potential in the appropriate wave equation) does not decay sufficiently fast at large distances, e.g. Schwarzschild, one identifies a second behaviour in the final stages the evolution.
For late times, a power law decay --- the so-called tail decay --- may dictate the dynamics~\cite{Price72,Gundlach94a}.

In fact, the exponentially damped oscillatory decay is not a feature exclusive of linear fields propagating on black-hole spacetimes, but
rather a generic behaviour of solutions of open dissipative systems described by wave equations subject to outgoing boundary conditions.
Such concept of QNM is essentially related to the notion of resonance in scattering theory and has acquired,
in recent years, a major role in other domains. In particular, it is remarkable the synergy with
recent developments in the optical study of nano-resonators (cf. \cite{LalYanVyn17}) as well as in the mathematical literature, where they
are usually referred to as `scattering resonances'
\cite{DiaZwo17,zworski2017mathematical}.

When considering closed (compact) conservative systems, with dynamics characterised in terms of self-adjoint operators,
  the notion of normal mode provides a powerful tool to analyse the system, in particular by making use of the
  completeness of normal eigenfunctions $\psi_n(x^k)$ to expand the solutions $\Psi(t,x^k)$
\beq
\label{eq:Spectral_Decomp_2}
\Psi(t,x^k)=\sum_{n=0}^\infty \eta_n\psi_n(x^k)e ^{ i \omega_nt} \ ,
\eeq
where the amplitudes coefficients $\eta_n$ are obtained  
  from the projection of the initial data onto the complete orthonormal system of eigenfunctions by using the scalar scalar inner product
  (this is guaranteed by the so-called `spectral theorem' associated to self-adjoint operators). 
In certain respects, quasi-normal modes represent in open systems the counterpart to normal modes in closed systems.
In this sense, it is natural to pose the question about the possibility of writing  an
expression of the type (\ref{eq:Spectral_Decomp_2}) for  solutions of initial value problems associated
with linear dissipative wave equations in terms of QNM expansions. More precisely,
  it is natural to try to assess the existence of appropriate space and time scales for which 
 such dissipative solutions admit well-defined  approximations in terms
  of QNM expansions.

However, in such dissipative scenarios, the relevant differential operator associated with the wave equation is no longer self-adjoint.
In particular, this amounts to a loss of  corresponding spectral theorem, so that quasi-normal modes do not generically  constitute
a complete set. Moreover, orthogonality is also generically lost, so that even if completeness is preserved, the straightforward projection
algortihm to determine the coefficients is no longer available.

On the other hand, in the context of stationary black-hole spacetimes, the time coordinate $t$ usually employed to parametrise the dynamical evolution of the fields actually foliates the spacetime into time-constant surfaces extending between the bifurcation sphere ${\cal B}$ and spatial infinity $i^0$. Within such Cauchy foliation, the dissipative character comes about only after one introduces the correct outgoing boundary conditions in the spatial asymptotic regions. As a consequence, the eigenfunctions $\psi_n(x^k)$ present an undesirable exponentially growth near the boundaries. This property constitutes one of the main drawbacks in the understanding of the technical and conceptual issues involved in such desired `resonant expansion'   representations of the form (\ref{eq:Spectral_Decomp_2}).
  
Indeed, in a recent article~\cite{Ansorg:2016ztf}, black-hole perturbation theory on the Schwarzschild background was revisited within the geometrical framework provided by a spacetime foliation in terms of horizon-penetrating hyperboloidal slices. The authors 
argue and demonstrate numerically that if the initial data are analytical in terms of a compactified coordinate in the appropriate hyperboloidal slices, a superposition of the form 
\beq
\label{eq:VSol_spectral}
\Phi(\tau, x^k) = \sum_{n=0}^{\infty} \eta_{n} \phi_n(x^k)e^{s_{n}\tau} + \int\limits_{-\infty}^{0} \eta(s)\phi( x^k;s)e^{s\tau} ds \ ,
\eeq
can be constructed for solutions corresponding to  initial value problems of linear wave equations in the Schwarzschild spacetime. 

Note that the spectral decomposition~\ref{eq:VSol_spectral} includes not only the (discrete) quasi-normal mode expansion, but also a contribution from the continuous spectrum along the negative real line $\Re(s)<0$. This term, responsible for the late-time tail decay, results from the existence of a branch cut along  $\Re(s)<0$ in the analytical extension of the corresponding propagator operator~\footnote{\label{f:spectral}
Given the elliptic operator $P_V=-\Delta + V$ obtained from the wave equation upon a Laplace transform, QNMs are associated with the poles in the `analytical extension' of the Green's function corresponding to $P_V$ into the whole complex plane (see appendix~\ref{App:AsymExp} for a discussion in a spectral setting). For effective potentials $V$ not decaying sufficiently fast at infinity --- for instance in the Schwarzschild case --- a branch cut starting at $s=0$ appears in addition to the QNM first-order poles.}.

More specifically, a semi-analytical algorithm was developed allowing to calculate all the elements of the spectral decomposition \eqref{eq:VSol_spectral}, i.e., the QNMs $s_n$ together with the functions $\phi_n( x^k)$ and $\phi( x^k;s)$ (depending only on the structure of the wave equation) as well as the amplitudes $\eta_n$ and $\eta(s)$ (read from the initial data). The study in~\cite{Ansorg:2016ztf} led to the following conjecture~
\footnote{\label{f:resonant_expansion}The symbol $\displaystyle \sum_{n=0}^{\infty}$ in expression (\ref{eq:VSol_spectral}) must be understood in a formal sense. More precisely, the semi-analytical algorithm in~\cite{Ansorg:2016ztf} does not allow to fully settle the convergence properties of the series in~(\ref{eq:VSol_spectral}). Specifically, we do not claim here QNM (extended with tails) completeness: conjecture \ref{cnjc:SpecDecomp} must be understood rather in terms of an `asymptotic expansion'. See appendix~\ref{App:AsymExp}.} :
\begin{conjecture}
\label{cnjc:SpecDecomp}
Given analytical initial data $\Phi_0$ and $\dot \Phi_0$ for the wave equation, the spectral decomposition (\ref{eq:VSol_spectral}) holds for all $\tau>\nu$ where $\nu$ is the mutual growth rate of $|\eta_n\phi_n|$ and $|\eta(s)\phi(s)|$, i.e., the quasinormal mode and branch cut excitation coefficients. 
\end{conjecture}
Even though the generic ideas introduced in~\cite{Ansorg:2016ztf} are aimed at a broader context, the semi-analytical algorithm developed there is tailored for the Schwarzschild spacetime. Nevertheless, the well-defined time-scale $\nu$ introduced in conjecture \ref{cnjc:SpecDecomp} is used in the context of of the AdS/CFT correspondence~\cite{Ammon:2016fru}. Yet, some technical calculations had to be addressed by a different method.

\medskip

In this paper, we extend the formulation of the spectral decomposition~\ref{eq:VSol_spectral} in the framework of hyperboloidal slices for fields propagating on a spherically symmetric, not necessarily vacuum, asymptotically flat spacetimes. Despite the restriction to spherically symmetric solutions, such first step generalisation already displays several features that enlightens the theoretical discussion and introduces technical challenges for the algorithm used in the calculation of all the elements within the spectral decomposition \eqref{eq:VSol_spectral}. Specifically, our main focus lies on the Reissner-Nordstr\"om case. Yet, one could envisage further scenarios --- for instance, solutions describing central black hole and a surrounding shell composed out of collision-less Einstein-Vlasov-matter~\cite{Andreasson:2005qy} or, relaxing the conditions on dimensionality and asymptotic structure, higher-dimensional spacetimes~\cite{Emparan:2008eg,Horowitz2012} and asymptotically AdS spacetimes~\cite{Ammon:2015wua,Nastase2015} --- for which the technicalities of the methods presented here could be appropriately adaptable.   

In the first part, this article is framed in the study of the geometrical aspects of the formalism as well as in the technical generalisation of the semi-analytical algorithm developed in~\cite{Ansorg:2016ztf} for the calculation of all elements in \eqref{eq:VSol_spectral}. In particular, while reviewing the construction of the spatially compactified hyperboloidal slices in a generic context, we explicitly identify in section \ref{sec:QNMFilter} the gauge degrees of freedom associated with the hyperboloidal foliation and the conformal compactification. This allows us to introduce a gauge optimally adapted to the study of black-hole perturbation theory within the present hyperboloidal framework, which we refer to as the {\em minimal gauge}. Then, we discuss in section \ref{sec:SpectralDec} the formal aspects for the construction of the desired spectral decomposition~\eqref{eq:VSol_spectral}. For this task, we first identify the correct conformally invariant wave equation to be solved. Next, a Laplace transform is applied to the wave equation in question and a spatial differential equation arises with an inhomogeneity determined by the initial data. This system is parametrised by the complex Laplace parameter $s$. The spectral decomposition~\ref{eq:VSol_spectral} is then obtained by the deformation of the inverse Laplace path integral which collects the contribution coming from the operator's poles and branch cut in the complex $s-$plane. Finally, section~\ref{sec:Algorithm} describes in details the generalised algorithm in the frequency domain to calculate the various ingredients of the spectral representation~(\ref{eq:VSol_spectral}). The algorithm is based on the expansion of the relevant functions into a Taylor series. Thus, the corresponding spatial differential equation gives rise to a recurrence relation determining the coefficients of the Taylor expansion. While in ~\cite{Ansorg:2016ztf}, the algorithm was restricted to a 3-term recurrence relation, we generalise it here into an arbitrary (but finite) $(m+2)$-terms recurrence relation. 

In a second stage, we discuss extensively the application of the formalism to the particular case of the Reissner-Nordstr\"om solution. In particular, we show in section~\ref{sec:RN} that the minimal gauge leads naturally to two choices for the conformal representations of the spacetime. It turns out that the two geometries have different spacetime limits in the extremal case~\cite{Paiva:1993bv,Geroch:1969ca}. When approaching extremality, one gauge leads to the usual extremal Reissner-Nordstr\"om black hole~\cite{HawEll73} whilst the second one shows a discontinuous transition to the near-horizon geometry~\cite{Carroll09,Bengtsson:2014fha}  (see \cite{Kunduri2013} for a review on near-horizon geometries) described by the Bertotti-Robinson metric~\cite{Bertotti59,Robinson59}. Interestingly, we actually observe that the gauge leading to the Bertotti-Robinson spacetime represents the precise counterpart, in the present geometrical framework, to the treatment introduced by Leaver~\cite{Leaver90}. Indeed, the geometrical approach based on conformally compactified spacetimes foliated by hyperboloidal slices straightforwardly recovers all factors introduced in~\cite{Leaver90} accounting for the correct boundary conditions leading to the QNMs.

We work with units such that the speed of light as well as Newton's constant of gravity are unity, $c = G = 1$.
\section{The geometrical framework}\label{sec:QNMFilter}
We begin by reviewing the construction of hyperboloidal slices in a spherically symmetric black-hole spacetimes and introducing a gauge which is best adapted to the present discussion of black-hole perturbation theory.

\subsection{Schwarzschild coordinates}
We start with a stationary (actually static) spherically symmetric line element in the form given by Schwarzschild coordinates $\{t,r,\theta, \varphi\}$
\beq
\label{eq:SpheSymMetric}
\d s^2 = -f(r) \d t^2 + f(r)^{-1}\d r^2 + r^2 \d\omega^2 \ ,
\eeq
with $d\omega^2 = \d\theta^2 + \sin(\theta)^2 \d\varphi^2$ the metric of the unit $2-$sphere. We assume the function $f(r)$ satisfies the following conditions:
\ben[(i)]
\item Asymptotic flatness expansion: 
\beq
f(r) \sim 1 - \frac{2M}{r} + o\left(\frac{1}{r}\right) \ ;
\eeq
\item Killing (black-hole) horizon at $r_{\cal H}$:
\beq
f(r_{\cal H}) = 0 \ ;
\eeq
\item $f(r)$ is a polynomial in $1/r$.
\een
While assumptions (i)-(ii) are natural for an asymptotically flat black-hole spacetime, the assumption (iii)
is motivated by the form of well-known spherically symmetric exact solutions. The tortoise coordinate $r^*=r^*(r)$ is fixed (up to a constant) by
\beq
\label{eq:TortoiseCoord}
\dfrac{\d r^*}{\d r}= \frac{1}{f(r)} \ .
\eeq
Along the slices of constant coordinate time $t$, the horizon surface $r=r_{\cal H}$ ($r^* \rightarrow -\infty$) corresponds to the bifurcation sphere ${\cal B}$, while $r \rightarrow +\infty$ ($r^* \rightarrow +\infty$) is the spatial infinity --- see fig.~\ref{fig:CoordinatesPenrose}.

\subsection{Ingoing Eddington-Finkelstein coordinates}\label{sec:EFCoordinates}
While in~\cite{Zenginoglu:2007jw,Zenginoglu:2011jz} the hyperboloidal coordinates follows directly from the coordinates $\{t,r,\theta,\varphi\}$,  here we first consider an intermediate step that enforces horizon-penetrating slices via the ingoing Eddington-Finkelstein coordinates. Besides, we can already at this stage discuss the compactification of the spacetime.
  
Spatially compactified and dimensionless ingoing Eddington-Finkelstein coordinates $\{ \bar{v},\sigma,\theta, \varphi\}$ are given by
\beq
\label{eq:EFCoord}
r = \lambda \frac{\rho(\sigma)}{\sigma},  \quad
t = \lambda \bar{v} - r^*(r(\sigma)) \ ,
\eeq
with  $\lambda$ a length scale of the spacetime, typically related here to the event horizon radius $r_{\cal H}$. The line element reads
\beq
\label{eq:EF_Metric}
\d s^2 =\frac{\lambda^2}{\sigma^2}\left[  -\sigma^2 F \d \bar{v}^2 - 2\beta \d \bar{v}\d \sigma + \rho^2 \d \omega^2 \right] \ ,
\eeq
with $\rho$, $F$ and $\beta$ functions of the coordinate $\sigma$. In particular, $F(\sigma)=f(r(\sigma))$ is the metric function, while
\beq 
\label{eq:beta}
\beta(\sigma) = \rho(\sigma) - \sigma \rho'(\sigma) \ ,
\eeq
is the radial component of the shift and therefore corresponding to a gauge freedom in the choice of the coordinate $\sigma$ --- see discussion in section \ref{sec:PolynomGauge}.

Finally, $\rho(\sigma)$ is the areal radius on the conformal representation of the spherically symmetric spacetime, whose conformal metric is given by
\beq
\label{eq:ConfMetric_1}
d\tilde{s}^2 = \Omega^2ds^2, \quad \Omega = \frac{\sigma}{\lambda} \ .
\eeq
As such, we impose $\rho$ to be a regular function on its domain adopting non-vanishing positive values. Moreover, we assume $\rho(\sigma)$ to be such that $\beta(\sigma)>0$ --- see eq.\eqref{eq:beta}.

The outgoing and ingoing null vectors in the conformal spacetime (normalized as $\tilde{g}_{ab}\tilde{l}^a\tilde{k}^b=-1$)
  are given, respectively, by
\beq
\label{eq:ConfNullVectors}
\tilde{l}^a = \zeta^{-1} \left[ \delta^a_{\bar{v}} - \frac{\sigma^2F}{2\beta} \delta^a_\sigma\right], \quad \tilde{k}^a = \frac{\zeta}{\beta} \delta^a_\sigma \ .
\eeq
The free boost-parameter $\zeta$ will be fixed in the next section. 

\subsection{Hyperboloidal coordinates}\label{sec:HypCoord}
We finally introduce the hyperboloidal coordinates $\{\tau, \sigma, \theta, \varphi\}$ via the height technique ~\cite{Zenginoglu:2007jw}
\beq
\label{eq:HyperCoord}
\tau = \bar{v} + h(\sigma) \ .
\eeq
In the new coordinates, the conformal line element is
\bea
\label{eq:ConfMetric_2}
\d \tilde{s}^2 = &-&\sigma^2 F \d\tau^2 + h'\left[ 2\beta - \sigma^2 F h'\right]\d\sigma^2 \nn \\
&-& 2\left[ \beta - \sigma^2 F h'\right] \d\tau\d\sigma + \rho^2\d\omega^2 \ .
\label{eq:Metric_HyperbCoord}
\eea
The height function $h(\sigma)$ must be specified in such a way that the spacelike surfaces $\tau=$ constant foliates future null infinity. This property is guaranteed by requiring $\tau$ to be a good parameter of the ingoing null vector, that is
\beq
\label{eq:ScriCond}
\tilde{k}^a\partial_a \tau = 1 \ .
\eeq 

\subsubsection{The height function}\label{sec:HeightFunc}

In the coordinates $\{\tau,\sigma,\theta, \varphi\}$, the components of the ingoing/outgoing null vectors are respectively
\bea
\tilde{k}^a &=&   \delta^a_\tau + \frac{1}{h'}\delta^a_\sigma. \\
\tilde{l}^a &=& \frac{h'}{2\beta^2} \left(2\beta - \sigma^2 F h'\right) \delta^a_\tau - \frac{\sigma^2F}{2\beta^2}h' \delta^a_\sigma.
\eea
Consistently with \eqref{eq:ScriCond}, the boost parameter $\zeta$ present in \eqref{eq:ConfNullVectors} has been fixed here such that $\tilde{k}^\tau = 1$. Then, the condition 
\beq
\lim_{\sigma\rightarrow 0}\tilde{k}^a  = \delta^a_\tau \Rightarrow  \lim_{\sigma\rightarrow 0} \frac{1}{h'} = 0 \ .
\eeq
guarantees that $\sigma=0$ is a null surface, corresponding actually to {\em future} null infinity (as opposed to the {\em past} character arising from the ingoing Eddington-Finkelstein coordinates is section~\ref{sec:EFCoordinates}).

Besides, we want to ensure that the components of the vector $\tilde{l}^a$ remain finite as $1/h' \rightarrow 0$. 
\begin{figure*}[]
\begin{center}
\includegraphics[width=8.0cm]{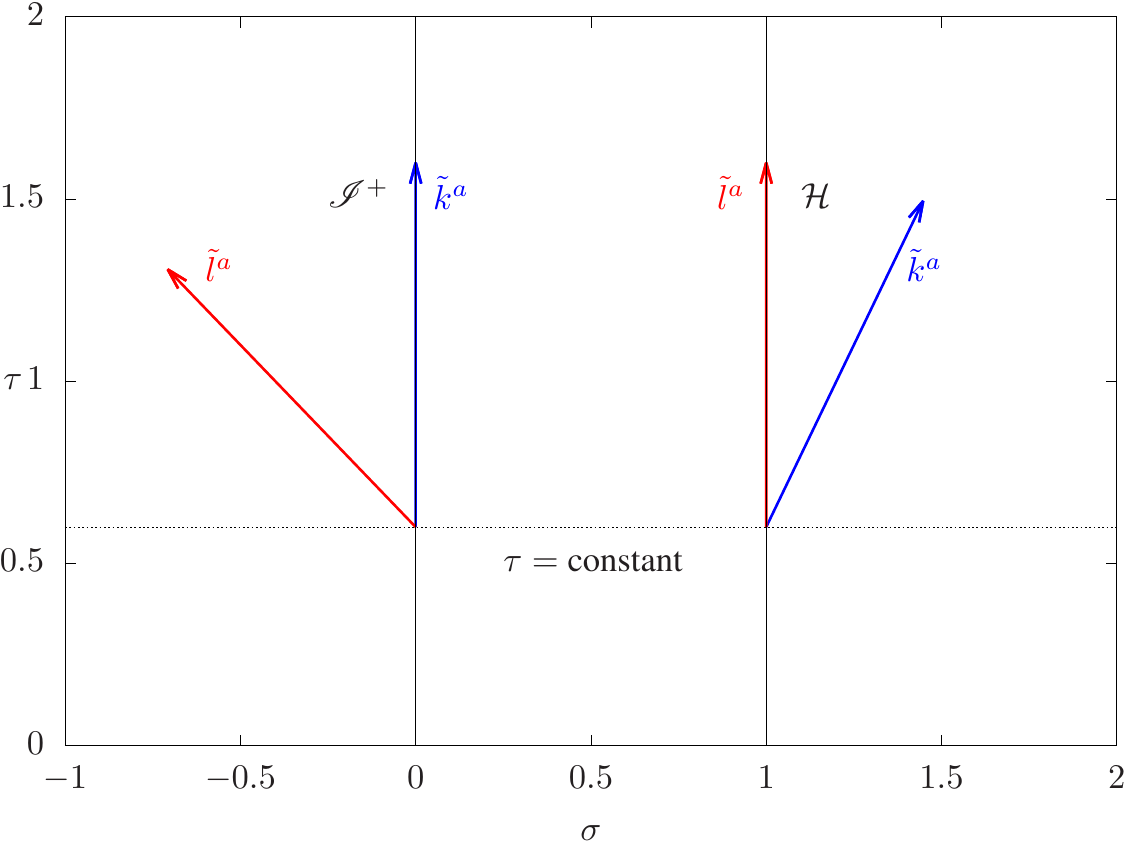}
\includegraphics[width=6.0cm]{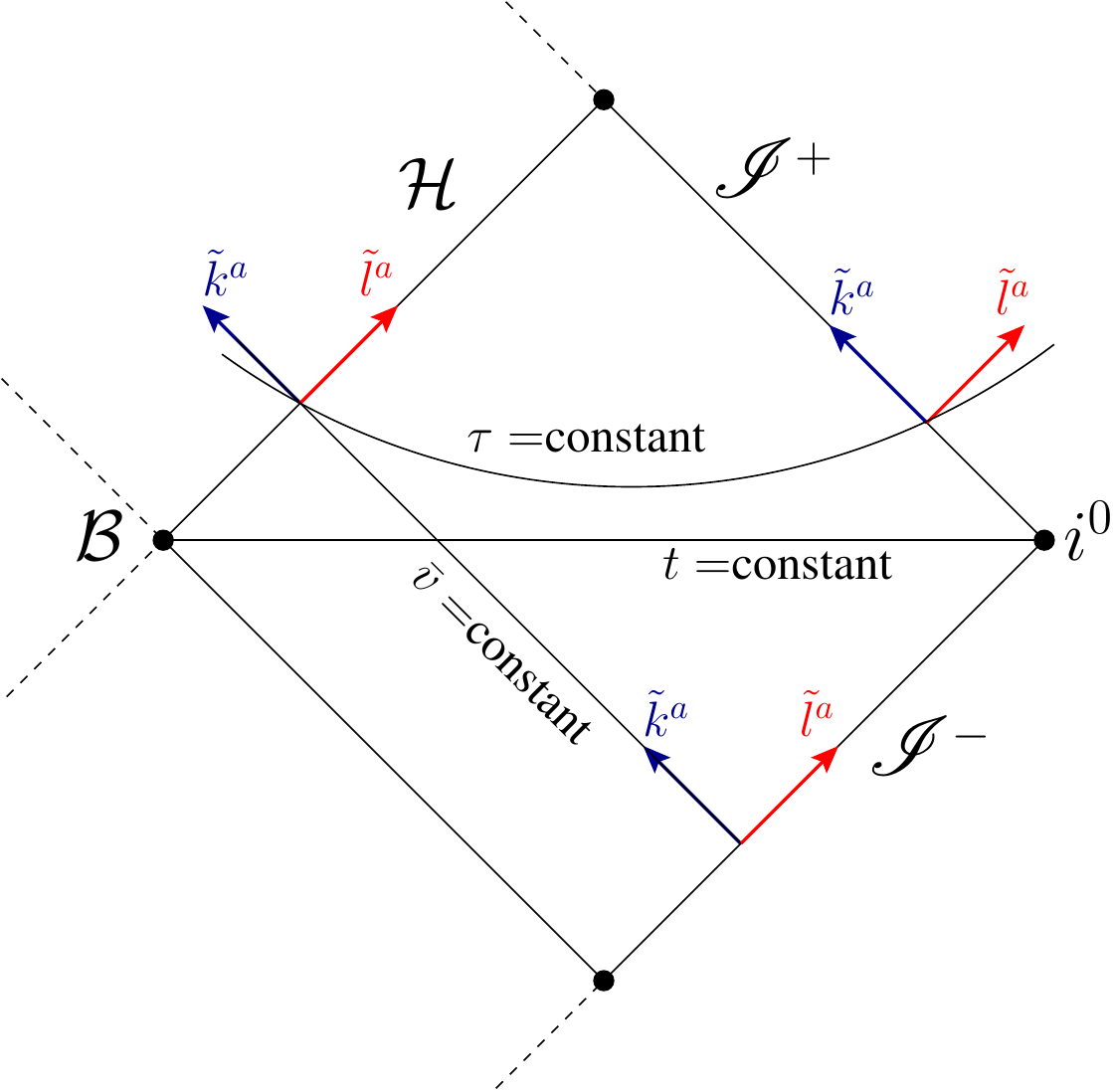}
\end{center}
\caption{Left panel: conformal null vectors on the spatially compactified spacetime with metric $\tilde{g}_{ab}$ in terms of the hyperboloidal coordinates $\{\tau, \sigma\}$. At $\sigma=0$ and $\sigma=1$ the light-cones point outwards the domain $\sigma\in[0,1]$. Such surfaces corresponds, respectively, to future null infinity and the black-hole horizon. Right panel: Carter-Penrose conformal diagram for the external region of an asymptotically flat, stationary black-hole spacetime. Time constant surface in the initial Schwarzschild coordinate $\{t,r\}$ extends between the bifurcation sphere ${\cal B}$ ($r=r_{\cal H}$) and spatial infinity $i^0$ ($r\rightarrow \infty$). Ingoing Eddington-Finkelstein coordinates $\{\bar{v}, \sigma\}$ ensure that $\sigma=\sigma_{\cal H}$ reaches the event horizon ${\cal H}$. Yet, $\sigma=0$ corresponds to past null infinity, where the ingoing null vector $\tilde{k}^a$ points to the interior of the domain. In the hyperboloidal coordinates $\{\tau, \sigma\}$, the height function ensures that $\sigma=0$ corresponds to future null infinity, with $\tilde{k}^a \propto \delta^a_\tau$.
}
\label{fig:CoordinatesPenrose}
\end{figure*}
Let us first assume the generic expansion
\beq
\label{eq:beta_expansion}
\beta(\sigma) = \beta_0 + \beta_1\sigma+ {\cal O}(\sigma^2) \ .
\eeq
The presence of the term $\beta_1$ in this expansion is, however, undesirable. Indeed, according to its definition \eqref{eq:beta}, the areal radius $\rho(\sigma)$ assumes the form
\beq
\rho(\sigma) = \beta_0 + \rho_1\sigma - \beta_1 \sigma \ln\sigma + {\cal O}(\sigma^2) \ .
\eeq
Thus, the condition $\beta_1 = 0$ is required for eliminating the logarithmic singularity at $\sigma=0$. It is also convenient to identify $\beta_0 = \rho_0$.

Then, we recall the assumption (i) for the function $f(r)$ which, together with eq.~\eqref{eq:EFCoord}, gives rise to the expansion
\beq
F(\sigma) = 1 - \frac{2M}{\lambda\rho_0}\sigma + {\cal O}(\sigma^2) \ .
\eeq
As $\sigma\rightarrow 0$, a finite value  for the component $\tilde{l}^\tau$ is obtained for
\beq
\label{eq:dh}
h'(\sigma) = \frac{2\rho_0}{\sigma^2} \left[ 1 + \frac{2M}{\rho_0\lambda}\sigma\right] + {\cal O}(1) \ .
\eeq
This result ensures the regularity of $\tilde{l}^\sigma$ as well. Integrating eq.~\eqref{eq:dh} leads to
\beq
h(\sigma) =  -2\rho_0\left[ \frac{1}{\sigma} - \frac{2M}{\lambda\rho_0}\ln\sigma\right] + A(\sigma) \ .
\eeq
The regular free function $A(\sigma)$ represents a freedom in the choice of the hyperboloidal foliation, to be fixed in the next section.

Finally, the condition $ \tilde{\nabla}_a \tau \tilde{\nabla}^a \tau  < 0$ (for having spacelike surfaces $\tau=$ constant) imposes
\beq
\label{eq:Spacelike_Condition}
0 < \sigma^2 h' < \frac{2\beta}{F} \ .
\eeq

\subsubsection{The minimal gauge}\label{sec:PolynomGauge}
In the previous sections, $\beta(\sigma)$ and $A(\sigma)$ were identified as gauge degrees of freedom. The former related to the definition of compact coordinate $\sigma$ --- and thus to the choice of the conformal representation of the spacetime \eqref{eq:ConfMetric_1} --- while the latter distinguishes different hyperboloidal foliations.

Given the freedom in $\beta$, we choose it be constant leading to a gauge where
\beq
\label{eq:Fix_rho}
\beta(\sigma) = \rho_0 \Rightarrow \rho(\sigma) = \rho_0 + \rho_1 \sigma \ .
\eeq

In terms of the coordinate $\sigma$, it is very convenient to fix the event horizon to the value $\sigma_{\cal H} = 1$. This particular choice constraint the relation between $\rho_0$ and $\rho_1$ to
\beq
\rho_0 = \frac{r_{\cal H}}{\lambda} - \rho_1 \ .
\eeq  

The freedom provided by $A(\sigma)$ can be used to specify further geometrical properties of the hyperboloidal slices, such as a constant mean curvature~\cite{Zenginoglu:2007jw,brill:2789,Malec:2009hg,Tuite:2013hza}. By allowing an angular dependence on the function $A$, one can even depart from spherical symmetry~\cite{Schinkel:2013tka,Schinkel:2013zm} in the coordinate description. We restrict ourselves to the simplest case
\beq
\label{eq:Fix_A}
A(\sigma) = 0 \ .
\eeq

We refer to the choices \eqref{eq:Fix_rho} and \eqref{eq:Fix_A} as the {\em minimal gauge}. Indeed, the degrees of freedom from the radial compactification and from the hyperboloidal foliation are reduced to a minimum and they consist merely in fixing the scaling parameter $\lambda$ and the value $\rho_1$. All relevant quantities in the line element \eqref{eq:Metric_HyperbCoord} are determined essentially by the properties of the function $F(\sigma)$. In particular, recalling assumption (iii), all components of the metric tensor become polynomials in $\sigma$.

\subsubsection{Comparison with Zenginoglu's scri-fixing prescription}
We finish this section by comparing the approach/notation we used for the height function technique against the original one introduced in~\cite{Zenginoglu:2007jw,Zenginoglu:2011jz}. First of all, the height function $h$ here follows from \eqref{eq:HyperCoord} --- i.e. a transformation from the ingoing Eddington-Finkelstein coordinate $\bar{v}$. Zenginoglu's height function $h_{\rm Z}$ is defined~\cite{Zenginoglu:2007jw,Zenginoglu:2011jz} in terms of the original Schwarzschild coordinate $t$. Taking into account the correct signs and dimension re-scalings, we have
\beq
h_{\rm Z} = - \left( r^* + \lambda\, h \right).
\eeq
In~\cite{Zenginoglu:2011jz}, the geometric framework is discussed in terms of the boost function
\beq
H:= \frac{\d h_{\rm Z}}{\d r^*}  = - 1 + \frac{\sigma^2 F}{\beta} h' \ .
\eeq
In particular, eq.~(2) in \cite{Zenginoglu:2011jz} specifies the conditions for the time constant slices to be horizon-penetrating and hyperboloidal. Due to our preliminary step in terms of ingoing Eddington-Finkelstein coordinates, all conditions imposed on $H$ as $r \rightarrow r_{\cal H}$ are automatically satisfied.

In a more general formulation, the work \cite{Zenginoglu:2007jw} discusses matching conditions that smoothly connect the asymptotic behaviour of the hyperboloidal slices to Cauchy surfaces on the interior of the spacetime~\cite{Zenginoglu:2010cq}. Moreover, the areal radius  $\rho$ of the conformal spacetime is regarded as the new compact coordinate and the conformal factor $\Omega(\rho)$ is a free function. Here, we choose a compact coordinate naturally adapted to the conformal factor via $\sigma = \lambda \Omega$, whereas the areal radius $\rho(\sigma)$ is a free function fixing the gauge. The motivation in \cite{Zenginoglu:2007jw} to introduce the matching conditions and to let a free conformal factor $\Omega(\rho)$ comes from the objective of applying the hyperboloidal approach to solve numerically the full non-linear Einstein's equation\footnote{See \cite{Hubner:1999th,Frauendiener2002,Bardeen:2011ip,Rinne:2009qx,Rinne:2013qc,Vano-Vinuales:2014koa,Morales:2016rgt,Hilditch:2016xzh,Vano-Vinuales:2017qij} for non-linear time evolutions in the context of hyperboloidal foliations.}. 

In the context of perturbation theory, however, where the background spacetime is known {\it a priori}, adapting the radial coordinates to the conformal factor and fixing the function $A(\sigma)=0$ simplifies significantly the equations under study. Thus, the {\em minimal gauge} reduces the complexity of the problem to a minimum, where $F(\sigma)$ is the only relevant function. In fact, the minimal gauge was employed beyond spherical symmetry in~\cite{Macedo:2014bfa} in the numerical evolution of the Teukolsky equation in the Kerr spacetime. Even though the qualitative results do not differ from other studies (e.g.~\cite{Zenginoglu:2008uc,Zenginoglu:2009ey,Racz:2011qu,Jasiulek:2011ce,Harms:2013ib}), the analytical structure of the spacetime metric and the wave equation becomes much simpler.

\section{Spectral decomposition for Black-hole perturbations}\label{sec:SpectralDec}

\subsection{Wave equation}\label{sec:WaveEquation}
\subsubsection{Master wave equation}
\label{sec:master_wave_eq}
Black-hole perturbation theory is usually formulated in the Schwarzschild coordinate system $\{t,r,\theta,\phi\}$. Introducing dimensionless coordinates $\bar{t} = {t}/{\lambda}$ and $x = {r^*}/{\lambda}$, one typically studies the wave equation
\beq
\label{eq:WaveEq}
-\Psi_{,\bar{t}\bar{t}} + \Psi_{,xx} - {\cal P}\Psi = 0 \ .
\eeq
The structure of equation \eqref{eq:WaveEq} is generic in a large class of fields propagating on a spherically background. For instance, scalar, electromagnetic or gravitational perturbations differ only in the forms for the potential ${\cal P}$. A solution to this equation is uniquely determined once we impose the initial data 
\beq
\label{eq:ID_Cauchy}
\Psi_0(x) = \Phi(0,x), \quad \dot\Psi_0(x) = \Psi_{,t}(0,x) \ ,
\eeq
together with (generally time-dependent) boundary conditions at inner $x=x_{\rm in}(t)$ and outer $x=x_{\rm out}(t)$ hypersurfaces. In particular, in the QNM setting, outgoing boundary conditions must be imposed at $x_{\rm out} \rightarrow +\infty$ and at $x_{\rm in} \rightarrow -\infty$. Note that in this particular Cauchy slicing foliated by $\bar{t} =$ constant, the limits $x_{\rm out} \rightarrow +\infty$ and at $x_{\rm in} \rightarrow -\infty$ correspond to spatial infinity and the bifurcation sphere, respectively.

The coordinate changes  \eqref{eq:EFCoord} and \eqref{eq:HyperCoord} can be applied directly to eq.~\eqref{eq:WaveEq}. Let us introduce the factorization
\beq
\label{eq:Potential2}
{\cal P}(r) = \frac{\lambda^2}{r^2}f(r){\rm P}(r) \ .
\eeq
Firstly, the introduction of the compact ingoing Eddington-Finkelstein coordinates \eqref{eq:EFCoord} leads to 
\bea
\label{eq:EF_WaveEq}
-2 \, {\rm v}_{,\bar{v}\sigma} + \left[\frac{\sigma^2 F}{\beta}\,{\rm v}_{,\sigma} \right]_{,\sigma}  - \frac{\beta}{\rho^2}{\rm P}\,{\rm v} = 0 \ ,
\eea
with ${\rm v}(\bar{v}, \sigma) = \Psi(t(\bar{v}, \sigma), r(\sigma) )$.

When passing from \eqref{eq:WaveEq} to \eqref{eq:EF_WaveEq}, one can factor out the quantity $\sigma^2 F(\sigma)$, which vanishes at null infinity $\sigma=0$ and at the horizon $F(\sigma_{\cal H}) = 0$. Yet, another factor in the form $\sigma^2 F(\sigma)$ is still present as the coefficient of ${\rm v}_{,\sigma\sigma}$ and the equation \eqref{eq:EF_WaveEq} degenerates at $\sigma = 0$ and $\sigma=\sigma_{\cal H}$. 

As a consequence of this latter feature, regularity conditions must be taken into consideration when solving the equation. We remind that the surface $\sigma=0$ corresponds here to the {\em past} null infinity, with the ingoing null vector $\tilde{k}^a$ pointing towards the interior of the domain $\sigma>0$ --- see eq.~\eqref{eq:ConfNullVectors}  (cf. also the right panel in figure \ref{fig:CoordinatesPenrose}). A unique solution to the wave equation
with initial data prescribed at the hypersurface $\bar{v}=0$ requires therefore not only data at such characteristic hypersurface, but also information at the hypersurface $\sigma=0$.

The identification of $\sigma=0$ with {\em future} null infinity comes only after the introduction of the hyperboloidal slices \eqref{eq:HyperCoord},
in which the wave equation reads
\bea
\label{eq:Hyper_WaveEq}
&& -2h'\left[1 - \frac{\sigma^2 F}{2\beta} h' \right] \, V_{,\tau \tau} - 2 \left[1 - \frac{\sigma^2 F}{\beta} h' \right] \, V_{,\tau \sigma} + \frac{\sigma^2 F}{\beta} V_{,\sigma\sigma} \nn \\
&& + \left[ \frac{\sigma^2 F}{\beta} \right]'  V_{,\sigma} +  \left[ \frac{\sigma^2 F h'}{\beta} \right]' V_{,\tau}
 - \frac{\beta}{\rho^2}{\rm P}\,V = 0 \ ,
\eea
with $V(\tau, \sigma) = {\rm v} (\bar{v}(\tau, \sigma), \sigma)$.

Indeed, according to the discussion in section \ref{sec:HypCoord}, the height function $h(\sigma)$ ensures that $\sigma=0$ corresponds
to {\em future} null infinity.  Then, despite the rather complicated form of Eq. (\ref{eq:Hyper_WaveEq}), it is straightforward to see
that the term proportional to $V_{,\sigma\sigma}$ still vanishes at both $\sigma=0$ and $\sigma=\sigma_{\cal H}$, i.e. at future null infinity and at the black hole horizon.
  
At the boundaries $\sigma =0$ and $\sigma=\sigma_{\cal H}$, the characteristics of the system never point inwards the domain $\sigma\in[0,\sigma_{\cal H}]$
 (cf. the left panel in figure \ref{fig:CoordinatesPenrose}). Therefore, the prescription of initial data 
\beq
\label{eq:ID_hyperboloid}
V_0(\sigma)=V(0,\sigma) \quad {\rm and}  \quad \dot{V}_0(\sigma) = V_{,\tau}(0,\sigma) \ ,
\eeq
is sufficient to fix uniquely the time solution for $\tau>0$ and no further boundary conditions are required.

Note however, that this procedure is based on a direct change of coordinates applied to the master equation \eqref{eq:WaveEq}. A more systematic geometrical approach should be based on a coordinate independent formulation according to the conformal transformation of the spacetime $g_{ab} = \Omega^{-2} \tilde{g}_{ab}$. In the next section we discuss this procedure for the scalar field. A complete conformal treatment of electromagnetic and gravitational perturbations is beyond the scope of this work.

\subsubsection{Conformal wave equation}\label{sec:ConformalWaveEq}
We start with the conformally invariant wave equation~\cite{Wald84} for a massless scalar field $U(t,r,\theta,\varphi)$ propagating in the background provided by the metric $g_{ab}$
\beq
\Box U - \dfrac{R}{6} U = 0 \ ,
\eeq
with $R$ the Ricci scalar associated to the metric $g_{ab}$. Here, we are particularly interested in spacetimes satisfying $R=0$.

Note that eq.~\eqref{eq:WaveEq} is recovered with the decomposition\footnote{\label{f:notation}For simplicity, the indices ${}_{\ell,m}$ were absent in eq.~\eqref{eq:WaveEq}, \eqref{eq:EF_WaveEq} and \eqref{eq:Hyper_WaveEq}.}  
\beq
\label{eq:ReScalScalarField}
U(t,r,\theta,\varphi) = \sum_{\ell, m} \frac{\Psi_{\ell,m}(t,r)}{r} Y_{\ell,m}(\theta, \varphi) \ .
\eeq
In particular, the potential is given by
\beq
{\rm P}(r) = \ell(\ell+1) + r\,f'(r) \ .
\eeq

In terms of the conformal metric $\tilde{g}_{ab}$, the conformally re-scaled scalar field $\tilde{U}$ satisfies
\beq
\label{eq:ConfScalarFieldEq}
\tilde{\Box} \tilde{U} -\dfrac{\tilde{R}}{6}\tilde{U} =0\ , \quad  \tilde{U} = \Omega^{-1} U \ ,
\eeq
with 
\bea
\tilde{R} = -\dfrac{6}{\Omega}\left[  \tilde{\Box} \Omega - 2 \frac{\tilde\nabla_a\Omega\tilde\nabla^a\Omega}{\Omega} \right] = - \frac{6\sigma}{\beta\rho^2} \left[ \dfrac{\rho^2 F}{\beta} \right]' \ ,
\eea 
the Ricci scalar associated with the metric $\tilde{g}_{ab}$. 
With the decomposition 
\beq 
\label{eq:Utilde_angular}
\tilde{U}(\tau, \sigma, \theta, \varphi) = \sum_{\ell, m} \Phi_{\ell,m}(\tau,\sigma) Y_{\ell,m}(\theta, \varphi) \ ,
\eeq 
we obtain
\bea
\label{eq:ConformalWaveEq}
&& -{2h'}\left[1 - \frac{\sigma^2 F}{2\beta} h' \right] \, \Phi_{,\tau \tau} - {2} \left[1 - \frac{\sigma^2 F}{\beta^2} h' \right] \, \psi_{,\tau \sigma} + \frac{\sigma^2 F}{\beta} \Phi_{,\sigma\sigma} \nn \\
&& +\dfrac{1}{\rho^2} \left[ \frac{\sigma^2 \rho^2 F}{\beta} \right]'  \Phi_{,\sigma} 
- \dfrac{1}{\rho}  \left[ 2\rho' - \dfrac{ \left( {\sigma^2 \rho^2 F h'}/{\beta}\right)'}{\rho} \right] \Phi_{,\tau} \nn \\
&& - \frac{\beta}{\rho^2}\tilde{\cal P}\, \Phi = 0 \ .
\eea
Here, the potential reads
\beq
\label{eq:ConfPotential}
\tilde{\cal P} = \ell(\ell+1) - \frac{\sigma}{\beta} \left[ \dfrac{\rho^2 F}{\beta} \right]' .
\eeq

Inserting eqs.~\eqref{eq:ReScalScalarField} and \eqref{eq:Utilde_angular} into \eqref{eq:ConfScalarFieldEq} --- and then using \eqref{eq:EFCoord} and \eqref{eq:ConfMetric_1} --- the conformally re-scaled scalar field $\Phi$ relates to $V$ via
\beq
\label{eq:RelScalarField}
V(\tau, \sigma)= \rho(\sigma)\Phi(\tau, \sigma) \ .
\eeq
If $\rho(\sigma) =$ constant, there is no distinction between changing the variable of the original master equation \eqref{eq:WaveEq} into \eqref{eq:Hyper_WaveEq}, as done in the previous subsection \ref{sec:master_wave_eq}, and writing the conformally invariant wave equation \eqref{eq:ConformalWaveEq}. Otherwise, the wave operator differs. In particular, in the minimal gauge\footnote{Note that the whole discussion in subsection \ref{sec:master_wave_eq} is independent of the gauge choice for $\rho=\rho(\sigma)$
  and $h=h(\sigma)$. Eqs.~\ref{eq:EF_WaveEq} and \ref{eq:Hyper_WaveEq}, as well as \eqref{eq:ConformalWaveEq},  apply therefore in the generic case, and not only in the {\em minimal gauge}.}, $\rho(\sigma)$ is an affine function of $\sigma$ (cf. Eq. (\ref{eq:Fix_rho})). In this setting, we privilege the conformally invariant wave equation as the natural choice, due to its geometrical formulation.

Deriving a conformally invariant master equation in a more generic setup including, for instance, electromagnetic and gravitational perturbation should be the topic of future works. Yet, we expect to relate a conformally re-scaled perturbation $\Phi(\tau, \sigma)$ with the well-known formulation in terms of a field $\Psi(\bar{t},x)$ satisfying a equation in the form of eq.~\eqref{eq:WaveEq} via a re-scaling
\beq
\label{eq:ConformalReScal}
\Psi(t(\tau,\sigma), r(\sigma)) \equiv V(\tau, \sigma)= \rho(\sigma)^n\Phi(\tau, \sigma) \ ,
\eeq
with $n$ depending on the type of field.

We apply this assumption to eq.~\eqref{eq:Hyper_WaveEq} in order to obtain a general form for the ``conformal'' wave equation
\bea
\label{eq:ConformalWaveEq_General}
&& -2h'\left[1 - \frac{\sigma^2 F}{2\beta} h' \right] \, \Phi_{,\tau \tau} 
- 2 \left[1 - \frac{\sigma^2 F}{\beta^2} h' \right] \, \Phi_{,\tau \sigma} 
+ \frac{\sigma^2 F}{\beta} \Phi_{,\sigma\sigma} \nn \\
&& +\dfrac{1}{\rho^{2n}} \left[ \frac{\sigma^2 \rho^{2n} F}{\beta} \right]'  \Phi_{,\sigma} 
-   \left[ 2n\frac{\rho'}{\rho} - \dfrac{ \left( {\sigma^2 \rho^{2n} F h'}/{\beta}\right)'}{\rho^{2n}} \right] \Phi_{,\tau} \nn \\
&& - \frac{\beta}{\rho^2}\tilde{\cal P}\, \Phi = 0 \ .
\eea
While the potential in eq.~\eqref{eq:ConfPotential} has a geometrical meaning in terms of the conformal Ricci scalar $\tilde{R}$, here we simply define it as
\beq
\label{eq:ConfPotential_General}
\tilde{{\cal P}} = {\rm P} - \dfrac{\rho^{2-n}}{\beta}\left[ \dfrac{\sigma^2F}{\beta} \left( \rho^n \right)'  \right]' \ .
\eeq

Eq.~\eqref{eq:RelScalarField} provides naturally the value $n=1$ to scalar perturbations. In sec.~\ref{sec:RN} we study a concrete example of electromagnetic and gravitational perturbations on the Reissner-Nordstr\"om spacetime. There, we identify the value $n=-1$ for such fields. It is tempting to speculate that the scaling $n=-1$ should follow naturally if one formulates {\it ab initio} the master equation for electromagnetic and gravitational perturbations within the conformal representation of the spacetime.

\subsection{Laplace Transformation}\label{sec:LapTrans}
We now proceed as in~\cite{Ansorg:2016ztf} and introduce the Laplace transformation
\beq
\label{eq:LaplaceTransform}
\hat{\Phi}(\sigma;s) := {\cal L}[\Phi(\tau,\sigma)](s) = \int_0^{\infty} e^{-s\tau}\Phi(\tau, \sigma)\d\tau, \  \Re(s)>0.
\eeq 
Applying the Laplace transformation to eq.~\eqref{eq:ConformalWaveEq_General} leads to the ordinary differential equation (ODE)
\beq
\label{eq:ODE}
{\mathbf A}(s) \hat{\Phi}(s) = B(s) \ ,
\eeq 
with 
\bea
\label{eq:ODE_OperatorA}
&{\mathbf A}(s) = \dfrac{\sigma^2 F}{\beta} \partial_{\sigma\sigma}
+ \bigg( \dfrac{1}{\rho^{2n}} \left[ \frac{\sigma^2 \rho^{2n} F}{\beta} \right]' - 2s\left[1 - \frac{\sigma^2 F}{\beta^2} h' \right] \bigg)\partial_{\sigma} \nn \\
&- \Bigg( \frac{\beta}{\rho^2}\tilde{\cal P} + s\left[2n\frac{\rho'}{\rho} - \frac{ \left( {\sigma^2 \rho^{2n} F h'}/{\beta}\right)'}{\rho^{2n}}\right] + 2 s^2 h'\left[1 - \frac{\sigma^2 F}{2\beta} h' \right]  \Bigg) \nn \\
&
\eea
and
\bea
\label{eq:ODE_SourceB}
 B(s) &=& -2h'\left[1 - \frac{\sigma^2 F}{2\beta} h' \right]\, \left( s\Phi_0 + \dot{\Phi}_0  \right)
 -  \left[1 - \frac{\sigma^2 F}{\beta} h' \right] \Phi_0{}_{,\sigma} \nn \\
&&  -   \left[ 2n\frac{\rho'}{\rho} - \frac{ \left( {\sigma^2 \rho^{2n} F h'}/{\beta}\right)'}{\rho^{2n}} \right] \Phi_0 \ ,
\eea
where $\Phi_0(\sigma) = \Phi(0,\sigma)$ and $\dot{\Phi}_0= \partial_\tau\Phi(0,\sigma)$ are the initial data for the wave equation \eqref{eq:ConformalWaveEq_General}.

\subsubsection{Comparison with Cauchy formulation}
Let us relate the operator ${\mathbf A}(s)$ on the hyperboloidal slice with
  the corresponding one on Cauchy slices $\bar{t}=\mathrm{constant}$.
First we write the Laplace transform of the field $\Psi(\bar{t}, x)$ with respect to the parameter $\bar{t}$ 
\footnote{\label{f:sbar_vs_s}In principle one should distinguish the spectral parameter $\bar{s}$ associated
    with the Laplace transform in terms of $\bar{t}$ from the parameter $s$ employed for the transformation
    in terms of $\tau$. However, such spectral parameters coincide since both $\bar{t}$ and $\tau$ are natural
    parameters of the timelike Killing vector $\xi=\lambda\partial_t$, namely $\xi(\bar{t})=\xi(\tau)=1$
    (actually $\xi=\partial_{\bar{t}}=\partial_\tau$).}
\beq
\hat{\Psi}(x;s) := {\cal L}[\Psi(\bar{t},x)](s) = \int_0^{\infty} e^{-s\bar{t}}\Psi(\bar{t}, x)\d\bar{t} \ .
\eeq
Applying the Laplace transformation to the wave equation \eqref{eq:WaveEq} leads to the standard formulation
\beq
{\cal D}(s) \hat{\Psi}(s)  = S(s) \ , 
\eeq
with
\beq
{\cal D}(s) = \partial_{xx} - (s^2 +{\cal P}), \quad S(s) = -\dot{\Psi}_0(x) - s \Psi_0(x) \ .
\eeq
As stated in eq.~\eqref{eq:ID_Cauchy}, $\Psi_0(x)$ and $\dot{\Psi}_0(x)$ are the initial data at
the {\em Cauchy} slice $\bar{t}=0$. Note that such functions are defined on a different time slice as compared to functions
$\Phi_0(\sigma)$ and $\dot{\Phi}(\sigma)$, which are defined in the hyperboloidal
slice $\tau=0$.

Taking into account the relation derived from \eqref{eq:EFCoord} and \eqref{eq:HyperCoord}
\beq
\tau = \bar{t}+x+h(\sigma(x)) \ ,
\eeq
the factor $e^{-s\tau}$ in the Laplace transform of $V(\tau,\sigma)$ writes as $e^{-s(x+h)}e^{-s\bar{t}}$. This result, together with the conformal rescaling \eqref{eq:ConformalReScal} motivate the introduction of the factor
\beq
\label{eq:Func_z}
Z(x;s):=\rho(\sigma(x))^n e^{s(x+h)} \ ,
\eeq
to relate the Laplace transforms with respect to $\tau$ and $\bar{t}$, and therefore the action of the
operators ${\mathbf A}(s)$ and ${\cal D}(s)$. Specifically, for functions $\psi(x;s)$ and $\phi(\sigma;s)$
related as
\beq
\psi(x;s) = Z(x;s) \phi(\sigma(x);s) \ ,
\eeq
 it holds
\beq
{\cal D}(s) \psi(s) = \frac{Z\sigma^2F}{\beta} {\mathbf A}(s) \phi(s) \ .
\eeq
In studies starting from the homogeneous equation ${\cal D}(s) \psi(s) = 0$, such a rescaling factor $Z$ is introduced after an asymptotic study of the ODE (and a bit of algebra) as a way of imposing the desired outgoing boundary conditions leading to QNMs\footnote{The concept of outgoing boundary conditions leading to QNMs is often expressed solely as $\psi(x;s)\sim e^{\mp sx}$ for $x\rightarrow \pm \infty$. Though necessary, such conditions are not sufficient --- see e.g. sec.~3.1.2 in~\cite{Nollert99} and references therein. Indeed, in \cite{Ansorg:2016ztf} regular solutions to the wave equation are constructed with arbitrary decay and frequencies times scales.}~\cite{Leaver85,Leaver90}.
In the present discussion, it follows clean and directly from the geometrical considerations of the hyperboloidal slices.

\subsection{Spectral decomposition}

\begin{figure}[t!]
\begin{center}
\includegraphics[width=8.0cm]{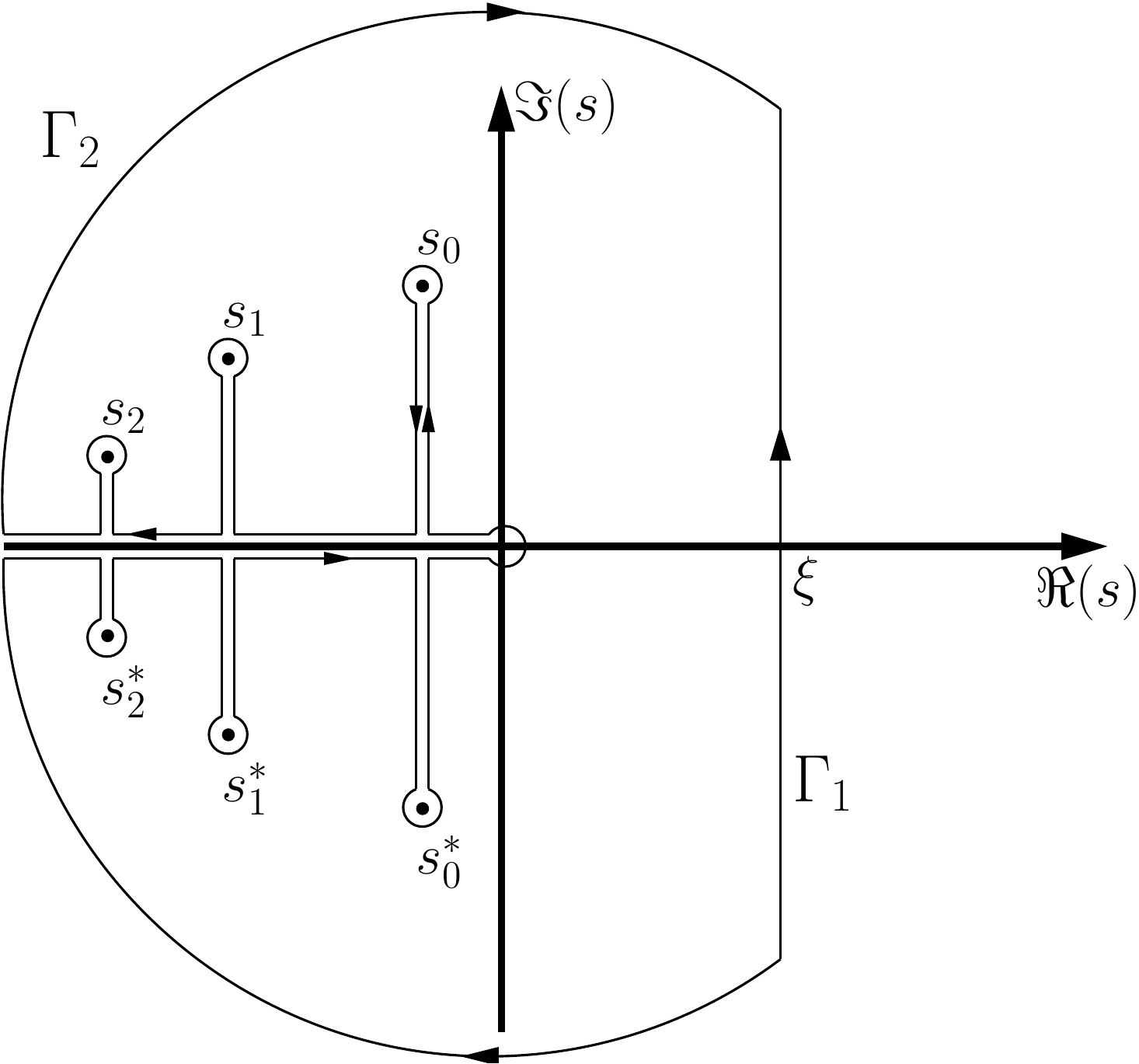}
\end{center}
\caption{Integration paths for the inverse Laplace transformation. The Bromwich integral (\ref{eq:BromInt}) is evaluated along the line
  $\Gamma_1$ in the right-plane $\Re(s)>0$ (taking its upper and lower limits to $+\infty$ and $-\infty$, respectively). 
    The spectral decomposition \eqref{eq:VSol_spectral} is formally obtained from the Cauchy theorem with the deformation
    of the path and integration along the curve $\Gamma_1-\Gamma_2$.
}
\label{fig:BromwichInt}
\end{figure}

The solution to the time evolution $\Phi(\tau, \sigma)$ is formally obtained via the inverse Laplace transformation (Bromwich Integral)
\beq
\label{eq:BromInt}
\Phi(\tau, \sigma) = \frac{1}{2\pi i}\int_{\Gamma_1} \hat{\Phi}(\sigma;s) e^{s\tau} \d s \ , 
\eeq
with the integration path
\beq
 \Gamma_1 = \left\{ s \in {\mathbb C} \, | \,  s = \xi + i \chi, \xi>0, \chi\in (-\infty, +\infty)\right\} \ .
 \eeq
 The spectral decomposition \eqref{eq:VSol_spectral} is obtained by (maximally) analytically extending the
 function $\hat{V}(\sigma;s)$  into the half-plane $\Re(s)<0$ and then appropriately deforming the path $\Gamma_1$ into that semi-plane --- see fig.~\ref{fig:BromwichInt}. 

When deforming the Bromwich integration path from $\Gamma_1$ to $\Gamma_2$, we gather a contribution from the QNMs $s_n$, the branch cut along the negative real axis $s\in\mathbb{R}^-$ and, in principle, the external semicircle, following the discussion in \cite{Leaver86c,Bachelot1993,Ansorg:2016ztf}. In particular, \cite{Ansorg:2016ztf} presents a detailed description of the algorithm and its application to the Schwarzschild case leading to
\[
\Phi(\tau, x^k) = \sum_{n=0}^{\infty} \eta_{n} e^{s_{n}\tau}\phi_n( x^k) + \int\limits_{-\infty}^{0} \eta(s)e^{s\tau}\phi( x^k;s) ds    
\] and conjecture \ref{cnjc:SpecDecomp}. 

In next section \ref{sec:Algorithm}, we present a more general procedure to calculate all the needed ingredients in the spectral decomposition, namely: i) the quasinormal mode frequencies $s_n$ together with the functions $\phi_n(x^k)$ and $\phi(x^k,x)$, and ii) the quasinormal and branch cut amplitudes, respectively, $\eta_n$ and  $\eta(s)$. The former are elements intrinsic to the wave equation, whereas the latter depend on the initial data. 

\section{Algorithm in frequency space}\label{sec:Algorithm}

\subsection{Taylor series expansions}\label{sec:TaylorExpansions}
In~\cite{Ansorg:2016ztf}, eq.~\ref{eq:ODE} was solved in the Schwarzschild background via a non-trivial Taylor series expansions around the horizon $\sigma =\sigma_{\cal H}$. Here, we generalize this procedure to an arbitrary asymptotically flat spherically symmetric spacetime. 
\subsubsection{Homogeneous Equations\\}\label{sec:AlgorithmHomEq}
We first consider the homogeneous Laplace transformed equation
\beq
\label{eq:HomODE}
\mathbf{A}(s) \phi(s) = 0 \ . 
\eeq
In a first step, we focus on solutions which are analytic in a neighbourhood around the horizon ${\cal H}$, gauge fixed in section \ref{sec:PolynomGauge} to $\sigma_{\cal H}=1$. Thus, we expand $\phi(s)$ in terms of a Taylor series
\beq
\label{eq:Ansatz_phi}
\phi(\sigma;s)  = \sum_{k=0}^{\infty} H_k u^k, \quad u = 1-\sigma \ .
\eeq
Singular points of this ODE are given by $\sigma=0$ and $\sigma=\infty$, as well as by the values $\sigma=\sigma_{{\rm H}_j}$, such that $F(\sigma_{{\rm H}_j})=0$. The smallest real root represents the event horizon ${\cal H}$, therefore we must have that $\sigma_{{\rm H}_0} = 1$
in the adopted gauge.

Given the unit circle
\beq\label{eq:circle_C} 
\mathbf{C}=\{\sigma\in\mathbb{C}: |1-\sigma|<1\} \ ,
\eeq 
a necessary condition for the convergence of the series \eqref{eq:Ansatz_phi} within $\mathbf{C}$ (and therefore up until the vicinity of future null infinity
at $\sigma=0$) requires that further (eventually complex) roots of $F(\sigma)$ lie outside of $\mathbf{C}$. In the particular case of   Reissner-Nordstr\"om, the (real) root $\sigma_1$ associated to the Cauchy horizon must satisfy $\sigma_{{\rm H}_1}\ge2$.

In the minimal gauge (see section \ref{sec:PolynomGauge}),  the coefficients of the operator $\mathbf{A}(s)$ are polynomials~\footnote{One might need multiply the whole equation by an appropriate power of $\rho$, in case $\rho(\sigma)$ is not a constant but presents a linear term in $\sigma$ --- see eq.~\eqref{eq:Fix_rho} } in $\sigma$. Therefore, the introduction of the Ansatz \eqref{eq:Ansatz_phi} into \eqref{eq:HomODE} gives rise to a $(m+2)$-term recurrence relation, i.e. a $(m+1)$-order recurrence relation 
\beq
\label{eq:HomRecRel}
\alpha_k H_{k+1} + \sum_{i=0}^{m}\beta_k^{(i)} H_{k-i} = 0\ , \quad k\ge m\ ,
\eeq
with $m \in {\mathbb N}$, $\alpha_{-1} = 0$ and the coefficients $\alpha_k$ and $\beta_k^{(i)}$ depending on the structure of the operator $\mathbf{A}(s)$ (see
Eq.~\eqref{eq:ODE_OperatorA} and (\ref{eq:RecRelCoeff}) later for an illustration in the Reissner-Nordstr\"om case).

The $m+1$ initial values $H_k$ ($k=0 \cdots m$) needed to iterate \eqref{eq:HomRecRel} must satisfy (a suitable normalisation is $H_0=1$)
\beq
\label{eq:InitialData_RecRel}
\alpha_k H_{k+1} + \sum_{i=0}^{k}\beta_k^{(i)} H_{k-i} = 0\ , \quad k = 0\cdots m-1\ .
\eeq
These constraints follow from extending the validity of the relation \eqref{eq:HomRecRel} for $k = 0\cdots m-1$ and imposing 
\beq
\label{eq:Reg_NegativeH}
H_{-k} = 0, \quad k = 1 \cdots m \ .
\eeq
Note that \eqref{eq:Reg_NegativeH} applied to \eqref{eq:HomRecRel} for $k<0$ automatically ensures $H_{k}=0$ for all $k<0$, i.e., the constraints \eqref{eq:InitialData_RecRel} encodes the information about the regularity of \eqref{eq:Ansatz_phi} at $\sigma=1$.

Let us now relax the conditions imposed by the constraints \eqref{eq:InitialData_RecRel} --- or equivalent, by \eqref{eq:Reg_NegativeH} --- and construct all the $m+1$ linearly independent sequences $\{H^{(\ell)}_k\}_{k=-m}^\infty$ ($\ell=0\cdots m$) satisfying
\bea
\label{eq:HomRecRel_General}
\label{eq:InitialData_RecRel_General}
& H^{(\ell)}_{-k} = \delta^{\ell}_k \ ,\quad k = 0 \cdots m \\ 
\label{eq:RecRel_General}
& \alpha_k H^{(\ell)}_{k+1} + \sum\limits_{i=0}^{m}\beta_k^{(i)} H^{(\ell)}_{k-i} = 0\ , \quad k\ge 0\ ,
\eea
with $\delta^{\ell}_k$ the Kronecker delta. For $\ell=0$, eqs.~\eqref{eq:HomRecRel_General} and \eqref{eq:RecRel_General} recovers \eqref{eq:HomRecRel}-\eqref{eq:Reg_NegativeH}, i.e., the sequence $\{H^{(0)}_k\}_{k=-m}^\infty$ corresponds to the Taylor coefficients leading to \eqref{eq:Ansatz_phi}.

\medskip
We consider now the asymptotic behaviour of~\eqref{eq:RecRel_General} for large $k$ values. From the findings in~\cite{Leaver85,Leaver90,Ansorg:2016ztf,Batic:2018nxk}, we consider the cases where $H_k$ behaves asymptotically as\footnote{The assumptions until the end of this section are corroborated by the scenarios studied in \cite{Ansorg:2016ztf,Ammon:2016fru,Macedo:2018gvw,Macedo2019}. A more rigorous statement relating such behaviours, the (polynomial in $k$) form of the coefficients $\alpha_k$, $\beta^{(i)}_k$ and the convergence region given by the unit circle $\mathbf{C}$ requires further work that goes beyond the scope of this paper.} 
\beq
\label{eq:Asympt_Ansatz}
H_k \sim e^{\xi k^p} k^{\zeta} A_k \ , \quad A_k =  1 + \sum_{j=1}^\infty \frac{\nu_j}{k^{jp}}
\eeq
with $p=1/2$ or $p=1$. 

For a fixed asymptotic parameter $p$, an algorithm to determine the coefficients $\xi$, $\zeta$ and $\{\nu_j\}_{j=1}^\infty$ consists in (i) inserting \eqref{eq:Asympt_Ansatz} back into the recurrence relation \eqref{eq:HomRecRel}, (ii) multiplying the result by $e^{-\xi k^p} k^{-\zeta}$, (iii) expanding the resulting expression in terms of $y=k^{-p}$ around $y=0$, and (iv) equating the coefficients of the asymptotic expansion order by order \cite{Ansorg:2016ztf}. 

From the perspective of the asymptotic expansion, the freedom to construct the $m+1$ linearly independent solutions is captured by the existence of $m+1$ asymptotic coefficients $\xi^{(\ell)}$, $\zeta^{(\ell)}$ and $\{\nu^{(\ell)}_j\}_{j=1}^\infty$ ($\ell = 0\cdots m$). Thus, the asymptotic behaviour of each sequence $\{H^{(\ell)}_k\}_{k=-m}^\infty$ is given by a linear combination 
\bea
H^{(\ell)}_k \sim && \lambda^{(\ell)}_{(0)}\, e^{\xi^{(0)} k^{p^{(0)}}} k^{\zeta^{(0)}} A^{(0)}_k + \cdots \nn \\
&&   + \lambda^{(\ell)}_{(\ell)} e^{\xi^{(m)} k^{p^{(m)}}} k^{\zeta^{(m)}} A^{(m)}_k.
\eea
with $\lambda^{\ell}_{\ell'}$ generic coefficients.

We also assume that $\left| e^{\xi^{(0)} } \right| > 1$, whereas, for $\ell = 1 \cdots m$, $\left| e^{\xi^{(\ell)} } \right| < 1$, i.e., the asymptotic $(0)$-mode grows for large $k$ while the $(\ell)$-modes ($\ell = 1 \cdots m$) decay. Moreover, we label the asymptotic parameters in such a way that 
$$0< \left| e^{\xi^{(m)} } \right| \leq \left| e^{\xi^{(m-1)} } \right| \leq \cdots \leq \left| e^{\xi^{(1)} } \right| < 1 \ . $$ 

Note that the sequences $\{H^{(\ell)}_k\}_{k=-m}^\infty$ are obtained according to \eqref{eq:RecRel_General}, by fixing their initial seeds via \eqref{eq:InitialData_RecRel_General}. When iterating \eqref{eq:RecRel_General} forwardly, we cannot control the asymptotic behaviour and, in general, the growing mode will dominate. Our aim is to introduce a new set of linear independent sequences $\{I^{(\ell)}_k\}_{k=-m}^\infty$ still satisfying the recurrence relation \eqref{eq:RecRel_General} but in a way that we remove the growing asymptotic behaviour\footnote{In \cite{Ansorg:2016ztf}, this procedure is done via a backward iteration of the recurrence relation. Appendix \ref{App:BackwardRecRel} discusses this strategy in the current context.}.

For this task, we first identify directly $H^{(0)}_k = I^{(0)}_k$, i.e., $I^{(0)}_k$ satisfies, in one hand, the constraints \eqref{eq:InitialData_RecRel} needed for the regularity properties of \eqref{eq:HomODE} at $\sigma = 1$ and, on the other hand, it generically grows asymptotically as $\sim e^{\xi^{(0)} k^{p^{(0)}}} k^{\zeta^{(0)}}$. Then, we successively filter the asymptotic modes by defining
\beq
\label{eq:LinearComb_AsymExp}
I^{(\ell)}_k = \sum_{\ell'=0}^{\ell} \kappa^{(\ell)}_{\ell'} H^{(\ell')}_k    , \ \quad (\ell=1 \cdots m)\ .
\eeq
For each $\ell = 1 \cdots m$, there are $(\ell+1)$ constants $\kappa^{(\ell)}_{\ell'}$ to be fixed. To determine such constants, we prescribe a truncation  $k_{\rm max}$ and consider \eqref{eq:LinearComb_AsymExp} at $(\ell+1)$ values $k=k_{\rm max} + l$ ($l=0\cdots\ell$). The asymptotic behaviour of $I^{(\ell)}_{k_{\rm max} + l}$ is then approximated via the algorithm from \cite{Ansorg:2016ztf} applied to the desired decaying behaviour $\sim e^{\xi^{(\ell)} k^{p^{(\ell)}}} k^{\zeta^{(\ell)}}$. This procedure provides us with  $(\ell+1)$ equations to fix $\kappa^{(\ell)}_{\ell'}$. Having imposed the desired decaying behaviour to the sequence $\{I^{(\ell)}_k\}_{k=-m}^\infty$, we normalise it to $I^{(\ell)}_0= 1$ by letting $I^{(\ell)}_k \rightarrow I^{(\ell)}_k/I^{(\ell)}_0$.

Summarising the discussion above, we have constructed $(m+1)$ linearly independent sequences $\{I^{(\ell)}_k\}_{k=-m}^\infty$ ($\ell = \cdots m$) which are solution to the recurrence relation \eqref{eq:RecRel_General} and satisfy: 
\ben[(1)]
\item There is {\em one} specific solution $I^{(0)}_k$ with
\beq
\label{eq:Property1}
I^{(0)}_k = 0 \quad {\rm for} \quad k<0 \ .
\eeq

\item As $k\rightarrow \infty,$ the coefficients $I_k^{(\ell)}$ decay for $\ell = 1, \ldots, m$. In contrast, the coefficients $I_k^{(0)}$ are assumed to diverge as $k\rightarrow \infty$. 

The solutions $I_k^{(\ell)}$ ($\ell = 1, \ldots, m$) are refereed to as {\em decaying} solutions.

\item For all $\ell = 0, \ldots, m$, the sequences are normalised to
\beq
\label{eq:Norm_Ik}
I^{(\ell)}_0 = 1 \ .
\eeq 

\item For the first few negative indices $-m \le k <  -\ell$, the decaying solutions $I^{(\ell)}_k$ ($\ell = 1, \ldots, m$) satisfy 
\beq
\label{eq:Property4}
I^{(\ell)}_{k} = 0 \ .
\eeq
\een

\subsubsection{Inhomogeneous equation}
We now proceed to the general algorithm to solve the Laplace transformed equation \eqref{eq:ODE}. According to the previous section, we seek a solution $\hat{\Phi}(\sigma;s)$ of the form
\beq
\label{eq:Series_hat_V}
\hat{\Phi}(\sigma;s) = \sum_{k=0}^{\infty} a_{k}(1-\sigma)^k \ .
\eeq
The coefficients $a_k$ satisfy the inhomogeneous recurrence relation
\beq
\label{eq:InHomRecRel}
\alpha_k a_{k+1} + \sum_{i=0}^{m}\beta_k^{(i)} a_{k-i} = q_k \ ,
\eeq
with $q_k$ the coefficients for the expansion of the source \eqref{eq:ODE_SourceB}
\beq
B(\sigma;s) = \sum_{k=0}^{\infty} q_{k}(1-\sigma)^k \ .
\eeq
As in~\cite{Ansorg:2016ztf}, we initially consider an inhomogeneity of polynomial kind, i.e., with
\beq
q_k = 0 \quad {\rm for} \quad k<0 \quad {\rm and} \quad k>K_{\rm max}.
\eeq
The analytic limit $K_{\rm max}\rightarrow \infty$ is taken in the end of this section. Finally, we restrict ourselves to regular solutions satisfying
\beq
\label{eq:BoundCond_ak}
\left\{
\begin{array}{ccc}
a_k = 0 & {\rm for} & k<0  \\ 
a_k \rightarrow 0 & {\rm as} & k\rightarrow \infty \ .
\end{array}
\right.
\eeq
We seek the solution $a_k$ to the inhomogeneous recurrence relation in the form
\beq
\label{eq:AnsatzSol_ak}
a_k = \sum_{\ell =0}^m c_{k,\ell} I_k^{(\ell)} \ ,
\eeq 
i.e., as a linear combination of the solutions $I_k^{(\ell)}$ to the homogeneous relation. In order to fix the coefficients $c_{k,\ell}$ we must impose, together with \eqref{eq:InHomRecRel}, $m$ further conditions. In particular, conditions 
\beq
\label{eq:Cond1}
\sum_{\ell =0}^m c'_{k,\ell} I_{k-p}^{(\ell)} = 0, \quad p = 0, \ldots, m-1 \ ,
\eeq
extend eq.~(64) in \cite{Ansorg:2016ztf} to the present general case. Eqs.~\eqref{eq:AnsatzSol_ak} and \eqref{eq:Cond1} constitute the generalisation of the Ansatz in \cite{Ansorg:2016ztf} for solution of the recurrence relation. 

In order to justify \eqref{eq:Cond1}, we introduce the discrete derivative
\beq
\label{eq:DiscreteDerivative_Appendix}
c'_{k,\ell} = c_{k+1,\ell} - c_{k,\ell} \ . 
\eeq
As a consequence, we find that
\bea
c_{k+1,\ell} &=& c_{k,\ell} + c'_{k,\ell} \nn \\
\label{e:ck+1_ck}
c_{k-i,\ell} &=& c_{k,\ell} - \sum_{j=1}^i c'_{k-j,\ell} \quad (i \ge 1) \ .
\eea
The first one recasts the definition (\ref{eq:DiscreteDerivative_Appendix}), whereas  the second one follows again from such definition by writing
\[
\sum_{j=1}^i c'_{k-j,\ell} = \sum_{j=1}^i c_{k-j+1,\ell} - \sum_{j=1}^i c_{k-j,\ell} \ ,
\]
and shifting the index $j'=j-1$ in the first sum of the right-hand-side to get
\[
\sum_{j=1}^i c'_{k-j,\ell} = \sum_{j'=0}^{i-1} c_{k-j',\ell} - \sum_{j=1}^i c_{k-j,\ell} = c_{k,\ell} - c_{k-i,\ell} \ .
\]
Combining these partial results with the Ansatz \eqref{eq:AnsatzSol_ak} we may write
\bea
a_{k+1} &=& \sum_{\ell =0}^m c_{k,\ell} I_{k+1}^{(\ell)} + \sum_{\ell =0}^m c'_{k,\ell} I_{k+1}^{(\ell)} \label{eq_akp1_Appendix} \\
a_{k-i} &=& \sum_{\ell =0}^m c_{k,\ell} I_{k-i}^{(\ell)} - \sum_{j=1}^i\bigg( \sum_{\ell =0}^m c'_{k-j,\ell} I_{k-i}^{(\ell)}\bigg). 
\eea
If we now impose, for {\em all} $k\in {\mathbb Z}$
\beq
\label{eq_akmi_Appendix}
a_{k-i} = \sum_{\ell =0}^m c_{k,\ell} I_{k-i}^{(\ell)} \ ,
\eeq
we must require for $j=1, \ldots, i$ that
\[
\sum_{\ell =0}^m c'_{k-j,\ell} I_{k-i}^{(\ell)} = \sum_{\ell =0}^m c'_{k',\ell} I_{k'-(i-j)}^{(\ell)} = 0 \ ,
\]
with $k' = k - j$. Thus, the $m$ supplementary conditions \eqref{eq:Cond1} necessary to determine the $c_{k,\ell}$ uniquely follows from the expression above for $i=1, \ldots, m$.

\medskip
A key element in the algorithm in \cite{Ansorg:2016ztf} is the introduction of a discrete Wronskian determinant associated with
two numerical sequences $\left\{a_k\right\}_{-\infty}^\infty$ and $\left\{b_k\right\}_{-\infty}^\infty$ (cf.  eq.~(55) in  \cite{Ansorg:2016ztf}). 
We are now apt to generalise such a discrete Wronskian determinant two a set of $m+1$ numerical sequences, as required in the
present generalized setting
\footnote{We assume that neither $\alpha_k$ nor $\beta^{(m)}_k$ vanish for a given integer $k$. Appendix C in \cite{Ansorg:2016ztf} discusses the modifications in the algorithm accounting for such cases in $2-$order recurrence relations. Its generalisation for a $m+1$-order recurrence relations is actually relevant in scenarios involving odd-dimensional spacetimes, indeed a subject of current work~\cite{Macedo2019}.
}.
We insert \eqref{eq_akp1_Appendix} and \eqref{eq_akmi_Appendix} into the original recurrence relation \eqref{eq:InHomRecRel} to get
\[
\sum_{\ell =0}^m c_{k,\ell}\left[ \alpha_k I^{(\ell)}_{k+1} + \sum_{i=0}^m \beta^{(i)}_k I^{(\ell)}_{k-i}\right] + \alpha_k \sum_{\ell =0}^m c'_{k,\ell} I^{(\ell)}_{k+1} = q_k \ .
\]
Since $I^{(\ell)}_k$ satisfies the homogeneous recurrence relation \eqref{eq:HomRecRel}, this expression reduces to
\beq
\label{eq:Cond2}
\alpha_k \sum_{\ell =0}^m c'_{k,\ell} I^{(\ell)}_{k+1} = q_k \ .
\eeq
The combination of  eqs.~\eqref{eq:Cond1} and \eqref{eq:Cond2} leads to
\beq
\label{eq:WronskianMatrix}
\hat{\cal W}_k
\left(
\begin{array}{c}
c'_{k,0} \\
c'_{k,1} \\
c'_{k,2} \\
\vdots \\
c'_{k,m}
\end{array}
\right) = 
\left(
\begin{array}{c}
q_k/\alpha_k \\
0 \\
0 \\
\vdots \\
0
\end{array}
\right) \ .
\eeq
with the matrix $\hat{\cal W}_k$ defined by
\beq
\hat{\cal W}_k := \left(
\begin{array}{cccc}
I_{k+1}^{(0)} & I_{k+1}^{(1)} &  \cdots & I_{k+1}^{(m)}  \\
I_{k}^{(0)} & I_{k}^{(1)}  & \cdots & I_{k}^{(m)} \\
I_{k-1}^{(0)} & I_{k-1}^{(1)}  & \cdots & I_{k-1}^{(m)} \\
\vdots & \vdots  & \ddots & \vdots \\
I_{k-(m-1)}^{(0)} & I_{k-(m-1)}^{(1)}  & \cdots & I_{k-(m-1)}^{(m)}
\end{array}
\right) \ .
\eeq
This matrix provides us with a general definition for the discrete Wronskian determinant, namely
\beq
\label{eq:WronskianDet}
W_k = \det \hat{\cal W}_k \ .
\eeq
Let us further define the sub-matrix $\hat{\cal W}_{k,\ell}$ as the one resulting from $\hat{\cal W}_k$ after we remove the first row and the column $(I^{(\ell)}_{k+1} \, I^{(\ell)}_{k} \, \cdots \, I^{(\ell)}_{k-(m+1)})^T$, i.e.
\beq
\label{eq:SubWronskianMatrix}
\underbrace{
\left(
\begin{array}{cccccc}
I_{k}^{(0)}   & \cdots & I_{k}^{(\ell - 1)}  & I_{k}^{(\ell +1)}  & \cdots & I_{k}^{(m)} \\
I_{k-1}^{(0)}   & \cdots & I_{k-1}^{(\ell - 1)}  & I_{k-1}^{(\ell +1)}  & \cdots & I_{k-1}^{(m)} \\
\vdots   & \ddots & \vdots & \vdots & \ddots & \vdots \\
I_{k-(m-1)}^{(0)}  & \cdots & I_{k-(m-1)}^{(\ell - 1)}  & I_{k-(m-1)}^{(\ell +1)}  & \cdots & I_{k-(m-1)}^{(m)}
\end{array}
\right)}_{=:\hat{\cal W}_{k,\ell} } \ .
\eeq
Analogously, 
\beq
\label{eq:SubWronskianDet}
W_{k,\ell} = \det \hat{\cal W}_{k,\ell} \ .
\eeq
With the help of eqs.~\eqref{eq:WronskianMatrix}-\eqref{eq:SubWronskianDet} we are now able to write down an explicit expression for the solution $a_k$. 

We begin by deriving a compact expression for the Wronskian determinant. First we define
\[
\hat{\cal W}_{k}^\alpha := \left(
\begin{array}{cccc}
\alpha_k I_{k+1}^{(0)} & \alpha_k I_{k+1}^{(1)} &  \cdots & \alpha_k I_{k+1}^{(m)}  \\
I_{k}^{(0)} & I_{k}^{(1)}  & \cdots & I_{k}^{(m)} \\
I_{k-1}^{(0)} & I_{k-1}^{(1)}  & \cdots & I_{k-1}^{(m)} \\
\vdots & \vdots  & \ddots & \vdots \\
I_{k-(m-1)}^{(0)} & I_{k-(m-1)}^{(1)}  & \cdots & I_{k-(m-1)}^{(m)}
\end{array}
\right) \ ,
\]
and
\[
\hat{\cal W}_{k}^{\alpha\beta} := \mathbb{I}^\beta\hat{\cal W}_{k}^\alpha \ ,
\]
with
\[
\mathbb{I}^\beta := \left(
\begin{array}{ccccc}
1 & \beta^{(0)}_k & \beta^{(1)}_k& \cdots & \beta^{(m-1)}_k  \\
0  & 1  & 0 & \cdots & 0 \\
0 & 0  &  1 & \cdots & 0 \\
\vdots & \vdots  & \vdots & \ddots & \vdots \\
0 & 0  & 0 & \cdots & 1
\end{array}
\right) \ .
\]
Note that
\beq
\label{eq:detWalpha}
\alpha_k W_k = \det \hat{\cal W}_{k}^\alpha = \det \hat{\cal W}_{k}^{\alpha\beta} \ .
\eeq
Now, since the $I_k^{(\ell)}$ satisfy the homogeneous recurrence relation, the first row of $\hat{\cal W}_{k}^{\alpha\beta}$ explicitly reads
\[
\left\{ \alpha_k I_{k+1}^{(\ell)} + \sum_{i=0}^{m-1} \beta_k^{(i)}I^{(\ell)}_{k-i}    \right\}^m_{\ell=0} = \left\{ -\beta_k^{(m)}I^{(\ell)}_{k-m} \right\}^m_{\ell=0}.
\]
Thus, we obtain
\begin{widetext}
\bea
\label{eq:detWalphabeta}
\det \hat{\cal W}_{k}^{\alpha\beta} &=& -\beta_k^{(m)} \det \left(
\begin{array}{cccc}
I_{k-m}^{(0)} &  I_{k-m}^{(1)} &  \cdots &  I_{k-m}^{(m)}  \\
I_{k}^{(0)} & I_{k}^{(1)}  & \cdots & I_{k}^{(m)} \\
I_{k-1}^{(0)} & I_{k-1}^{(1)}  & \cdots & I_{k-1}^{(m)} \\
\vdots & \vdots  & \ddots & \vdots \\
I_{k-(m-1)}^{(0)} & I_{k-(m-1)}^{(1)}  & \cdots & I_{k-(m-1)}^{(m)}
\end{array}
\right) = -\beta_k^{(m)} (-1)^m W_{k-1} \ .
\eea
\end{widetext}
In the above expression, the determinant $W_{k-1}$ comes about once we swap the 1st and 2nd row, thereafter the 2nd and 3rd row and so on, until the first row is shifted to being the very last one. At each swap, we gain a factor $-1$ . Since there are in total $m$ swaps in this $(m+1)\times(m+1)$ dimensional matrix, a final factor $(-1)^m$ appears as well.

Combining \eqref{eq:detWalpha} with \eqref{eq:detWalphabeta} gives the simple result
\[
{\alpha_k}W_k = (-1)^{m+1}{\beta_k^{(m)}}W_{k-1} \ ,
\]
and therefore the desired result
\beq
 W_k = W_{-1} \prod_{j=0}^k(-1)^{m+1} \frac{\beta_j^{(m)}}{\alpha_j}. \label{eq:WronskianDeterm_Sol}
\eeq
We proceed now to find the solution $a_k$. Thanks to property \eqref{eq:Property1} in assumption (1), the normalisation \eqref{eq:Norm_Ik} in assumption (3) and \eqref{eq:Property4} in assumption (4), we obtain
\beq
\label{eq:detW_-1}
W_{-1} = W_{-1,0} = \prod_{\ell = 1}^{m} I^{(\ell)}_{-\ell} \ .
\eeq
Moreover, for $k\leq -2$, $W_k = 0$ and therefore the system \eqref{eq:WronskianMatrix} cannot be inverted. 
Still, we want to enforce \eqref{eq:Cond1} for {\em all} $k$. In particular, eq.~\eqref{eq:Cond1} reads for $k<0$ 
\beq
\hat{\cal W}_{k,0} \left(
\begin{array}{c}
c'_{k,1} \\
c'_{k,2} \\
\vdots \\
c'_{k,m}
\end{array}
\right) = \left(
\begin{array}{c}
0 \\
0 \\
\vdots \\
0
\end{array}
\right) \ .
\eeq
If we assume that $\hat{\cal W}_{k,0}$ is invertible ($W_{k,0}\neq 0$), we may conclude\footnote{The passage from the strict inequality $k<0$ to $k\leq 0$ follows from the definition of $c'_{k,\ell}$ in eq.~\eqref{eq:DiscreteDerivative_Appendix}. } that for $\ell = 1 \cdots m$
\beq
c'_{k,\ell} = 0 \quad (k<0) \Rightarrow c_{k,\ell} = \bar{c}_\ell = {\rm const.} \quad (k \leq 0)
\eeq
The constants $\bar{c}_\ell$ can be determined from \eqref{eq:AnsatzSol_ak}. Indeed, taking $m$ negative $k$ values yields a system of $m$ equations to determine the $\bar{c}_{\ell}'s$. Since $a_k = 0$ for $k<0$ and assuming the system is invertible we obtain simply
\beq
c_{k,\ell} = 0 \quad {\rm for} \quad \ell = 1, \ldots, m \quad {\rm and} \quad k\leq 0\ .
\eeq
For $\ell = 0$ though, the $c_{k,0}$ with $k\leq0$ are arbitrary and undetermined.

We now concentrate on the solution of \eqref{eq:WronskianMatrix} for $k\ge 0$ with ``initial conditions" $c_{0,\ell}=0$ for $\ell = 1 \cdots m$ and a free, undetermined parameter $c_{0,0}$. The solution of this system emerges through the Cramer's rule as
\beq
c'_{k,\ell} = (-1)^\ell \frac{q_k}{\alpha_k}\frac{W_{k,\ell}}{W_k}.
\eeq
With $c_{0,\ell}=0$ for $\ell = 1 \cdots m$ we get from \eqref{e:ck+1_ck}
\bea
c_{k,0} &=& c_{0,0} + \sum_{j=0}^{k-1} \frac{q_j}{\alpha_j}\frac{W_{j,0}}{W_j} \\
c_{k,\ell} &=& (-1)^{\ell} \sum_{j=0}^{k-1} \frac{q_j}{\alpha_j}\frac{W_{j,\ell}}{W_j}, \quad \ell = 1, \ldots, m.
\eea
It then follows, from \eqref{eq:AnsatzSol_ak}, that
\bea
a_k &=& I^{(0)}_k\Bigg(c_{0,0} + \sum_{j=0}^{k-1} \frac{q_j}{\alpha_j}\frac{W_{j,0}}{W_j} \Bigg) \nn \\
&& + \sum_{\ell = 1}^m (-1)^{\ell}I_k^{(\ell)} \sum_{j=0}^{k-1} \frac{q_j}{\alpha_j}\frac{W_{j,\ell}}{W_j} \ .
\eea
The constant $c_{0,0}$ is determined by imposing \eqref{eq:BoundCond_ak}. Since $q_k = 0$ for $k>K_{\rm max}$ and, according to assumption (2), $I_k^{(0)}$ diverges in this limit, we must have
\beq
c_{0,0} = - \sum_{j=0}^{K_{\rm max}} \frac{q_j}{\alpha_j}\frac{W_{j,0}}{W_j} \ .
\eeq
Hence 
\beq
a_k = - I^{(0)}_k\sum_{j=k}^{K_{\rm max}} \frac{q_j}{\alpha_j}\frac{W_{j,0}}{W_j}  + \sum_{\ell = 1}^m (-1)^{\ell}I_k^{(\ell)} \sum_{j=0}^{k-1} \frac{q_j}{\alpha_j}\frac{W_{j,\ell}}{W_j} \ .
\eeq
In this expression, the $K_{\rm max}$ can be taken arbitrarily large and we account for that by considering the formal series obtained by letting $K_{\rm max}\rightarrow \infty$. Besides we can explicitly substitute the expression for $W_j$ according to \eqref{eq:WronskianDeterm_Sol} to get
\beq
\label{eq:Sol_ak}
a_k =\frac{1}{W_{-1}}\Bigg( - I^{(0)}_k\sum_{j=k}^{\infty} q_j \Xi_{j,0} 
+ \sum_{\ell = 1}^m (-1)^{\ell}I_k^{(\ell)} \sum_{j=0}^{k-1} \ q_j\Xi_{j,\ell}\Bigg) \ ,
\eeq
with
\beq
\label{eq:Xi}
\Xi_{k,\ell} = W_{k,\ell}\,\frac{(-1)^{m+1}}{\alpha_k}\prod_{j=0}^k \frac{\alpha_j}{\beta_j^{(m)}} \ .
\eeq

\subsection{Quasi-normal modes and amplitudes}
In accordance to \cite{Ansorg:2016ztf}, the quasi-normal modes $s_n$ are specific values in the complex $s-$plane for which the procedure described in the previous section fails. According to \eqref{eq:Sol_ak}, this occurs whenever $W_{-1}$ vanishes. 
Specifically, and thanks to \eqref{eq:detW_-1}, this occurs when
\bea
\label{e:QNM_condition}
I^{(1)}_{-1} = 0  \ .
\eea
This condition, together with property (4) in section \ref{sec:AlgorithmHomEq}, leads to $I^{(1)}_{k} = 0$ for {\em all} $k<0$. Thus, the two solutions $I^{(0)}_k$ and $I^{(1)}_k$ become linearly dependent. In fact, they coincide here due to the normalisation (3).

Having identified the condition in the present setting for the location of the quasi-normal modes, which depends only on the wave equation in question and is a property of the spacetime alone, we proceed to calculate the quasi-normal amplitudes and the jump function along the branch cut, resulting from the specific choice of initial data.

\subsubsection{Discrete amplitudes}
Let us first recall the simplified notation introduced in \cite{Ansorg:2016ztf} for the recurrence relation \eqref{eq:InHomRecRel} 
\beq
{\cal A}(s)\cdot \{a_k\} = \{q_k\} \ ,
\eeq
with the operator ${\cal A}(s)$ defined by
\beq
\left[ {\cal A}(s)\cdot \{a_k\} \right]_k := \alpha_k a_{k+1} + \sum_{i=0}^{m}\beta_k^{(i)} a_{k-i} \ .
\eeq
Assuming that $W_{-1}$ has simple poles at discrete values $s_n$, we may write
\beq
\label{eq:ak_QNM}
a_k(s) = \frac{\eta_n I^{(0)}_k}{s-s_n} +g_k(s) \ .
\eeq
Furthermore, we introduce a second operator ${\cal C}_n$ via
\beq
\label{eq:OperatorC}
{\cal C}_n(s) = \frac{{\cal A}(s) - {\cal A}(s_n)   }{s - s_n} \ .
\eeq
In the limit $s\rightarrow s_n$, the decompositions \eqref{eq:ak_QNM} and \eqref{eq:OperatorC} together with the condition for homogeneous solutions,
(\ref{eq:HomRecRel})  lead to
\beq
\label{eq:RecRel_eta}
{\cal A}(s_n)\cdot\{g_k\} = \{q_k\} - \eta_n {\cal C}_n(s_n)\cdot\{ I^{(0)}_k\} \ .
\eeq
For $s\neq s_n$, the $k=0$ solution to \eqref{eq:RecRel_eta} according to \eqref{eq:Sol_ak} would be
\beq
g_0 = -\frac{1}{W_{-1}} \sum_{j=0}^{\infty} \left( q_j - \eta_n C_j  \right)\Xi_{j,0} \ ,
\eeq
with 
\beq
C_j := \left[ {\cal C}_n(s)\cdot \{I^{(0)}_k\}\right]_j \ .
\eeq
Since  $W_{-1}\rightarrow 0$ in the limit $s\rightarrow s_n$, a finite value for $g_0$ is obtained only if
\beq
\label{eq:QNAmplitude}
\displaystyle
\eta_n =  \frac{\sum\limits_{j=0}^{\infty} q_j \Xi_{j,0} }{\sum\limits_{j=0}^{\infty} C_j \Xi_{j,0}} \ .
\eeq
For a given set of initial data, the $\eta_n$ are the amplitudes associated to each quasinormal mode $s_n$ and eq.~\eqref{eq:QNAmplitude} generalises the result from~\cite{Ansorg:2016ztf}.
\subsubsection{Branch cut amplitude}
We end this section by considering the situation in which we approach the negative real axis in the complex $s-$plane. Depending on whether we approach the axis from above or from below we might encounter different solutions for the inhomogeneous recurrence relation. We begin by defining
\beq
a_k^{\pm} = \lim_{\varepsilon \rightarrow 0} a_k(s\pm i |\varepsilon|) \ ,
\eeq
and we assume the symmetry condition (with upper asterisk ${}^*$ denoting complex conjugation).
\beq
a_k^-=[a_k^+]^* \ ,
\eeq
that follows from $V(\tau,\sigma)$ being a real valued function.

Since the difference $d_k = a_k^+ - a_k^-$ satisfies the homogeneous recurrence relation and has $d_k = 0$ for $k<0$, it must be proportional to
$ I^{(0)}_k$, so we may write
\beq
d_k = a_k^+ - a_k^- = d_0\, I^{(0)}_k \ . 
\eeq
According to the $k=0$ value in \eqref{eq:Sol_ak}, we have 
\beq
\label{eq:d_0}
d_0 = - \sum_{j=0}^{\infty} q_j  \left(  \frac{\Xi^+_{j,0}}{W_{-1}^+} -  \frac{\Xi^-_{j,0}}{W_{-1}^-} \right) \ .
\eeq
Apart from properties (1)-(4) mentioned previously, let us also assume that when getting to the negative real axis from above or from below\footnote{In the cases studied, this assumption has always been found to be realised.}: 
\ben
\setcounter{enumi}{4}
\item Only the values of $I^{(1)}_k$ are different;

\item The recurrence relation coefficients $\alpha_k$ and $\beta_k^{(\ell)}$ do {\em not} differ.
\een
Then, the contribution at the branch cut is due to
\beq
\label{eq:D_j}
D_j = \frac{W^+_{j,0}}{W_{-1}^+} -  \frac{W^-_{j,0}}{W_{-1}^-} \ ,
\eeq
which can be re-written in terms of $D^{(1)}_0$ in~\footnote{The last equality is valid because $D^{(1)}_j$ satisfies the homogeneous recurrence relation with $D^{(1)}_k=0$ for $k<0$.}
\beq
\label{eq:D^1_k}
D^{(1)}_j = \frac{I^{(1)}_{j}{}^+}{I^{(1)}_{-1}{}^+} -  \frac{I^{(1)}_{j}{}^-}{I^{(1)}_{-1}{}^-} = D^{(1)}_0\, I^{(0)}_{\blue{j}} \ ,
\eeq
with
\beq
D^{(1)}_0 = 2 i\, \Im\left( \frac{I^{(1)}_0{}^+}{I^{(1)}_{-1}{}^+}\right) \ .
\eeq
To show this, we first write the determinant $W_{j,0}$ in eq.~\eqref{eq:SubWronskianDet} explicitly as
\beq
W_{j,0} = \epsilon_{i_1i_2\cdots i_m} I^{(1)}_{j-(i_1-1)}I^{(2)}_{j-(i_2-1)}\cdots I^{(m)}_{j-(i_m-1)} \ ,
\eeq
with $\epsilon_{i_1i_2\cdots i_m}$ the Levi-Civita symbol. In the above expression, a sum is assumed for each index $i_p = 1, \ldots, m$ $(p=1, \ldots, m)$.

Then, with the help of assumption (5) together with eqs.~\eqref{eq:detW_-1} and \eqref{eq:D^1_k} we have
\bea
D_j &=& \frac{D^{(1)}_0}{\prod\limits_{\ell = 2}^{m} I^{(\ell)}_{-\ell}} \epsilon_{i_1i_2\cdots i_m} I^{(0)}_{j-(i_1-1)} I^{(2)}_{j-(i_2-1)}\cdots I^{(m)}_{j-(i_m-1)} \nn \\
&=& \frac{D^{(1)}_0}{\prod\limits_{\ell = 2}^{m} I^{(\ell)}_{-\ell}} W_{j,1} \ .
\eea
Substituting the above expression back into eq.~\eqref{eq:d_0} and recalling the definition of the branch cut amplitude from \cite{Ansorg:2016ztf} in terms of $d_0$ as $\eta(s):=i\, d_0/(2\pi)$, we obtain
\beq
\eta(s)=   \Im\left( \frac{1}{\pi W^+_{-1} }\right)  \sum_{j=0}^{\infty} q_j \Xi_{j,1} \ .
\eeq

\section{Reissner-Nordstr\"om spacetime}\label{sec:RN}

We apply the program presented in the previous sections to the Reissner-Nordstr\"om solution. A stationary charged black hole spacetime is characterised by the metric function
\beq
\label{eq:f(r)}
f(r) = 1 - \frac{2M}{r} + \frac{Q^2}{r^2} = \left(1-\frac{r_+}{r}\right)\left(1-\frac{r_-}{r}\right) \ .
\eeq
Here, $M$ and $Q$ are, respectively, the mass and charge of the black hole, while $r_\pm$ are the coordinate values of the horizons given by
\beq
r_\pm = M\left[ 1\pm \sqrt{1 - \epsilon^2} \right], \quad \epsilon = \frac{Q}{M} \ ,
\eeq
$r_+$ corresponding to the event horizon and $r_-$ to the Cauchy horizon.
The Schwarzschild spacetime is recovered when $\epsilon=0$, while the extreme black-hole solution is obtained with $\epsilon = 1$. Note that the approach to extremality can lead to different geometries~\cite{Geroch:1969ca,Bengtsson:2014fha}.

A convenient alternative parametrisation to this spacetime is given in terms of
\beq
\label{eq:parameter_kappa}
\kappa = \frac{r_-}{r_+} \Rightarrow \epsilon= 2\frac{\sqrt{\kappa}}{1+\kappa} \ .
\eeq
The Schwarzschild and extremal black-hole limits correspond to $\kappa = 0$ and $\kappa = 1$, respectively. This new parameter simplifies the expressions we are going to study. In particular, the coordinate locations of the horizons are re-written as
\beq
r_+ = \frac{2M}{1+\kappa}, \quad r_- =  \frac{2M\kappa}{1+\kappa} \ .
\eeq

The propagation of scalar, electromagnetic and linear gravitational fields in the Reissner-Nordstr\"om spacetime is dictated by the master wave equation \eqref{eq:WaveEq} introduced in section \ref{sec:WaveEquation} in terms of the potential (cf.~\cite{Moncrief74a,Moncrief74b,Moncrief75,Zerilli74})  
\bea
\label{eq:Potential_RN}
{\rm P}(r) =  \ell(\ell+1) +\frac{r_+}{r}\left( \mu - \kappa\,q\frac{r_+}{r}\right) \ ,
\eea
with
\bea
\displaystyle
\mu &=& \frac{1}{r_+}\times \left\{ 
\begin{array}{c}
2M  \\
-3M\left[1 - \sqrt{1+ \epsilon^2\, A  }\right]  \\
-3M\left[1 + \sqrt{1+\epsilon^2\, A }\right] 
\end{array}
\right.\nn \\ 
&=&
\left\{ 
\begin{array}{cc}
1+\kappa &  \\
-\frac{3}{2}\left[1+\kappa - \sqrt{(1+\kappa)^2 +4 \kappa A }\right] \\
-\frac{3}{2}\left[1+\kappa + \sqrt{(1+\kappa)^2 +4 \kappa A }\right]
\end{array}
\right.,
\label{eq:Coeff_m} \\
q &=& \left\{ 
\begin{array}{c}
2 \\
-4  \\
-4
\end{array}
\right. \ .
\label{eq:Coeff_q}
\eea
Specifically, the three (ordered) alternative expressions correspond respectively to scalar, electromagnetic and gravitational perturbations.
In the expressions above, $A$ stands for $A = \left(\frac{2}{3}\right)^2(\ell+2)(\ell-1)$.

We are interested, however, in the conformally re-scaled wave equation~\eqref{eq:ConformalWaveEq_General} in section \ref{sec:MetricWaveRN} with the potential re-defined as \eqref{eq:ConfPotential_General}.
For the study of such equation, it will be more convenient to
re-parametrise $\mu$ and $q$ as
\bea
q&=& 3n -1  \nn \\
\label{eq:PotParam}
\mu &=& n(1+\kappa) - (1-n)\mathfrak{m}_{\pm} \\
{\mathfrak m}_\pm &=& \
\dfrac{1+\kappa \pm 3\sqrt{(1+\kappa)^2+4\kappa A}}{4} \ . \nn
\eea
As discussed in section \ref{sec:ConformalWaveEq}, the value $n=+1$ for scalar perturbations is a consequence of their natural conformal re-scaling.
For electromagnetic ($\mathfrak m_-$) or gravitational ($\mathfrak m_+$) perturbations, the value $n=-1$ is justified in section \ref{sec:MetricWaveRN}.

\subsection{The hyperboloidal foliation}\label{sec:HypFoliatRN}
The hyperboloidal slices are constructed according to the procedure discussed in sections \ref{sec:EFCoordinates} and \ref{sec:HypCoord}. Interestingly, in the context of the minimal gauge, two choices for the conformal representation of the spacetime appear as natural. The first one follows closely the steps
in \cite{Schinkel:2013tka,Schinkel:2013zm,Macedo:2014bfa,Ansorg:2016ztf} and it consists in fixing the areal radius $\rho(\sigma)$ to a constant --- see eq.~\eqref{eq:Fix_rho}. The second one maps the Cauchy horizon $r_-$ into a coordinate $\sigma_-$ that does not depend on the charge parameter $\kappa$. As we are going to discuss, the latter provides the geometrical counterpart to the framework introduce by Leaver~\cite{Leaver90} and it allows us to discuss the near-horizon limit of the extremal black hole.

\subsubsection{Areal radius fixing}\label{sec:AreaRadFix}
As discussed in section \ref{sec:PolynomGauge}, the minimal gauge is completely fixed by the free parameters $\lambda$ and $\rho_1$. In this
  section we discuss the first conformal representation case corresponding to
fixing the areal radius $\rho(\sigma)$ to a constant value $\rho_0$. With this aim, and keeping the notation in \cite{Ansorg:2016ztf}, we choose
\beq
\lambda = 2r_+, \quad \rho_1 = 0 \ ,
\eeq
which implies $\rho_0 = 1/2$.
This choice is equivalent to compactifying the spacetime and defining the height function via
\beq
\label{eq:Coord_rho_fix}
r=\frac{r_+}{\sigma}, \quad h(\sigma) = - \frac{1}{\sigma} + (1+\kappa)\ln(\sigma) \ .
\eeq
As a consequence, the conformal line element reads
\bea
\d \tilde{s}^2 &=&  \frac{1}{4}\d\omega^2
-\sigma^2\left( 1-\sigma \right)\left( 1-\kappa \sigma\right)\d\tau^2 \nn \\
&&+ \left[ 1+\sigma(1+\kappa) \right]\left[ 1+ \kappa (1+\kappa)(1-\sigma)\right] \d \sigma^2 \nn \\
&&+ \bigg( 1 - 2\sigma^2\left[ 1+ \kappa(1+\kappa)(1-\sigma)\right]\bigg)\d\tau\d\sigma \ .
\label{eq:Metric_RN_rhofix}
\eea
As prescribed, the event horizon ${\cal H}$ is given by $\sigma = 1$ while $\sigma = 0$ locates $\scri$. Furthermore, the Cauchy horizon $r_-$ is mapped to the value 
$\sigma_-= \kappa^{-1}$. In particular, we are interested in the exterior region $\sigma \in [0,1]$. 

The transformation \eqref{eq:Coord_rho_fix} and, consequently, the line element \eqref{eq:Metric_RN_rhofix} is well defined in the whole range of the parameter $\kappa \in [0,1]$. In particular, the limit $\kappa\to 1$ correspond to (the standard) extremal Reissner-Nordstr\"om (see below
  and e.g. \cite{HawEll73}).
However, the algorithm from section \ref{sec:Algorithm}, based on Taylor expansions, is only valid if $\sigma_-\ge 2$ --- see discussion about convergence radius in eq.~\eqref{eq:circle_C}. Specifically, in this particular coordinate system, methods based on a Taylor expansion around $\sigma=1$ applies only for $\kappa \in [0,1/2]$.

\subsubsection{Cauchy horizon fixing}\label{sec:CauchyHorzFix}
One way to avoid the limitation imposed in the range of the parameter $\kappa$ is to fix the location of the Cauchy horizon to a value that does not depend on $\kappa$. It is convenient to write $\sigma_-$ in terms of a constant $c$, as
\beq
\sigma_-= c^{-1} \ .
\eeq
Together with the choice $\lambda = 2r_+$, this fixes the parameters in the minimal gauge to
\beq
\rho_0 = \frac{1-\kappa}{2(1-c)} \ ,\quad \rho_1 = \frac{\kappa - c}{2(1-c)} \ ,
\eeq
which then leads to the compactification and height function
\beq
\label{eq:Coord_CauchyHorz_fix}
\begin{array}{c}
\displaystyle
r = 2r_+\frac{\rho(\sigma)}{\sigma} \ , \quad \rho(\sigma) = \frac{1-\kappa +\sigma (\kappa - c)}{2(1-c)} \ ,  \\
\displaystyle
h = -\frac{1-\kappa}{\sigma(1-c)} + (1+\kappa)\ln\sigma \ .
\end{array}
\eeq
By fixing the Cauchy horizon at $\sigma_-=c^{-1}>2$, we ensure that the algorithm based on Taylor expansions around the
event horizon $\sigma=1$ is always valid.

\subsubsection{Extremality and singular limits}
\label{s:sing_limits}
After having introduced the elements of these particular hyperboloidal coordinates, we comment on a
  geometric aspect relevant in the discussion below, namely the metric type
  of null infinity, with a focus on the extremal limit.

  With this aim,  we evaluate the metric type the
  vector normal to the hypersurface defined by $\Omega(\sigma)=0$, with $\Omega(\sigma) = \sigma/\lambda$, as
\bea
\label{eq:norm_dOmega}
\tilde{\nabla}^a\Omega \tilde{\nabla}_a \Omega &=& \frac{\sigma^2F(\sigma)}{\lambda^2\beta^2} \nn \\
&=& \frac{\sigma^2(1-\sigma)(1-c\sigma)(1-c)^2}{r_+^2\bigg(1-\kappa +\sigma(\kappa-c) \bigg)^2} \ .
\eea
In the compactification scheme of subsection \ref{sec:AreaRadFix}, with a gauge enforcing the constancy of the areal radius through
  $c=\sigma_-^{-1}=\kappa$, one obtains
  \bea
   \lim_{\sigma\rightarrow 0}  \tilde{\nabla}^a\Omega \tilde{\nabla}_a \Omega = 0 \ \ , \  \ c = \kappa \ ,
   \eea
 for all values of $\kappa$, including the extremal case $\kappa = 1$. Null infinity is therefore a null hypersurface
   both in the subextremal and extremal, in agreement with the standard extremal limit of Reissner-Nordstr\"om.
   The situation is more delicate for choices $c\neq \kappa$, such as in subsection \ref{sec:CauchyHorzFix}.
   In this case, for $\kappa<1$ the limit of (\ref{eq:norm_dOmega}) at $\sigma\to 0$
\beq
\lim_{\sigma\rightarrow 0}  \tilde{\nabla}^a\Omega \tilde{\nabla}_a \Omega = 
0 \ , \kappa\in [0,1) \ , \ c \neq \kappa \ .
  \eeq
We find that null infinity for subextremal Reissner-Nordstr\"om is a null
hypersurface, consistently with the previous gauge $c = \kappa$. On the other hand, in case of (naively) calculating 
the limit $\sigma\to 0$ of the expression (\ref{eq:norm_dOmega}) for $\kappa = 1$ one would get
\beq
\label{eq:norm_dOmega_kappa1}
\lim_{\sigma\rightarrow 0}  \tilde{\nabla}^a\Omega \tilde{\nabla}_a \Omega = 
r_+^{-2}=M^{-2}  \ ,  \ \kappa = 1 , c\neq \kappa \ ,
\eeq
suggesting that, in this conformal scheme, the conformal boundary for the extremal case $\kappa = 1$ is timelike~\footnote{Such
    timelike null infinity is typical of anti-de Sitter (AdS) spacetimes.}. However, we note that the change of coordinates
  in  eqs.~\eqref{eq:Coord_CauchyHorz_fix} is actually ill-defined for $\kappa=1$, since the transformation reduces to $r(\sigma) = r_+ =$
  constant. This suggests some kind of singular behaviour in the process of considering the limit of spacetimes as $\kappa\to 1$,
  when $c \neq \kappa$.

  In order to clarify this issue, and since the timelike character of null infinity does not depend on the particular value of $c$ (at least as long as $c\neq \kappa$), let us consider for concreteness the line element of the conformal metric in the gauge of section \ref{sec:CauchyHorzFix} defined
  by transformations (\ref{eq:Coord_CauchyHorz_fix}). In addition, we choose for simplicity $c=0$ or,
equivalently, $\sigma_-\to \infty$. Then
\bea
\label{e:line_element_c=0}
d\tilde{s}^2 &=&\frac{1-\kappa}{4\rho^2} \Bigg( -(1-\kappa) \, \sigma^2(1-\sigma)  \d\tau^2 \nn \\
&&-   \bigg( 1-2\sigma^2 -2\kappa(1-\sigma)\left[(2-\kappa(1-\sigma)\right]\bigg) \d\tau\d\sigma  \nn \\
&& + \left[1 +\sigma - \kappa(1-\sigma)\right] \d\sigma^2  \Bigg) + \rho^2 \d\omega^2.
\label{eq:Metric_RN_CauchyHorzfix}
\eea
The pathological behaviour for $\kappa = 1$ in this gauge is apparent in the line element.
However, a regularization is obtained through the introduction of an appropriate  re-scaling of the time coordinate
\beq
\label{eq:Time_Transf}
T=(1-\kappa)\tau \ .
\eeq
The combined transformations \eqref{eq:Coord_CauchyHorz_fix} and \eqref{eq:Time_Transf}, for $\kappa<1$, lead to~\cite{Bengtsson:2014fha}
\bea
d\tilde{s}^2 &=&\frac{1}{4\rho^2} \Bigg( -\sigma^2(1-\sigma)  \d T^2 \nn \\
&&-   \bigg( 1-2\sigma^2 -2\kappa(1-\sigma)\left[(2-\kappa(1-\sigma)\right]\bigg) \d\tau\d\sigma  \nn \\
&& + (1-\kappa)\left[1 +\sigma - \kappa(1-\sigma)\right] \d\sigma^2  \Bigg) + \rho^2 \d\omega^2 \ .
\eea
This expression makes now also sense in the limit $\kappa\to 1$, defining the line element
\bea
d\tilde{s}^2_{\rm ext} &=&  -(1-\sigma)  \d T^2 + \d T\d\sigma + \sigma^2 \d\omega^2 \ .
\label{eq:Metric_RN_CauchyHorzfix}
\eea
The corresponding physical manifold $\d s^2 = \Omega^{-2} \d\tilde{s}^2$, called the Bertotti-Robinson metric, is the near-horizon limit of the extremal black hole. Its topology is a direct product of $1+1$ AdS spacetime with a sphere of constant radius, indeed with a timelike
null infinite in accordance with the expression in (\ref{eq:norm_dOmega_kappa1}). We are therefore in the presence 
of two different extremal limits of Reissner-Nordstr\"om, an illustration of the subtleties to be considered when discussing the limits
of spacetimes \cite{Geroch:1969ca,Paiva:1993bv}.

\subsubsection{Geometric insights into Leaver's treatment}
\label{s:Leaver}
The previous discussion on singular limits of spacetimes is not academical. On the contrary, it is directly relevant to
Leaver's treatment of QNMs in Reissner-Nordstr\"om~\cite{Leaver90} since, as we show below, the latter corresponds to the
  gauge choice in subsection \ref{sec:CauchyHorzFix}, namely the fixing of the Cauchy horizon coordinate position with $\sigma_-=\infty$.
  Since, as seen in \ref{s:sing_limits},
  such coordinates suffer a singular extremal limit, the question about the extremal limit QNM calculation following
  \cite{Leaver90} is raised.

Let us justify the connection between Leaver's treatment in \cite{Leaver90} and the coordinates introduced in \ref{sec:CauchyHorzFix}.
  For this,  let us consider the coordinate $u$  in the Taylor expansion \eqref{eq:Ansatz_phi},
  a coordinate that naturally emerges in the geometric discussion in terms of hyperboloidal slices.
  Indeed,  under the  choice $c=0$ leading to the line element (\ref{e:line_element_c=0})  and fixing the Cauchy horizon at $\sigma_- = \infty$,
  such coordinate $u$ is written in terms of the original coordinate $r$ as
\beq
\label{eq:Coord_u}
u  = \frac{r - r_+}{r-r_-} \ .
\eeq
This is precisely the coordinate introduced by~\cite{Leaver90}, here identified within a natural geometric setting.
The key point in the present discussion refers to the fact that the treatment
  in \cite{Leaver90} (based on Cauchy slices) requires the introduction of regularization factors in the Taylor expansion
  in order to deal with the QNM asymptotic conditions. Remarkably, and as it happened in \cite{Ansorg:2016ztf},
  such factors arise in the present discussion directly as
  a consequence of the geometric treatment in terms of hyperboloidal slices.
  As a matter of fact, the factor $Z$ in Eq.~\eqref{eq:Func_z} reads
\bea
\label{eq:Z_factors}
Z(r) &=& \rho^{n}e^{s \left[ \frac{r^*}{\lambda} +h \right]   }  \nn \\
      &=&C  \rho^{n}\left( \frac{r_+}{r}\right)^{s_\kappa} e^{-s_\kappa r_\kappa} \left( 1- \kappa \frac{r_+}{r}\right)^{-s_\kappa }  u^b \ ,
      \label{eq:Z_Leaver}
\eea
where $C$ is a constant. The parameters 
\beq
\label{eq:LeaverExtraFactor}
s_\kappa = \frac{(1+\kappa)s}{2}, \, r_\kappa = \frac{r}{(1+\kappa)r_+}, \, b = \frac{s}{2(1-\kappa)} \ ,
\eeq
are adapted to the normalisation used in \cite{Leaver90}. 

For $n=-1$, the factors in (\ref{eq:Z_factors}) coincide exactly with the ones appearing multiplying the Taylor expansion in \cite{Leaver90}. More specifically,  the first factor
  (an appropriate power of $\rho$) was introduced in \cite{Leaver90} to reduce the number of terms in the recurrence relation --- see section \ref{sec:MetricWaveRN} --- whereas the rest were introduced to incorporate the boundary conditions appropriate for QNMs.
    
  In contrast, in the present discussion the factor $\rho^{n}$ is motivated as a  direct consequence of using the {\em conformal} wave equation, whereas the rest of the factors are a straightforward consequence of the  use of hyperboloidal slices.

  Note that \cite{Leaver90} considers only electromagnetic and gravitational perturbations. Therefore we are led to the conclusion that the appropriate
  conformal re-scaling factor introduced in section~\ref{sec:ConformalWaveEq} is $n=+1$ for scalar field and $n=-1$ for electromagnetic and linear gravitational fields 
 
  To summarize the discussion in this section \ref{sec:HypFoliatRN}, we identify two relevant outcomes of the present geometric approach:
  \begin{itemize}
  \item[(i)] It is the simultaneous combination of a conformal approach and the use of the hyperboloidal
  slices that renders a  geometric explanation of the factors in Leaver's Taylor expansion approach to quasi-normal modes.
\item[(ii)] Such geometric perspective provides an insight into the extremal limit corresponding to Leaver's quasi-normal modes
  calculation. Indeed, according to the discussion above, the approach to extremality in \cite{Leaver90} (i.e. $c=0$)
  would correspond to the Bertotti-Robinson spacetime rather than the standard extremal Reissner-Nordstr\"om.
  A natural question to assess is the relation between the limit values of QNMs in such approach to
    $\kappa\to 1$ and the QNMs directly calculated in the standard extremal  Reissner-Nordstr\"om spacetime \cite{Onozawa:1995vu}.
  \end{itemize}

\subsection{Laplace analysis}\label{sec:MetricWaveRN}
We now proceed to write explicitly all the elements for the Laplace analysis of the Reissner-Nordstr\"om spacetime in the hyperboloidal minimal gauge. We consider here the two cases discussed in the previous section, namely \ref{sec:AreaRadFix} and \ref{sec:CauchyHorzFix}.
\subsubsection{Areal radius fixing}
Here, the areal radius is fixed to $\rho_0 = 1/2$ and the Cauchy horizon depends on the charge parameter as $\sigma_- = \kappa^{-1}$. According to discussion in section \ref{sec:WaveEquation}, in this particular gauge there is no distinction between considering a wave equation based on a conformally invariant framework \eqref{eq:ConformalWaveEq_General} or simply applying a coordinate transformation into the well-known formulation \eqref{eq:WaveEq}.

The Laplace operator \eqref{eq:ODE_OperatorA} and the source function \eqref{eq:ODE_SourceB} read
\bea
&&{\mathbf A}(s) = {\sigma}^{2} \left( 1-\kappa \sigma \right)  \left(1- \sigma \right) \partial_{\sigma\sigma} +  
 \Bigg(  \left( 2- 3\,\sigma \right)\sigma \nn\\
 && + \kappa \sigma^2 \left( 4\sigma-3 \right) 
 +  s \left[ 1 - 2\sigma^2  - 2\kappa \left(1 +  \kappa \right) {\sigma}^{2}(1-\sigma) \right]  \Bigg) \partial_\sigma \nn \\
&& - \Bigg(  \ell(\ell+1) +\sigma( \mu - \kappa q \sigma) 
+ s \sigma \left[  2 +  \left(2- 3\sigma \right)\kappa(1+ \kappa)  \right] \nn \\
&& + \left[ 1 + \left( 1-\sigma \right)\kappa(1+\kappa)  \right]  \left[1 +\sigma(1+ \kappa) \right] {s}^{2} \Bigg), \\
&& B(s) = - \left[ 1 + \left( 1-\sigma \right)\kappa(1+\kappa)  \right]  \left[1 +\sigma(1+ \kappa) \right] \bigg( s\Phi_0 + \dot{\Phi}_0\bigg) \nn \\
&& + \bigg( 1 - 2\sigma^2\left[1   +  \kappa \left(1 +  \kappa \right) (1-\sigma)\right] \bigg)\Phi_0{}_{,\sigma} \nn \\
&& - \sigma \left[  2 +  \left(2- 3\sigma \right)\kappa(1+ \kappa)  \right]\Phi_0 \ .
\eea

With the expansion around the horizon \eqref{eq:Ansatz_phi}, we obtain a 3-order recurrence relation. According to our notation --- see eqs.~\eqref{eq:HomRecRel} and \eqref{eq:InHomRecRel}, we have $m=2$ and coefficients
\bea
\label{eq:RecRelCoeff}
&\alpha_k& =  \left( k+1 \right)   \left[ s + (1- \kappa)(1+k) \right], \nn \\
&\beta^{(0)}_k& = -\ell(\ell+1) -\mu  -2 (k+s +1)(k+s) + s \kappa^2 \left( 1+ 2k \right)   \nn \\
&& + \kappa\left[ (k+s)(3k-s+1) + q+ 2k \right]    \\
&\beta^{(1)}_k& = \mu-1 +(k+s)^2 - \left[ (k+s)(s+3k) + 2q- 3 \right] \kappa   \nn \\
&& - (s \kappa +  4k + 3s  )s {\kappa}^{2} \nn \\
&\beta^{(2)}_k& = \kappa \left( \kappa s+k+s+1 \right)  \left( \kappa s+k+s-2
 \right) +\kappa q \ .  \nn 
\eea

A study of the asymptotic behaviour to the solutions to homogeneous recurrence relation shows --- cf. the discussion around \eqref{eq:Asympt_Ansatz}.
\beq
\label{eq:AsympBeh}
I^{(0)}_k \sim  e^{\xi \sqrt{k} } k^{\zeta}, \,\,  I^{(1)}_k \sim e^{-\xi \sqrt{k}}  k^{\zeta}, \,\, I^{(2)}_k \sim  \upsilon^ k k^{\varsigma}
\eeq
with
\bea
\label{eq:AsympCoeff_exp}
\displaystyle
&\displaystyle \xi = 2\sqrt{s}, \quad \upsilon = \frac{\kappa}{\kappa-1} \ , \\
\label{eq:AsympCoeff_pow}
&\displaystyle \zeta =  \frac{1+\kappa}{2}s - \frac{3}{4},  \quad \varsigma = -\left(1 +  \frac{\kappa^2}{1-\kappa}s\right) \ .
\eea
As expected, assumption (2) in section \ref{sec:TaylorExpansions} is valid only for $\kappa \in [0, 1/2]$. Indeed, for $\kappa > 1/2$, $I^{(2)}_k$ is no longer a decaying solution and its growth reflects the fact that the Cauchy horizon $\sigma_-=\kappa^{-1}<2$ has entered the unit circle \eqref{eq:circle_C}.

\subsubsection{Cauchy horizon fixing}\label{sec:TaylorCauchyHorizon}
In this case, the Cauchy horizon is fixed at $\sigma_- = \infty$ and the areal radius reads
\[
\rho(\sigma) = \frac{1-\kappa(1-\sigma)}{2} \ .
\]
Since $\rho(\sigma)$ is no longer a constant, eqs.~\eqref{eq:Hyper_WaveEq} and \eqref{eq:ConformalWaveEq_General} differs in this gauge. As stated previously, from the theoretical perspective, we regard the conformally invariant wave equation as the most natural choice, due to its geometrical formulation. Thus, the Laplace operator \eqref{eq:ODE_OperatorA} reads\footnote{In the Cauchy fixing gauge, expressing explicitly the dependence of the parameter $m$ and $q$ in terms of $n$ --- cf.~\eqref{eq:PotParam} --- simplifies the form of the operator ${\mathbf A}(s)$. This accounts for the appearance of parameters $n$ and $\mathfrak{m}_{\pm}$ in eq.~\eqref{eq:OperA_HorzFix}. In particular, terms going as $\rho^{-1}$ vanish for $n=1$.}
\bea
&&{\mathbf A}(s) = \left( 1- \kappa \right)\sigma^2  \left(1- \sigma \right) \partial_{\sigma\sigma} 
+\Bigg( \sigma \left[ 2-\sigma \left( 1+2n  \right)  \right]  \nn \\
&&  - \dfrac { \left( 1-n \right) {\sigma}^{2}  }{\rho} - 2s( \sigma^2-2\rho^2)  \Bigg)\partial_\sigma 
 -s^2(\sigma+ 2 \rho ) \nn \\
 && -2s\left[ \sigma-  \kappa \rho\, n  -  \dfrac {  \kappa(1-  n)\sigma^2}{2\rho}  \right]  -(1-\kappa)\Bigg( \ell(\ell +1) \nn \\
&& + \sigma+ (1- n)\left[ 1 -  \dfrac{1-\kappa + \sigma(1+{\mathfrak m}_{\pm})   }{2\rho}\right]  \Bigg) \ , 
\label{eq:OperA_HorzFix}
\eea
with the source given by
\bea
&& B(s) =-(\sigma+ 2 \rho ) \bigg( s\Phi_0 + \dot{\Phi}_0\bigg) - 2( \sigma^2-2\rho^2)\,\Phi_{0,\sigma} \nn \\
&& -2\left[ \sigma-  \kappa \rho\, n  -  \dfrac {  \kappa(1-  n)\sigma^2}{2\rho}  \right] \Phi_0 \ . \label{eq:SourceB_HorzFix}
\eea

The choice for eq.~\eqref{eq:ConformalWaveEq_General} (instead of \eqref{eq:Hyper_WaveEq}) is also justified from a practical point of view, as the Taylor expansion \eqref{eq:Ansatz_phi} leads to a simpler recurrence relation. Indeed, the recurrence relation for scalar perturbation is actually of order 2 ($m=1$) with coefficients
\bea
&\alpha_k& =  \left( k+1 \right)  \left[ s+(1- \kappa)(1+k) \right] \nn \\
&\beta_k^{(0)}& = -(1-\kappa)\left[1+ \ell \left( \ell+1 \right) +2k\left( k+ s+1\right) \right] \nn \\
&& +s\left[ \kappa - 2\left( k+ s+1\right)\right] \\
&\beta_k^{(1)}& = \left[ k+ s(1+\kappa) \right]  \left[ s + k(1-\kappa) \right] \ . \nn
\eea
The asymptotic behaviours for the solutions of the homogeneous recurrence relation are
\beq
\label{eq:AsympBehavScalarLeaver}
I^{(0)}_k \sim  e^{\xi \sqrt{k(1-\kappa)} } k^{\zeta}, \,\,  I^{(1)}_k \sim e^{-\xi \sqrt{k(1-\kappa)}}  k^{\zeta} \ ,
\eeq 
with $\xi$ and $\zeta$ still given by \eqref{eq:AsympCoeff_exp} and \eqref{eq:AsympCoeff_pow}.

For electromagnetic/gravitational perturbation ($n=-1$), we obtain the 3-order recurrence relation ($m=2$)
\bea
&\alpha_k& =  \left( k+1 \right)  \left[ s+(1- \kappa)(1+k)  \right]  \nn \\
&\beta_k^{(0)}& = - \left( 1-\kappa \right)\left[\ell \left( \ell+1 \right) +\mu\right]  - 2\left( k+ s+ 1\right) \left[s + k(1-\kappa)\right] \nn\\ 
 &&     + \kappa \left( k-3 \right) \left[   (1-k)(1-\kappa) - s  \right] \nn  \\
&\beta_k^{(1)}& =  \left[ -1+\mu+\kappa \ell \left( \ell+1 \right) +12 \kappa \right]  \left( 1-\kappa \right) \nn \\
&& + \left( k+s \right) ^{2} \left( 1+\kappa \right) + 2\kappa\left( s-5 \right) \left[k +  s \left( 1-\kappa \right) \right]  \\
&& - \left( k+2s-5 \right)  \left( 2k-s \right) {\kappa}^{2} \nn \\
&\beta_k^{(2)}& = -\kappa \left[ s(1+\kappa)+k-3 \right]  \left[ s+ (k-3)(1-\kappa) \right] \nn \ .
\eea
Apart from \eqref{eq:AsympBehavScalarLeaver}, there exists here a third solution to the homogeneous recurrence relation whose asymptotic behaviour is
\beq
\label{eq:AsympBehavEMGravLeaver}
I_k^{(2)} \sim \kappa^k k^{-6} \ .
\eeq

\subsection{Results}

We end this section with the application of the algorithm from sec.~\ref{sec:Algorithm}  to the Reissner-Nordstr\"om spacetime. As a representative example, we show here results for scalar and gravitational fields, whereas the discussion in appendix is exemplified by an electromagnetic perturbation. The qualitative discussion, though, does not depend on the specific choice of the parameters in the effective potential.

We first calculate the quasinormal modes (QNM) frequencies as the zeros of the Wronskian determinant $W_{-1}$ \eqref{eq:detW_-1}. The algorithm for this procedure is described in \cite{Ansorg:2016ztf} (cf. section IV.D.2 in that reference). The results were obtained by dividing the interval $\kappa=[0,0.5]$ (areal radius fixing gauge) or $\kappa=[0,0.99]$ (Cauchy horizon fixing gauge) into an equidistant $\kappa$-grid of size $\sim10^{-2}-10^{-3}$.

\begin{figure}[h!]
\begin{center}
\includegraphics[width=8.5cm]{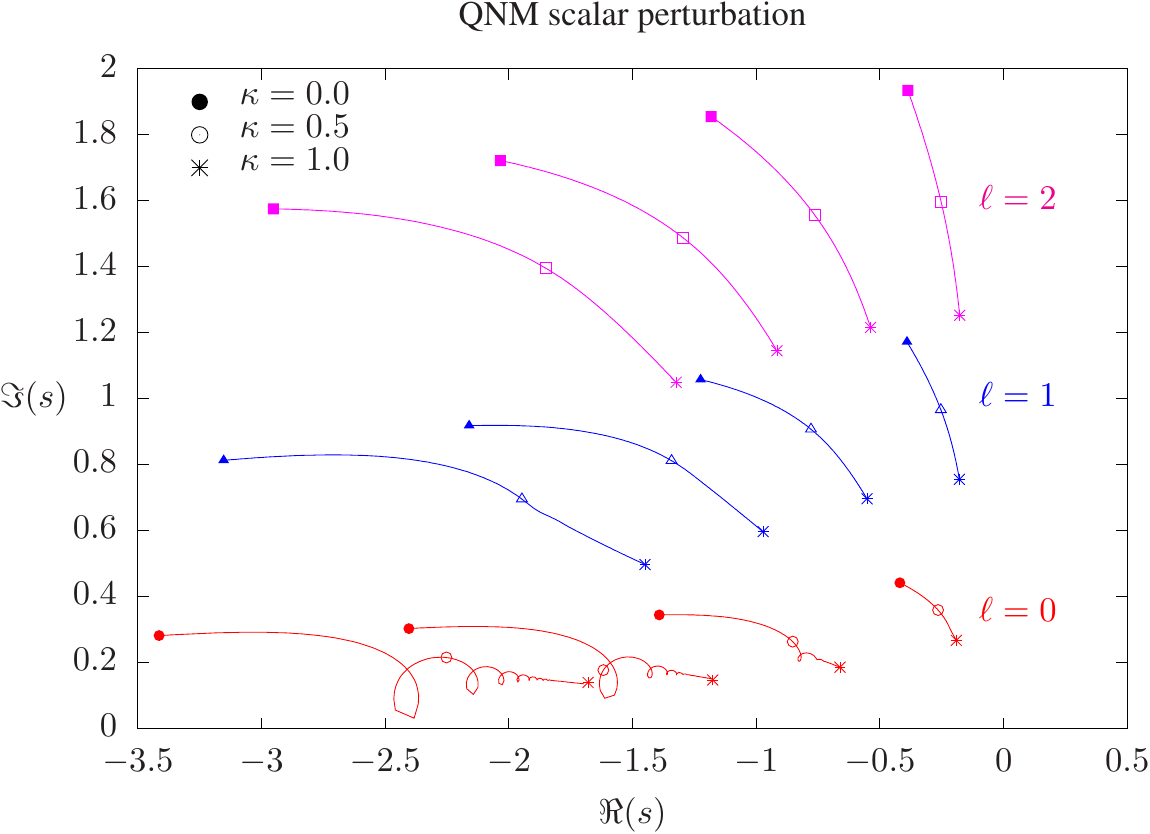}
\end{center}
\caption{Quasinormal modes for scalar perturbations for $\ell=0$ (red circle), $\ell = 1$ (blue triangle) and $\ell=2$ (magenta square). The solution was obtained in the ``areal radius fixing" gauge ($\kappa\in[0,0.5]$) and the ``Cauchy horizon fixing'' gauge ($\kappa\in[0,0.99]$). The Schwarzschild value ($\kappa =0$) is given by solid points, while the empty points mark the limit of validity for the method in the ``areal radius fixing" gauge ($\kappa=0.5$). Even though, the extremal limit in the ``Cauchy horizon fixing" gauge leads to the Robinson-Bertotti spacetime, a continuous extrapolation of the values for $\kappa \lesssim1$ into the extremal value $\kappa=1$ is consistent with the results obtained in the literature~\cite{Onozawa:1995vu} (star points).}
\label{fig:QNM_Scalar}
\end{figure}

Fig.~\ref{fig:QNM_Scalar} displays the results for scalar perturbations with $\ell=0$ (red circle), $\ell=1$ (blue triangle) and $\ell=2$ (magenta square). The solid points are the values for the Schwarzschild solution $\kappa=0$, whereas the empty points mark the value $\kappa=0.5$ for which the specific algorithm from section~\ref{sec:Algorithm} ceases to be valid in the ``areal radius fixing" gauge. As expected, the results for $\kappa\in[0,0.5]$ coincides in both gauges. Moreover, in the ``Cauchy horizon fixing" gauge we can calculate further until $\kappa \lesssim1$. The value $\kappa=1$ is not available in this gauge because the limiting process is discontinuous and it leads to the near-horizon geometry. It is interesting to notice that the looping structure in the higher QNMs for the scalar field with angular mode $\ell=0$ is not an artefact of the numerical resolution, but it actually reflects the parametric dependence of the QNMs on $\kappa$\footnote{We do not have an explanation for this behaviour and a clarification would require further work.}.

The limitation in the calculation of the QNMs in the ``areal radius fixing" gauge for $\kappa\leq 0.5$ is a consequence of the method --- see discussion around eq.~\eqref{eq:circle_C} --- and not an intrinsic issue of QNMs. Other methods --- such as the one\footnote{The work~\cite{Ammon:2016fru} describes an alternative method to calculate not only the QNMs, but also the discrete amplitudes $\eta_n$. The method, however does not allow us to calculate the branch cut amplitude $\eta(s)$.} employed in~\cite{Ammon:2016fru} --- can be used to calculate the modes in the full range $\kappa \in [0,1]$.  In particular, this alternative method was used, on the one hand, to confirm the looping structure in the higher QNMs for the scalar field with angular mode $\ell=0$. On the other hand, the method from~\cite{Ammon:2016fru} allows us to calculate the QNMs for the extremal Reissner-Nordstr\"om case $\kappa=1$ (star points in Fig.~\ref{fig:QNM_Scalar}). 

The results are in accordance with the literature~\cite{Onozawa:1995vu} and it is consistent with a continuous extrapolation of the values obtained in the ``Cauchy horizon fixing" gauge. This allows us to answer the question posed in point (ii) at the end
  of section \ref{s:Leaver}: the discontinuous spacetime limit at $\kappa\to 1$ does not reflect in a discontinuous limit in the QNMs. This naturally raises the issue
of a possible ``QNM-isospectrality'' between extremal Reissner-Nordstr\"om and the near-horizon Bertotti-Robinson metric.

\begin{figure*}[t!]
\begin{center}
\includegraphics[width=8.5cm]{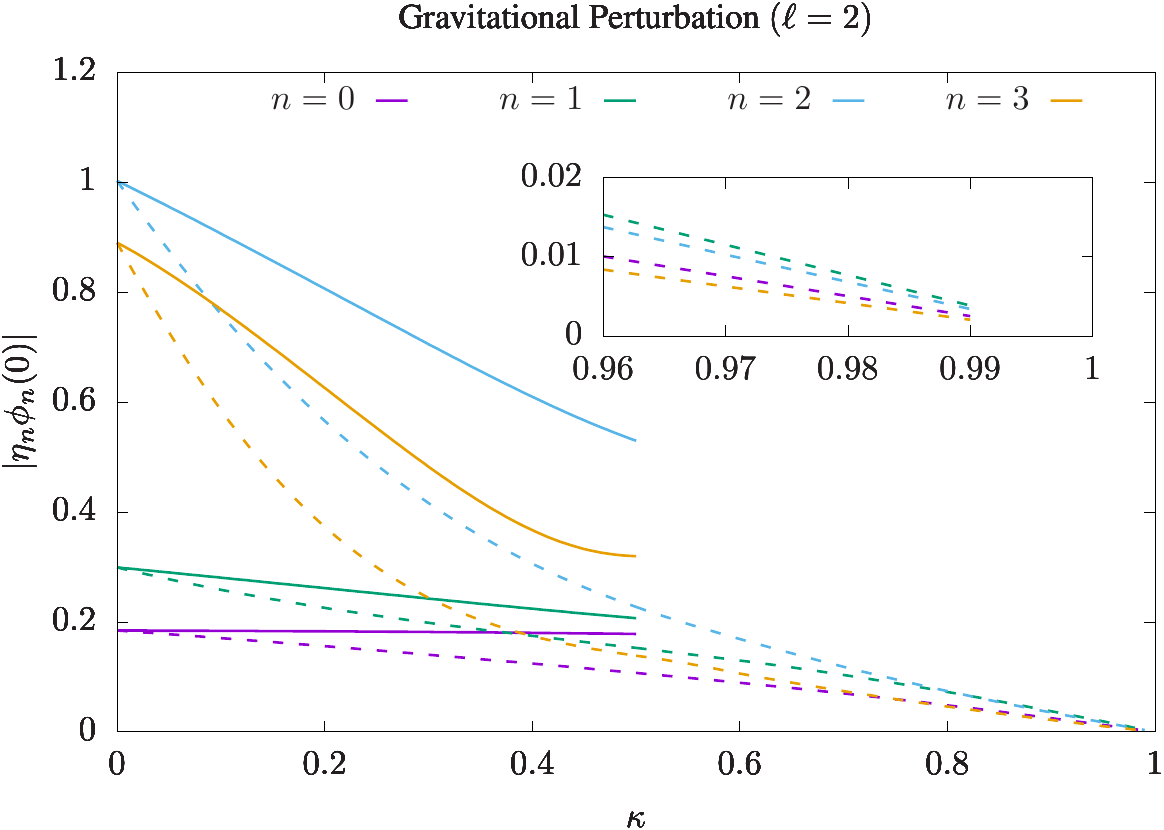}
\includegraphics[width=8.5cm]{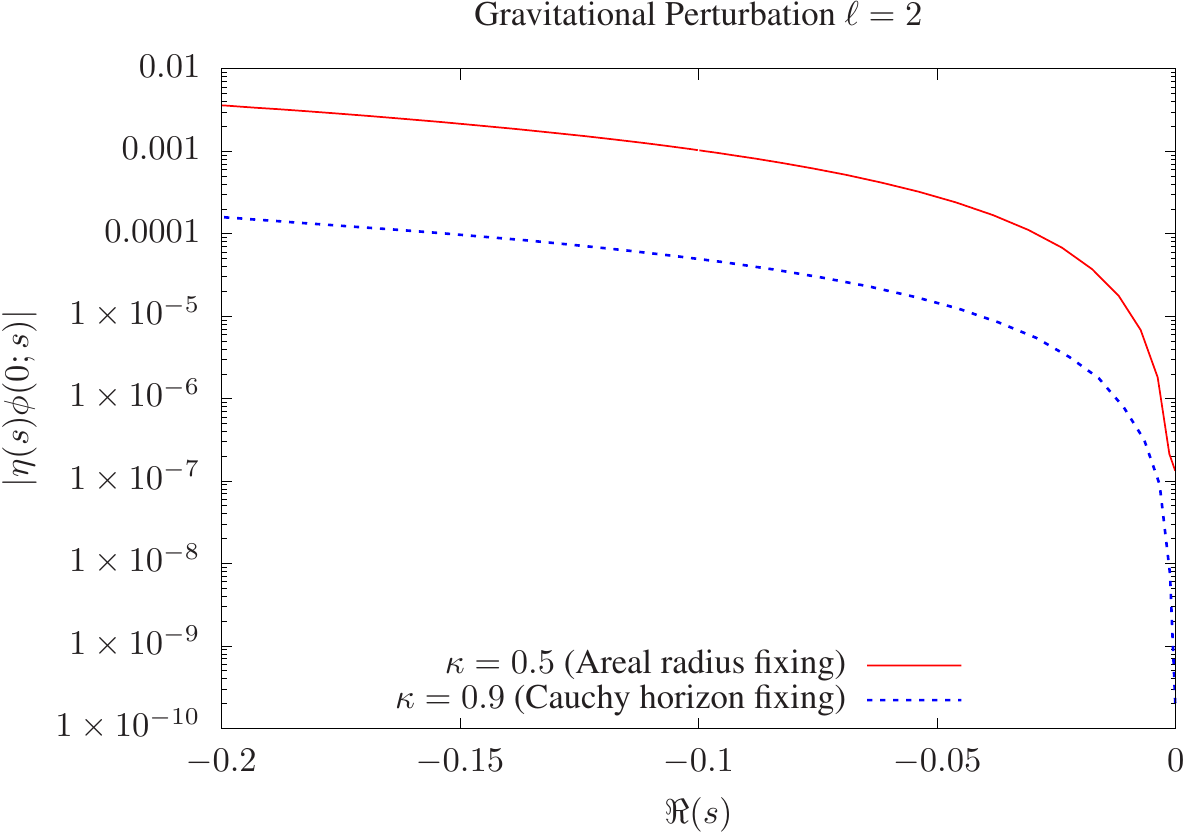}
\end{center}
\caption{Left panel: quasinormal modes amplitudes $\eta_n \phi_n(\sigma)$ at $\sigma =0$ as a function of the charge parameter $\kappa$ for the first four QNMs. The continuous lines displays the results in the ``areal radius fixing" gauge ($\kappa\in[0,0.5]$), while the dotted lines in the ``Cauchy horizon fixing" gauge ($\kappa\in[0,0.99]$). The inset focuses on the behaviour around $\kappa = 1$ and it shows that the amplitudes vanish in the limit to the Robinson-Bertotti spacetime. Right panel: branch cut amplitude at $\sigma = 0$ for $\kappa = 0.5$ (``areal radius fixing" gauge) and $\kappa = 0.9$ ``Cauchy horizon fixing" gauge. Both behave as $~s^2$, leading to a late time decay in the form $\tau^{-3}$ --- see fig.~\ref{fig:SpecTimeEvol}.
}
\label{fig:Amplitudes}
\end{figure*}

Once QNM frequencies are obtained, we address the resonant expansion aspects of the algorithm. That is, we proceed to the calculation of the amplitudes related to each quasi-normal mode $\eta_n$ and to the branch cut $\eta(s)$ for given initial data. As a representative example, we consider gravitational perturbation with angular mode $\ell=2$ and the initial data
\beq
\label{eq:ID_example}
V_0(\sigma) = \sigma(1-\sigma) \quad \dot{V}_0(\sigma) = 0 \ .
\eeq
In particular, we show in fig.~\ref{fig:Amplitudes} the invariants $\eta_n \phi_n(\sigma)$ and $\eta(s)\phi(s;\sigma)$, both calculated at $\scri$ ($\sigma =0$).

The left panel shows the dependence of the quasi-normal mode amplitude $\eta_n$ as a function of $\kappa$ for the ``areal radius fixing" gauge for $\kappa\in[0,0.5]$ (continuous lines) and the ``Cauchy horizon fixing" gauge $\kappa\in[0,0.99]$ (dotted lines). In particular, the inset displays the value around $\kappa \lesssim1$ where we observe a tendency towards zero. A full understanding of this limiting behaviour would require the control of boundary data. Indeed, since the limiting spacetime in ``Cauchy horizon fixing" gauge has a timelike boundary involving the topology of AdS$_2$, a unique solution to the wave equation is no longer obtained only with the initial data~\eqref{eq:ID_example} and boundary conditions at $\sigma = 0$ are required as well. In~\cite{Ammon:2016fru}, the QNM amplitudes of asymptotically AdS spacetimes are obtained after reformulating the corresponding wave equations to include the Dirichlet boundary conditions. 

\begin{figure}[b!]
\begin{center}
\includegraphics[width=8.5cm]{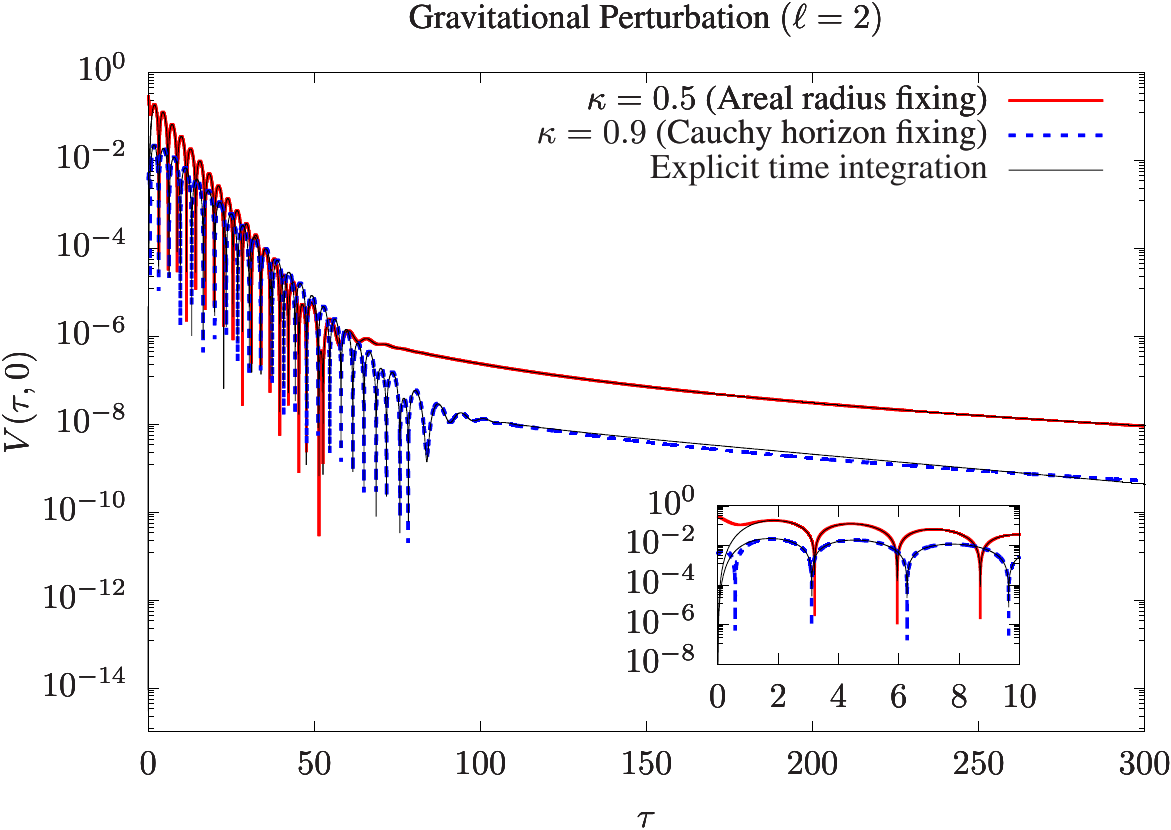}
\end{center}
\caption{Time dependence according to the spectral decomposition~\eqref{eq:VSol_spectral}. The straight red line is obtained for $\kappa=0.5$  in the ``areal radius fixing" gauge, whereas the dotted blue line is for $\kappa=0.9$ in the "Cauchy horizon fixing" gauge. The black straight line corresponds to the explicit time evolution with the code~\cite{Macedo:2014bfa}. The inset focus on the match between both methods around $\tau=0$. 
}
\label{fig:SpecTimeEvol}
\end{figure}

Then, the right panel of fig.~\ref{fig:Amplitudes} shows $|\eta(s)\phi(s;\sigma)|$ around $s=0$ for $\kappa = 0.5$ (``areal radius fixing" gauge) and $\kappa =0.9$ (``Cauchy horizon fixing" gauge). The behaviour of the branch cut amplitude around the origin is responsible for the late-time tail decay. Indeed, assuming that
\beq
\eta(s)\phi(s;\sigma) \sim s^{\gamma} \ ,
\eeq
we obtain
\bea
V_{\rm tail}(\tau,\sigma) &=& \int_{-\infty}^0 \eta(s)\phi(s;\sigma) e^{s\tau} d s \nn \\
&\sim& \int_{-\infty}^0  s^{\gamma} e^{s\tau} d s 
\sim \tau^{-(\gamma +1)} \ .
\eea
From the plot in Fig.~\ref{fig:Amplitudes}, one can read the behaviour $\gamma = 2$, which should lead to the expected tail decay $\tau^{-3}$ along future null infinity~\cite{Gundlach94a}. Indeed, fig.~\ref{fig:SpecTimeEvol} shows the complete time dependence according to the spectral decomposition ~\eqref{eq:VSol_spectral} for $\kappa=0.5$ (``areal radius fixing" gauge) and $\kappa=0.9$ (``Cauchy horizon fixing" gauge). Specifically, we have considered only the first $4$ dominant QNMs in the spectral decomposition~\eqref{eq:VSol_spectral}. Moreover, the integration along the branch cut is performed here just in the interval $s\in[-0.2,0)$ which is enough to focus in the late times. For comparison, the figure also brings the evolution obtained via an explicit time integration of the wave equation with the code~\cite{Macedo:2014bfa}, the inset displaying the match between both methods for $\tau \sim 0$. Notice the remarkable agreement between the spectral decomposition \eqref{eq:VSol_spectral} and the time evolution --- see last paragraph in appendix~\ref{App:AsymExp} for a discussion concerning the assessment of conjecture \ref{cnjc:SpecDecomp}.

\section{Discussion}\label{sec:Discussion} 

In this article we have studied the spectral decomposition of solutions to dissipative linear wave equations formulated on spherically symmetric, stationary and asymptotically flat spacetimes containing a black hole. Focus has been placed on a particular geometric frame that exploits the use of conformal compactifications together with a foliation on hyperboloidal hypersurfaces.  
Specifically, we extend the results in~\cite{Ansorg:2016ztf} by (i) defining the so-called minimal gauge, as the simplest hyperboloidal gauge adapted to a spherically symmetric, stationary and asymptotically flat, black-hole spacetime and (ii) generalising \blue{} the semi-analytical algorithm based on a Taylor expansion that enables us to construct the ingredients $\{s_n,\phi_n, \eta_n\}_{n=1}^\infty$ as well as $\{\phi(s), \eta(s)\}_{s\in\mathbb{R}^-}$ of the spectral decomposition \eqref{eq:VSol_spectral}. Such extensions involve non-trivial technical generalizations allowing us now to deal with multiple horizon settings and higher-order recurrence relations in the underlying algorithm.
Moreover, we have detailed the discussion on how this geometrical framework gives rise naturally to the regularisation factors needed in the usual Cauchy-based foliation to incorporate the appropriate boundary conditions leading to quasi-normal modes.

As a particular example, we have applied the formalism to the Reissner-Nordstr\"om spacetime. In a first stage we have focused on the geometric content of the framework, showing that the minimal gauge leads to two natural different conformal hyperboloidal compactifications, referred in this paper as ``areal radial fixing" and ``Cauchy horizon fixing". The former fixes the areal radius of the conformal spacetime $\rho$ to a given constant value, leading to a system in which the coordinate value of the Cauchy horizon depends on the charge parameter $\kappa$. The latter, instead, fixes the coordinate location of the Cauchy horizon to a value independent of the charge parameter. 

While both gauges reduce in the Schwarzschild limit $\kappa = 0$ to the coordinate system used in~\cite{Ansorg:2016ztf}, they lead to different geometries in the limit to extrematility. For $\kappa=1$, the ``areal radial fixing" gauge provides the usual extremal Reissner-Nordstr\"om black-hole limit, while the ``Cauchy horizon fixing" gauge has the near-horizon geometry as limit. This solution --- the so-called Bertotti-Robinson spacetime --- is the direct product of AdS$_2$ with a sphere of constant radius. Even though the distinct spacetime limits were previously discussed in~\cite{Carroll09,Bengtsson:2014fha}, they were recovered in this work via different choices for the conformal compactification of the spacetime. It would be insightful to study the topic of spacetime limits in a more generic context within the realm of conformal methods in General Relativity~\cite{Kroon:2016ink}. 

On the top of that, we also  identified the ``Cauchy horizon fixing" gauge as the geometrical counterpart of Leaver's treatment of quasi-normal modes in Reissner-Nordstr\"om~\cite{Leaver90}. It is well-known that Leaver's methodology is not valid at $\kappa=1$ (cf.~e.g.~\cite{Leaver90,Onozawa:1995vu}). Due to the understanding of the limiting process in the ``Cauchy horizon fixing" gauge, we conclude that the limitation of Leaver's algorithm is not just technical, but rather a consequence of the geometry employed. 

Note that the original algorithm~\cite{Leaver90} was modified in~\cite{Onozawa:1995vu} to treat the extremal case exclusively. The strategy\footnote{The same strategy was recently used in the context of the extremal Kerr black hole~\cite{Richartz:2015saa} as well.} was to perform the Taylor expansion around the point $r_0 = r_+/2$ instead of $r_0 = r_+$. Even though the conclusion is that ``the numerically computed values are consistent with the values earlier obtained by Leaver", we would like to stress that the comparison between both works is not as straightforward as expected. 
Indeed, from the geometrical point of view, the spacetime studied in Onazawa et. al~\cite{Onozawa:1995vu} corresponds to the extremal limit achieved within the ``areal radial fixing" gauge. Leaver's approach --- being based on the ``Cauchy horizon fixing" gauge --- has a discontinuous limit to extremality into the Bertotti-Robinson spacetime.

An interesting open question is to determine whether such Bertotti-Robinson spacetime admits some natural conditions in the AdS boundary leading to the same QNMs as the ones in the usual extremal limit. The assessment of such QNM-isospectrality between
extremal Reissner-Nordstr\"om and the corresponding near-horizon geometry could offer insight into correlations between geometrical properties at the horizon and at null infinity~\cite{Jaramillo:2011re,Jaramillo:2011rf,Jaramillo:2012rr,Gupta:2018znn} in black hole extremal settings.

In a second stage, we {have} applied the algorithm developed in section \ref{sec:Algorithm} to construct the solution --- in general, after an initial transitory --- to the dissipative wave equation in terms of the spectral decomposition \eqref{eq:VSol_spectral}. In this context,  the ``areal radial fixing" gauge has a technical limitation. Since the coordinate value of the Cauchy horizon depends on the charge parameter $\kappa$, the convergence radius of the Taylor expansion reduces as the Cauchy horizon approaches the event horizon. In particular, the algorithm is valid for the values $\kappa \in [0, 1/2]$. Working in the ``Cauchy horizon fixing" gauge is a natural first solution to this limitation. As already mentioned, however, this option changes the geometry of the limiting extremal spacetime.

An alternative solution would be to follow~\cite{Onozawa:1995vu,Richartz:2015saa} and adapt the algorithm to a Taylor expansion around $r_0 = r_+/2$ for {\em all} values of $\kappa$ (and not only to the extremal case). In our compactified radial coordinate, this corresponds to an expansion around the regular point $\sigma_0 = 1/2$. In this case, assumption (I) in section \ref{sec:TaylorExpansions} is not valid anymore as one obtains, in fact, two independent solutions satisfying eq.~\eqref{eq:Property1}. The exploration of this particular possibility requires a specifically dedicated study.

We conclude by pointing out that the geometrical framework and/or the semi-analytical algorithm develop here is a promising and potentially valuable tool in several fields. In the new era of gravitational wave astronomy, we intend to further explore the extraction of highly-accurate wave forms for binary black-hole systems in the large-mass-ratio regime~\cite{Mitsou:2010jv,Bernuzzi:2010xj,Zenginoglu:2011zz,Bernuzzi:2011aj,Harms:2014dqa,Harms:2015ixa,Thornburg:2016msc} apart from studying some models extending beyond General Relativity~\cite{Cardoso:2003jf,Cardoso:2017njb}. 
On the other hand, resonant expansions in optical systems have proved to be an efficient tool in the study of near-field properties of nano-resonators
  \cite{LalYanVyn17}. The algorithm developed here, conveniently adapted to the dispersive case with a frequency-dependent permittivity, could shed light
  into the ambiguities
  in the determination of the QNM expansion coefficients in such optical setting~\footnote{We note that in the optical context, tails play typically
      no role in the spectral expansion 
    (\ref{eq:VSol_spectral}), due to the compact support of permittivity devitations from the background. In this particular respect, the optical case simplifies.}. Finally, on a purely mathematical ground, the issues here raised in the calculation
  of QNMs in singular spacetime limits define a non-trivial problem on ``resonance isospectrality'' in the setting of the spectral analysis of
  non-selfadjoint operators\cite{Sjost18}.
   
\section*{Acknowledgements}
RPM and JLJ share a profound scientific and personal debt with Marcus Ansorg. In the particular context of this article, and despite all adversities, his ingenious strength was fundamental for obtaining the main results in sec.~\ref{sec:Algorithm}. This work is dedicated, in deep gratitude, to his memory.  RPM would like to thank the warm hospitality of the Institut de Math\'ematiques de Bourgogne. This work was partially supported by the European Research Council Grant No. ERC-2014- StG 639022-NewNGR ``New frontiers in numerical general relativity" and it utilised Queen Mary's Apocrita HPC facility, supported by QMUL Research-IT. JLJ thanks the support of the FABER project in the PARI programmes (Region Bourgogne Franche-Comt\'e), the FEDER funding from the European Union, as well as the I-QUINS project (I-SITE BFC).

\section{Appendix}

\subsection{Spectral elements and resonant asymptotic expansions}\label{App:AsymExp}
\label{App:asymp_expansions}
We sketch here briefly the basic spectral elements needed to characterise QNMs, with a focus on the notion
of resonant expansion, in an attempt of better seizing conjecture \ref{cnjc:SpecDecomp}.

Let us consider the wave equation in $\mathbb{R}\times \mathbb{R}^n$, with $n$ odd
\bea
\label{e:flat_wave_eq}
  \left\{
  \begin{array}{l}
    \left(\partial^2_t - \Delta + V(\mathbf{x})\right)\Phi(t,\mathbf{x}) = 0 \ , \\
    \Phi(0,\mathbf{x}) = \Phi_0(\mathbf{x}) \ , \\
    \partial_t \Phi(0,\mathbf{x}) = \Phi_1(\mathbf{x}) \ ,
  \end{array}
  \right.
  \eea
  with $\Delta$ the Laplacian in $\mathbb{R}^n$, $V(\mathbf{x})$ a bounded potential and $\Phi_0(\mathbf{x})$ and $\Phi_1(\mathbf{x})$ prescribed
  initial data\footnote{Assumptions are required on the initial data functional spaces are required, 
    e.g. $\Phi_0(\mathbf{x})\in H^1(D)$ and
    $\Phi_1(\mathbf{x})\in L^2(D)$, with $D$ an appropriate domain $D\subset \mathbb{R}^n$.}. Defining the elliptic  operator
  in $\mathbb{R}^n$
  \bea
  P_V=-\Delta + V \ ,
  \eea
  Eq. (\ref{e:flat_wave_eq}) becomes, under Laplace
  transformation
  \bea
  \label{e:Laplace_3D}
  \left(P_V + s^2\right)\hat{\Phi}(s, \mathbf{x}) = S(s,  \mathbf{x}) \ ,
  \eea
  with $S(s,  \mathbf{x})$ a source determined in terms of the initial data, as
  $S(s,  \mathbf{x})= s\Phi_0(\mathbf{x}) + \Phi_1(\mathbf{x})$.

  A key element in the spectral analysis of Eq. (\ref{e:flat_wave_eq}) is provided
  by the notion of the resolvent $R_V(s)$ of $P_V$. Such operator is defined, between
  the appropriate domain and codomain functional spaces, as the inverse of
  the operator on the left-hand-side of Eq. (\ref{e:Laplace_3D}), that is\footnote{The resolvent $R_{V}(s)$ is intimately related
    to the Green function $G_s(\mathbf{x}-\mathbf{y})$ of
    $\left(P_V +s^2\right)$, characterised as $\left(P_V +s^2\right)G_s(\mathbf{x}-\mathbf{y}) = \delta^{(n)}(\mathbf{x}-\mathbf{y})$,
    so that the Green function is the integral kernel of the resolvent operator
\bea
R_V(s)(\Phi)(\mathbf{x}) =
\int_{\mathbb{R}^n} G_\omega(\mathbf{x}-\mathbf{y}) \Phi(\mathbf{y})  d^n\mathbf{y}\ .
\eea
The resolvent of $P_V$ is usually defined
    as  $R_V(\omega) = \left(P_V - \omega^2\;\mathrm{Id}\right)^{-1}$. We have chosen rather the version (\ref{e:resolvent}), with $s=i\omega$, since
    it is better adapted to the discussion in section \ref{e:Intro} in terms of the Laplace transform.}
  \bea
  \label{e:resolvent}
  R_V(s) = \left(P_V + s^2\;\mathrm{Id}\right)^{-1} \ .
  \eea
  For appropriate potentials $V$, the resolvant can be shown to be analytical in the right half-plane $\Re(s)>0$ of the complex spectral parameter.
  Then, under the suitable assumptions on $V$, a meromorphic extension of $R_V(s)$ exists in the left half-plane $\Re(s)<0$.
  Scattering resonances, i.e. QNMs frequencies $s_n$, are then defined as poles in such meromorphic extension.
  For potentials $V$ not decaying sufficiently fast at infinity, the  extension
  of $R_V(s)$ to the whole $s$-complex plane presents a more complicated analytical structure. In particular, as commented in footnote \ref{f:spectral},
  in the Schwarzschild case a branch cut starting at $s=0$ appears in addition to the QNM poles, giving rise to a 
  tail contribution at late times. This is also de case in the Reissner-Nordstr\"om case studied here
  (cf. \cite{DiaZwo17,zworski2017mathematical} and references therein for a spectral analysis discussion
  of scattering resonances, as well as  \cite{ChiLeuMaa98} for a more heuristic account, in particular adressing tails and initial transient issues).
  
  The elements introduced above allow us to address a central point in this work, namely 
  conjecture \ref{cnjc:SpecDecomp} concerning the spectral decomposition (\ref{eq:VSol_spectral}) of the scattered field.
  A claim on QNM {\em completeness} (together with tail functions) would require: (i) the identification
  of  some appropriate functional space for the scattered fields, and (ii) a claim on the convergence of the series in~(\ref{eq:VSol_spectral})
  in such linear space. No claim is made here in this respect.
  It is known that QNM completeness is a property of very particular potentials (cf e.g. \cite{Beyer:1998nu}
  where QNM completeness is shown for a P\"oschl-Teller potential) and does not hold in general. On the other hand,
  for suitable potentials, sound results  hold for resonant or QNM expansions, if understood as {\em asymptotic expansions}.
  It is in this sense that, in principle, the series in (\ref{eq:VSol_spectral}) must be interpreted.

  To illustrate this point on asymptotic expansions, we refer to the results on resonant expansions
  of scattered waves by Lax and Phillips \cite{LaxPhi89}
  and Vainberg \cite{Vainb73}. For concreteness, for bounded potentials $V$ under the appropriate hypotheses (namely on their
  support) the following kind of result
  can be shown (cf. e.g. \cite{zworski2017mathematical,DiaZwo17} for full details and background):
  for any $a>0$ the scattered field solution to Eq. (\ref{e:flat_wave_eq}) can be written as a
    'resonant expansion' in terms of QNMs 
\bea
  \label{e:expansion_resonant_states}
  \Phi(t, \mathbf{x}) = \sum_{\mathrm{\Re}(s_j)>-a} u_j(\mathbf{x})e^{s_jt}  + E_a(t) \ ,
  \eea
  where $\{s_j\}^\infty_{j=1}$ are the resonances of $P_V$ ---namely the poles of the meromorphic extension of $R_V(s)$, as discused above---
  and $u_j$ are the corresponding resonant states, determined also in terms of the resolvent $R_V(s)$ as~\footnote{Note that
    the resonant state $u_j(\mathbf{x})$ is obtained through the application of the resolvent $R_V(s)$ on the source $S(s,\mathbf{x})$
    in (\ref{e:Laplace_3D}), consistently with its link with the Green function. Note, comparing (\ref{eq:VSol_spectral}) and
    (\ref{e:expansion_resonant_states}) that the 'meaningful' quantity is indeed the combination $u_j = \eta_j \phi_j$,
    and not $c_j$ and $\phi_j$ separately. This justifies the term {\em invariant} for the amplitudes in Fig.~\ref{fig:Amplitudes}.
  }
  \bea
  \label{e:resonant_states}
  u_j = i\;\mathrm{Res}_{s=s_j} \left(R_V(s)\Phi_1  + s R_V(s)\Phi_0\right) \ .
  \eea
  The QNM sum in (\ref{e:expansion_resonant_states}) is finite if the resonant frequencies $s_n$ do not accumulate near the real axis and, crucially
  for our discussion on asymptotic series,
  constants $C_a$ and $T_a$ exist such that
    the error $E_a(\tau)$ can be bound for $\tau\geq T_a$ as
    \bea
        \label{e:bound_E_a}    
  \!\!\!\!\!\!\!\!\!\!\!\!\!\!||E_a(t)||_{H^1}\leq C_a e^{-a\tau}\left(||\Phi_0||_{H^1} + ||\Phi_1||_{L^2} \right) \ , \ \tau\geq T_a \ .
  \eea
    The key point we want to stress here is that the bound (\ref{e:bound_E_a}) on the error $E_a(\tau)$ is {\em not uniform} in $a$. This means that we can
    indeed incorporate more QNMs into the resonant expansion by enlarging the ``band'' in the left half-complex plane defined by each {\em fixed} $a$,
    but nothing guarantees that the constant $C_a$ does not explode in the process. As a consequence, in general actual convergence cannot be shown
    and we must treat the QNM expansion rather as an asymptotic series.

    Once we have insisted in the, a priori, asymptotic character of the QNM expansion in conjecture~\ref{cnjc:SpecDecomp},
    let us bring attention to the remarkable result
    shown in Fig.~\ref{fig:SpecTimeEvol}: the extraordinary agreement ---at all time scales--- between the explicit time evolution of the initial data
    and its 'spectral expansion' (\ref{eq:VSol_spectral}) poses the prospect of
    an actual 'full' convergence in some appropriate space and, therefore,  QNM and tail completeness in the Reissner-Nordstr\"om case. Such a possibility
    should be studied with tools different from the ones in the present work.
    
\subsection{Backward recurrence relation}\label{App:BackwardRecRel}
A key element in the algorithm from \cite{Ansorg:2016ztf} when constructing the decaying solutions in assumption (2) is the usage of a {\em backward} iteration of the homogeneous \eqref{eq:HomRecRel}. Here, we review how to construct the decaying solutions with such a technique in the present context and discuss the instability issues arising from the backward iteration.

According to the algorithm from sec.~\ref{sec:TaylorExpansions}, inserting the Ansatz \eqref{eq:Asympt_Ansatz} back into the recurrence relation \eqref{eq:HomRecRel} and multiplying the result by $e^{-\xi k^p} k^{-\zeta}$, leads to
\bea
\label{eq:RecRelAk}
&& 0=\alpha_k e^{-\xi k^p[1- (1+k^{-1})^p]} (1+k^{-1})^{\zeta}A_{k+1} \nn \\
&& + \sum_{i=0}^{m}\beta_k^{(i)} e^{-\xi k^p[1- (1+ik^{-1})^p]} (1-ik^{-1})^{\zeta}A_{k-i}  \ .
\eea 
For a fixed asymptotic parameter $p$, we expand the above expression in terms of $y=k^{-p}$ around $y=0$, with the help of the \texttt{Series} command in \texttt{Mathematica}. The quantities $\xi$, $\zeta$ and $\{\nu_j\}_{j=1}^{2J_{\rm max}}$ are determined by equating the coefficients of the expansion order by order.

The dependences of $\xi$ and $\zeta$ on the physical parameters of the problem --- $s$, $\kappa$ and $\ell$ in the case of Reissner-Nordstr\"om --- are easily obtained from the lowest coefficients of the expansion. Examples of $\xi$, $\zeta$ for the Reissner-Nordstr\"om case are given by eqs.~\eqref{eq:AsympBeh}-\eqref{eq:AsympCoeff_pow}, \eqref{eq:AsympBehavScalarLeaver} and \eqref{eq:AsympBehavEMGravLeaver}. 

There are two numerical parameters controlling the approximation: $J_{\rm max}$ and $k_{\rm max}$. The former is the truncation value that approximates $A_k$ ---cf. eq.~\eqref{eq:Asympt_Ansatz}, whereas the latter fixes all Taylor expansions introduced in sec.~\ref{sec:Algorithm}. 

Once a numerical resolution $k_{\rm max}$ and $J_{\rm max}$ is fixed, we approximate the values $I_{k_{\rm max} + 1}, I_{k_{\rm max}}, \ldots, I_{k_{\rm max}-m+1}$ and iterate the eq.~\eqref{eq:HomRecRel} backwards to obtain 
\beq
\displaystyle
I_{k-m} = -\dfrac{\alpha_k I_{k+1}+ \sum\limits_{i=0}^{m-1}  \beta^{(i)}_k I_{k-i}}{\beta_k^m}, \quad k = k_{\rm max}, \ldots, 0 \ .
\eeq 
\begin{figure*}[t!]
\begin{center}
\includegraphics[width=7.5cm]{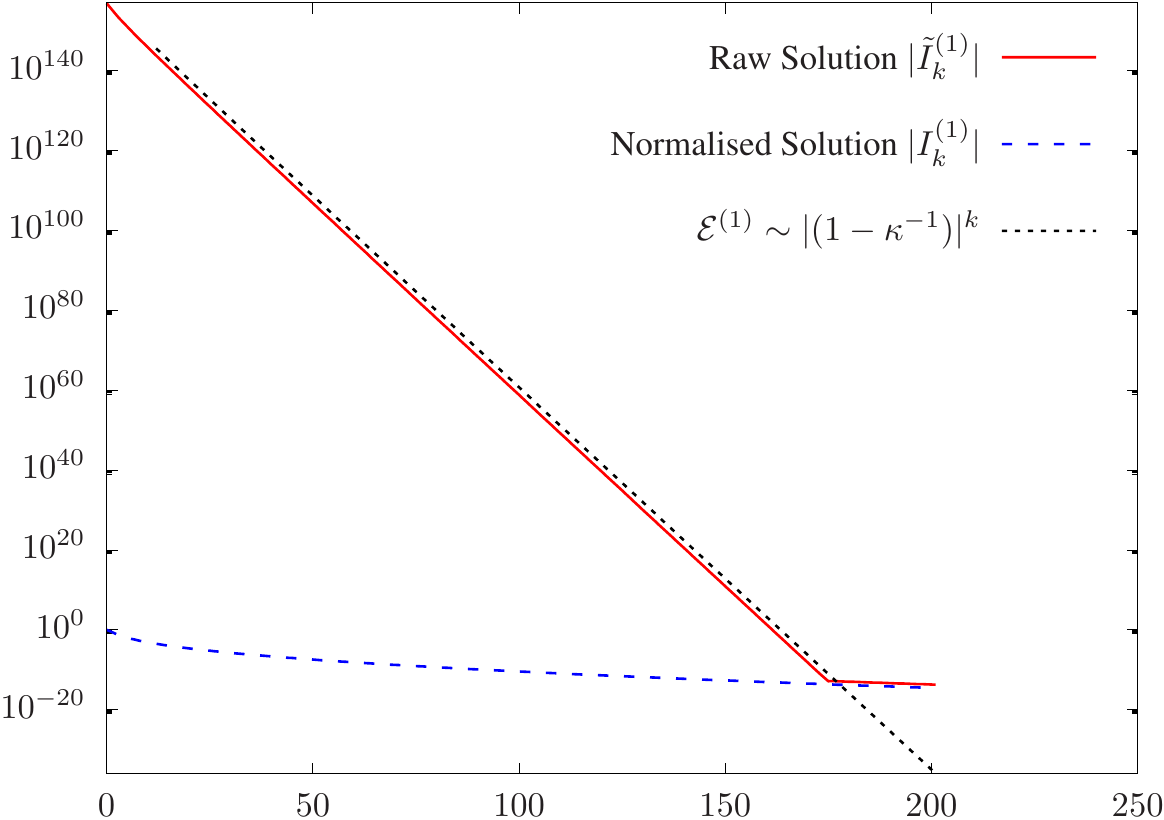}
\includegraphics[width=7.5cm]{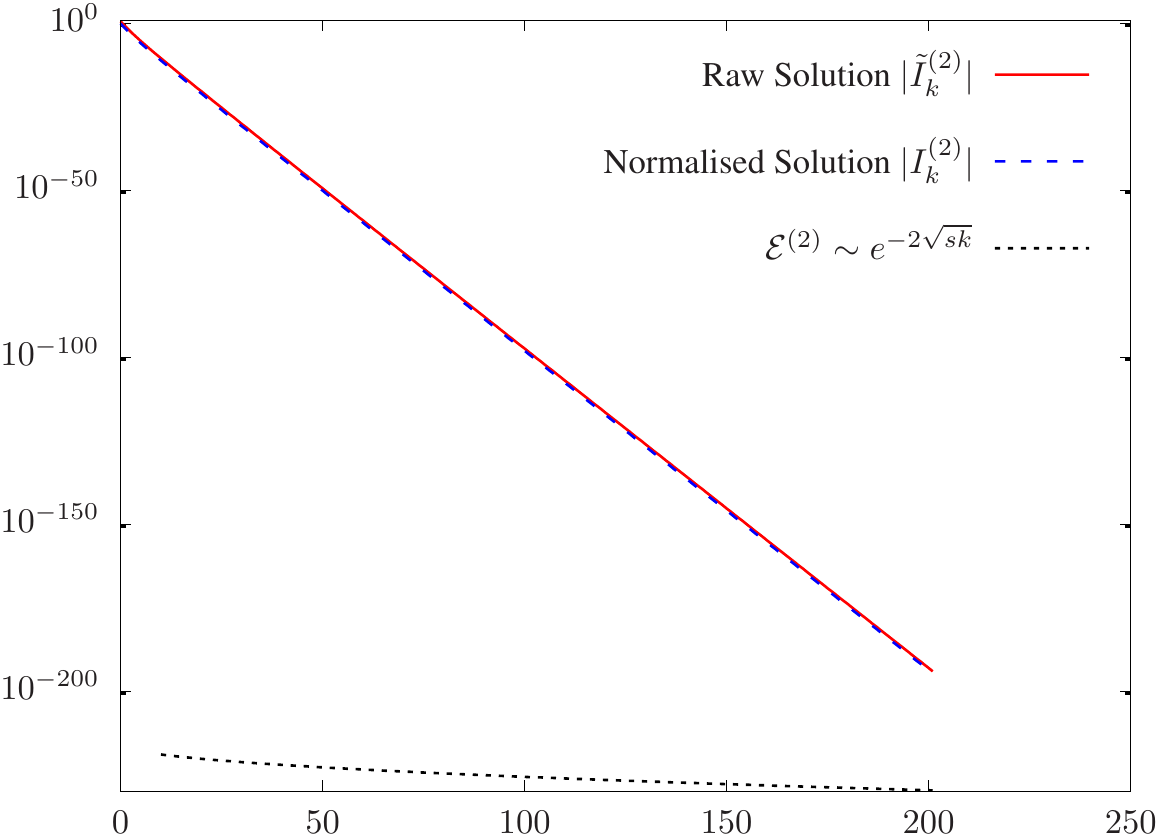}
\end{center}
\caption{Unstable behaviour from the backward iteration of recurrence relation \eqref{eq:RecRelCoeff}. The solutions are obtained for an electromagnetic perturbation with parameters $s=1+i$, $\kappa = 0.1$, $\ell=3$ and numerical resolution $N_{\rm max} = 200$, $J_{\rm max} = 10$. Left panel: the raw solution $\tilde{I}^{(1)}_k \sim e^{-2\sqrt{sk}}$ (red straight line) is unstable since the numerical error ${\cal E}^{(1)}$ (dotted black line) excites $I^{(2)}-$modes growing as $|1-\kappa^{-1}|^{k}$ for $k<N_{\rm max}$. Right panel: the raw solution $\tilde{I}^{(2)}_k \sim |1-\kappa^{-1}|^{k}$ (red straight line) is not affected by the numerical error ${\cal E}^{(2)} \sim e^{-2\sqrt{sk}}$ (dotted black line). The instability is filtered via the linear combination \eqref{eq:LinComb_DecaySol}, which normalises the solution according to \eqref{eq:Property4} (blue dashed line).
}
\label{fig:DecaySol_Insta}
\end{figure*} 

Unfortunately, the existence of several decaying solution may lead to an unstable backward iteration. Indeed, the decaying solutions actually {\em grow} as we iterate the recurrence relation backwards to lower values of $k$. Therefore any numerical error in one decaying solution will excite modes related to another decaying solution. Depending on the power $p$ of the exponencial behaviour, the numerical error may grow and contaminate the solution. 

The phenomena is easily understood with a concrete example. Let us consider the $3-$order recurrence relation with coefficients \eqref{eq:RecRelCoeff}. Here we treat an electromagnetic perturbation for $\kappa = 0.1$. We fix the Laplace parameter to $s= 1 + i$, the angular mode $\ell =3$ and the numerical resolution to $k_{\rm max}=200$ and $J_{\rm max} = 10$. 

The right-hand-side of eq.~\eqref{eq:RecRelAk}, calculated at $k=k_{\rm max}$, provides an error estimative ${\cal E}^{(1)}$ and ${\cal E}^{(2)}$ for the approximated asymptotic values of the solutions $I^{(1)}_{k_{\rm Max}}$ and $I^{(2)}_{k_{\rm Max}}$ --- see eqs.~\eqref{eq:AsympBeh}-\eqref{eq:AsympCoeff_pow}. 
The error ${\cal E}^{(1)}$ excites $I^{(2)}$-modes that behave as $\upsilon^k$ ($|\upsilon|<1$). For $k<N_{\rm max}$, the error growth is faster than the behaviour of $I^{(1)}_{N_{\rm max}}$ itself. Thus the instability occurs --- see the left panel of fig.~\ref{fig:DecaySol_Insta}. 
On the contrary, the error ${\cal E}^{(2)}$ excites $I^{(1)}$-modes behaving merely as $e^{-\xi\sqrt{k}}$, which does not affect the the solution $I^{(2)}_k$ --- see right panel of fig.~\ref{fig:DecaySol_Insta}. 

The figure displays these unstable solutions which we denote here by $\tilde{I}^{(\ell)}_k$. However, we can construct new solutions $I^{(\ell)}_k$ and filter the instability of the raw solutions $\tilde{I}^{(\ell)}_k$ via the linear combination 
\beq 
\label{eq:LinComb_DecaySol}
I^{(\ell)}_k = \sum_{\ell'=\ell}^m \varkappa^\ell_{\ell'}\tilde{I}^{(\ell')}_k \ .
\eeq
The coefficients of the linear combination \eqref{eq:LinComb_DecaySol} are fixed by imposing \eqref{eq:Norm_Ik} and \eqref{eq:Property4}. Fig.~\ref{fig:DecaySol_Insta} depicts the filtered solution $I^{(\ell)}_k$ constructed via the backward recurrence relation.

\bibliographystyle{apsrev4-1-noeprint.bst}
\bibliography{bibitems}

\begin{thebibliography}{81}%
\makeatletter
\providecommand \@ifxundefined [1]{%
 \@ifx{#1\undefined}
}%
\providecommand \@ifnum [1]{%
 \ifnum #1\expandafter \@firstoftwo
 \else \expandafter \@secondoftwo
 \fi
}%
\providecommand \@ifx [1]{%
 \ifx #1\expandafter \@firstoftwo
 \else \expandafter \@secondoftwo
 \fi
}%
\providecommand \natexlab [1]{#1}%
\providecommand \enquote  [1]{``#1''}%
\providecommand \bibnamefont  [1]{#1}%
\providecommand \bibfnamefont [1]{#1}%
\providecommand \citenamefont [1]{#1}%
\providecommand \href@noop [0]{\@secondoftwo}%
\providecommand \href [0]{\begingroup \@sanitize@url \@href}%
\providecommand \@href[1]{\@@startlink{#1}\@@href}%
\providecommand \@@href[1]{\endgroup#1\@@endlink}%
\providecommand \@sanitize@url [0]{\catcode `\\12\catcode `\$12\catcode
  `\&12\catcode `\#12\catcode `\^12\catcode `\_12\catcode `\%12\relax}%
\providecommand \@@startlink[1]{}%
\providecommand \@@endlink[0]{}%
\providecommand \url  [0]{\begingroup\@sanitize@url \@url }%
\providecommand \@url [1]{\endgroup\@href {#1}{\urlprefix }}%
\providecommand \urlprefix  [0]{URL }%
\providecommand \Eprint [0]{\href }%
\providecommand \doibase [0]{http://dx.doi.org/}%
\providecommand \selectlanguage [0]{\@gobble}%
\providecommand \bibinfo  [0]{\@secondoftwo}%
\providecommand \bibfield  [0]{\@secondoftwo}%
\providecommand \translation [1]{[#1]}%
\providecommand \BibitemOpen [0]{}%
\providecommand \bibitemStop [0]{}%
\providecommand \bibitemNoStop [0]{.\EOS\space}%
\providecommand \EOS [0]{\spacefactor3000\relax}%
\providecommand \BibitemShut  [1]{\csname bibitem#1\endcsname}%
\let\auto@bib@innerbib\@empty
\bibitem [{\citenamefont {Chandrasekhar}(1983)}]{Chandrasekhar83}%
  \BibitemOpen
  \bibfield  {author} {\bibinfo {author} {\bibfnamefont {S.}~\bibnamefont
  {Chandrasekhar}},\ }\href@noop {} {\emph {\bibinfo {title} {The Mathematical
  Theory of Black Holes}}},\ Chandrasekhar83\ (\bibinfo  {publisher} {Oxford
  University Press},\ \bibinfo {address} {Oxford, England},\ \bibinfo {year}
  {1983})\BibitemShut {NoStop}%
\bibitem [{\citenamefont {Kokkotas}\ and\ \citenamefont
  {Schmidt}(1999)}]{Kokkotas99a}%
  \BibitemOpen
  \bibfield  {author} {\bibinfo {author} {\bibfnamefont {K.~D.}\ \bibnamefont
  {Kokkotas}}\ and\ \bibinfo {author} {\bibfnamefont {B.~G.}\ \bibnamefont
  {Schmidt}},\ }\href@noop {} {\bibfield  {journal} {\bibinfo  {journal}
  {Living Rev. Relativ.}\ }\textbf {\bibinfo {volume} {2}},\ \bibinfo {pages}
  {2} (\bibinfo {year} {1999})},\ \bibinfo {note}
  {http://www.livingreviews.org/lrr-1999-2}\BibitemShut {NoStop}%
\bibitem [{\citenamefont {Nollert}(1999)}]{Nollert99}%
  \BibitemOpen
  \bibfield  {author} {\bibinfo {author} {\bibfnamefont {H.-P.}\ \bibnamefont
  {Nollert}},\ }\href@noop {} {\bibfield  {journal} {\bibinfo  {journal}
  {Class. Quantum Grav.}\ }\textbf {\bibinfo {volume} {16}},\ \bibinfo {pages}
  {R159} (\bibinfo {year} {1999})}\BibitemShut {NoStop}%
\bibitem [{\citenamefont {Berti}\ \emph {et~al.}(2009)\citenamefont {Berti},
  \citenamefont {Cardoso},\ and\ \citenamefont {Starinets}}]{Berti:2009kk}%
  \BibitemOpen
  \bibfield  {author} {\bibinfo {author} {\bibfnamefont {E.}~\bibnamefont
  {Berti}}, \bibinfo {author} {\bibfnamefont {V.}~\bibnamefont {Cardoso}}, \
  and\ \bibinfo {author} {\bibfnamefont {A.~O.}\ \bibnamefont {Starinets}},\
  }\href {\doibase 10.1088/0264-9381/26/16/163001} {\bibfield  {journal}
  {\bibinfo  {journal} {Class. Quantum Grav.}\ }\textbf {\bibinfo {volume}
  {26}},\ \bibinfo {pages} {163001} (\bibinfo {year} {2009})}\BibitemShut
  {NoStop}%
\bibitem [{\citenamefont {Konoplya}\ and\ \citenamefont
  {Zhidenko}(2011)}]{Konoplya:2011qq}%
  \BibitemOpen
  \bibfield  {author} {\bibinfo {author} {\bibfnamefont {R.~A.}\ \bibnamefont
  {Konoplya}}\ and\ \bibinfo {author} {\bibfnamefont {A.}~\bibnamefont
  {Zhidenko}},\ }\href {\doibase 10.1103/RevModPhys.83.793} {\bibfield
  {journal} {\bibinfo  {journal} {Rev. Mod. Phys.}\ }\textbf {\bibinfo {volume}
  {83}},\ \bibinfo {pages} {793} (\bibinfo {year} {2011})}\BibitemShut
  {NoStop}%
\bibitem [{\citenamefont {Price}(1972)}]{Price72}%
  \BibitemOpen
  \bibfield  {author} {\bibinfo {author} {\bibfnamefont {R.}~\bibnamefont
  {Price}},\ }\href@noop {} {\bibfield  {journal} {\bibinfo  {journal} {Phys.
  Rev. D}\ }\textbf {\bibinfo {volume} {5}},\ \bibinfo {pages} {2419} (\bibinfo
  {year} {1972})}\BibitemShut {NoStop}%
\bibitem [{\citenamefont {Gundlach}\ \emph {et~al.}(1994)\citenamefont
  {Gundlach}, \citenamefont {Price},\ and\ \citenamefont
  {Pullin}}]{Gundlach94a}%
  \BibitemOpen
  \bibfield  {author} {\bibinfo {author} {\bibfnamefont {C.}~\bibnamefont
  {Gundlach}}, \bibinfo {author} {\bibfnamefont {R.}~\bibnamefont {Price}}, \
  and\ \bibinfo {author} {\bibfnamefont {J.}~\bibnamefont {Pullin}},\
  }\href@noop {} {\bibfield  {journal} {\bibinfo  {journal} {Phys. Rev. D}\
  }\textbf {\bibinfo {volume} {49}} (\bibinfo {year} {1994})}\BibitemShut
  {NoStop}%
\bibitem [{\citenamefont {Lalanne}\ \emph {et~al.}(2018)\citenamefont
  {Lalanne}, \citenamefont {Yan}, \citenamefont {Vynck}, \citenamefont
  {Sauvan},\ and\ \citenamefont {Hugonin}}]{LalYanVyn17}%
  \BibitemOpen
  \bibfield  {author} {\bibinfo {author} {\bibfnamefont {P.}~\bibnamefont
  {Lalanne}}, \bibinfo {author} {\bibfnamefont {W.}~\bibnamefont {Yan}},
  \bibinfo {author} {\bibfnamefont {K.}~\bibnamefont {Vynck}}, \bibinfo
  {author} {\bibfnamefont {C.}~\bibnamefont {Sauvan}}, \ and\ \bibinfo {author}
  {\bibfnamefont {J.-P.}\ \bibnamefont {Hugonin}},\ }\href {\doibase
  10.1002/lpor.201700113} {\bibfield  {journal} {\bibinfo  {journal} {Laser \&
  Photonics Reviews}\ }\textbf {\bibinfo {volume} {12}},\ \bibinfo {pages}
  {1700113} (\bibinfo {year} {2018})}\BibitemShut {NoStop}%
\bibitem [{\citenamefont {Dyatlov}\ and\ \citenamefont {Zworski}()}]{DiaZwo17}%
  \BibitemOpen
  \bibfield  {author} {\bibinfo {author} {\bibfnamefont {S.}~\bibnamefont
  {Dyatlov}}\ and\ \bibinfo {author} {\bibfnamefont {M.}~\bibnamefont
  {Zworski}},\ }\href@noop {} {\bibinfo  {journal} {book in preparation;
  http://math.mit.edu/~dyatlov/res/}\ }\BibitemShut {NoStop}%
\bibitem [{\citenamefont {Zworski}(2017)}]{zworski2017mathematical}%
  \BibitemOpen
\bibfield  {journal} {  }\bibfield  {author} {\bibinfo {author} {\bibfnamefont
  {M.}~\bibnamefont {Zworski}},\ }\href@noop {} {\bibfield  {journal} {\bibinfo
   {journal} {Bulletin of Mathematical Sciences}\ }\textbf {\bibinfo {volume}
  {7}},\ \bibinfo {pages} {1} (\bibinfo {year} {2017})}\BibitemShut {NoStop}%
\bibitem [{\citenamefont {Ansorg}\ and\ \citenamefont
  {Panosso~Macedo}(2016)}]{Ansorg:2016ztf}%
  \BibitemOpen
  \bibfield  {author} {\bibinfo {author} {\bibfnamefont {M.}~\bibnamefont
  {Ansorg}}\ and\ \bibinfo {author} {\bibfnamefont {R.}~\bibnamefont
  {Panosso~Macedo}},\ }\href {\doibase 10.1103/PhysRevD.93.124016} {\bibfield
  {journal} {\bibinfo  {journal} {Phys. Rev.}\ }\textbf {\bibinfo {volume}
  {D93}},\ \bibinfo {pages} {124016} (\bibinfo {year} {2016})}\BibitemShut
  {NoStop}%
\bibitem [{\citenamefont {Ammon}\ \emph {et~al.}(2016)\citenamefont {Ammon},
  \citenamefont {Grieninger}, \citenamefont {Jimenez-Alba}, \citenamefont
  {Macedo},\ and\ \citenamefont {Melgar}}]{Ammon:2016fru}%
  \BibitemOpen
  \bibfield  {author} {\bibinfo {author} {\bibfnamefont {M.}~\bibnamefont
  {Ammon}}, \bibinfo {author} {\bibfnamefont {S.}~\bibnamefont {Grieninger}},
  \bibinfo {author} {\bibfnamefont {A.}~\bibnamefont {Jimenez-Alba}}, \bibinfo
  {author} {\bibfnamefont {R.~P.}\ \bibnamefont {Macedo}}, \ and\ \bibinfo
  {author} {\bibfnamefont {L.}~\bibnamefont {Melgar}},\ }\href {\doibase
  10.1007/JHEP09(2016)131} {\bibfield  {journal} {\bibinfo  {journal} {JHEP}\
  }\textbf {\bibinfo {volume} {09}},\ \bibinfo {pages} {131} (\bibinfo {year}
  {2016})}\BibitemShut {NoStop}%
\bibitem [{\citenamefont {Andreasson}(2005)}]{Andreasson:2005qy}%
  \BibitemOpen
  \bibfield  {author} {\bibinfo {author} {\bibfnamefont {H.}~\bibnamefont
  {Andreasson}},\ }\href {\doibase 10.12942/lrr-2005-2} {\bibfield  {journal}
  {\bibinfo  {journal} {Living Rev. Rel.}\ }\textbf {\bibinfo {volume} {8}},\
  \bibinfo {pages} {2} (\bibinfo {year} {2005})}\BibitemShut {NoStop}%
\bibitem [{\citenamefont {Emparan}\ and\ \citenamefont
  {Reall}(2008)}]{Emparan:2008eg}%
  \BibitemOpen
  \bibfield  {author} {\bibinfo {author} {\bibfnamefont {R.}~\bibnamefont
  {Emparan}}\ and\ \bibinfo {author} {\bibfnamefont {H.~S.}\ \bibnamefont
  {Reall}},\ }\href {\doibase 10.12942/lrr-2008-6} {\bibfield  {journal}
  {\bibinfo  {journal} {Living Rev. Rel.}\ }\textbf {\bibinfo {volume} {11}},\
  \bibinfo {pages} {6} (\bibinfo {year} {2008})}\BibitemShut {NoStop}%
\bibitem [{\citenamefont {{Horowitz}}(2012)}]{Horowitz2012}%
  \BibitemOpen
  \bibfield  {author} {\bibinfo {author} {\bibfnamefont {G.~T.}\ \bibnamefont
  {{Horowitz}}},\ }\href@noop {} {\emph {\bibinfo {title} {Black Holes in
  Higher Dimensions, by Gary T.~Horowitz, Cambridge, UK: Cambridge University
  Press, 2012}}}\ (\bibinfo {year} {2012})\BibitemShut {NoStop}%
\bibitem [{\citenamefont {Ammon}\ and\ \citenamefont
  {Erdmenger}(2015)}]{Ammon:2015wua}%
  \BibitemOpen
  \bibfield  {author} {\bibinfo {author} {\bibfnamefont {M.}~\bibnamefont
  {Ammon}}\ and\ \bibinfo {author} {\bibfnamefont {J.}~\bibnamefont
  {Erdmenger}},\ }\href
  {http://www.cambridge.org/de/academic/subjects/physics/theoretical-physics-and-mathematical-physics/gaugegravity-duality-foundations-and-applications}
  {\emph {\bibinfo {title} {{Gauge/gravity duality}}}}\ (\bibinfo  {publisher}
  {Cambridge Univ. Pr.},\ \bibinfo {address} {Cambridge, UK},\ \bibinfo {year}
  {2015})\BibitemShut {NoStop}%
\bibitem [{\citenamefont {Nastase}(2015)}]{Nastase2015}%
  \BibitemOpen
  \bibfield  {author} {\bibinfo {author} {\bibfnamefont {H.}~\bibnamefont
  {Nastase}},\ }\href
  {http://www.cambridge.org/us/academic/subjects/physics/theoretical-physics-and-mathematical-physics/introduction-adscft-correspondence}
  {\emph {\bibinfo {title} {{Introduction to the AdS/CFT Correspondence}}}}\
  (\bibinfo  {publisher} {Cambridge Univ. Pr.},\ \bibinfo {address} {Cambridge,
  UK},\ \bibinfo {year} {2015})\BibitemShut {NoStop}%
\bibitem [{\citenamefont {Paiva}\ \emph {et~al.}(1993)\citenamefont {Paiva},
  \citenamefont {Reboucas},\ and\ \citenamefont {MacCallum}}]{Paiva:1993bv}%
  \BibitemOpen
  \bibfield  {author} {\bibinfo {author} {\bibfnamefont {F.~M.}\ \bibnamefont
  {Paiva}}, \bibinfo {author} {\bibfnamefont {M.~J.}\ \bibnamefont {Reboucas}},
  \ and\ \bibinfo {author} {\bibfnamefont {M.~A.~H.}\ \bibnamefont
  {MacCallum}},\ }\href {\doibase 10.1088/0264-9381/10/6/013} {\bibfield
  {journal} {\bibinfo  {journal} {Class. Quant. Grav.}\ }\textbf {\bibinfo
  {volume} {10}},\ \bibinfo {pages} {1165} (\bibinfo {year}
  {1993})}\BibitemShut {NoStop}%
\bibitem [{\citenamefont {Geroch}(1969)}]{Geroch:1969ca}%
  \BibitemOpen
  \bibfield  {author} {\bibinfo {author} {\bibfnamefont {R.~P.}\ \bibnamefont
  {Geroch}},\ }\href {\doibase 10.1007/BF01645486} {\bibfield  {journal}
  {\bibinfo  {journal} {Commun. Math. Phys.}\ }\textbf {\bibinfo {volume}
  {13}},\ \bibinfo {pages} {180} (\bibinfo {year} {1969})}\BibitemShut
  {NoStop}%
\bibitem [{\citenamefont {Hawking}\ and\ \citenamefont
  {Ellis}(1973)}]{HawEll73}%
  \BibitemOpen
  \bibfield  {author} {\bibinfo {author} {\bibfnamefont {S.~W.}\ \bibnamefont
  {Hawking}}\ and\ \bibinfo {author} {\bibfnamefont {G.~F.~R.}\ \bibnamefont
  {Ellis}},\ }\href@noop {} {\emph {\bibinfo {title} {The large scale structure
  of space-time}}}\ (\bibinfo  {publisher} {Cambridge University Press},\
  \bibinfo {year} {1973})\BibitemShut {NoStop}%
\bibitem [{\citenamefont {Carroll}\ \emph {et~al.}()\citenamefont {Carroll},
  \citenamefont {Johnson},\ and\ \citenamefont {Randall}}]{Carroll09}%
  \BibitemOpen
  \bibfield  {author} {\bibinfo {author} {\bibfnamefont {S.~M.}\ \bibnamefont
  {Carroll}}, \bibinfo {author} {\bibfnamefont {M.~C.}\ \bibnamefont
  {Johnson}}, \ and\ \bibinfo {author} {\bibfnamefont {L.}~\bibnamefont
  {Randall}},\ }\href
  {http://iopscience.iop.org/article/10.1088/1126-6708/2009/11/109,
  year={2009},} {\bibfield  {journal} {\bibinfo  {journal} {Journal of High
  Energy Physics}\ }\textbf {\bibinfo {volume} {2009}},\ \bibinfo {pages}
  {109}}\BibitemShut {NoStop}%
\bibitem [{\citenamefont {Bengtsson}\ \emph {et~al.}(2014)\citenamefont
  {Bengtsson}, \citenamefont {Holst},\ and\ \citenamefont
  {Jakobsson}}]{Bengtsson:2014fha}%
  \BibitemOpen
  \bibfield  {author} {\bibinfo {author} {\bibfnamefont {I.}~\bibnamefont
  {Bengtsson}}, \bibinfo {author} {\bibfnamefont {S.}~\bibnamefont {Holst}}, \
  and\ \bibinfo {author} {\bibfnamefont {E.}~\bibnamefont {Jakobsson}},\ }\href
  {\doibase 10.1088/0264-9381/31/20/205008} {\bibfield  {journal} {\bibinfo
  {journal} {Class. Quant. Grav.}\ }\textbf {\bibinfo {volume} {31}},\ \bibinfo
  {pages} {205008} (\bibinfo {year} {2014})}\BibitemShut {NoStop}%
\bibitem [{\citenamefont {Kunduri}\ and\ \citenamefont
  {Lucietti}(2013)}]{Kunduri2013}%
  \BibitemOpen
  \bibfield  {author} {\bibinfo {author} {\bibfnamefont {H.~K.}\ \bibnamefont
  {Kunduri}}\ and\ \bibinfo {author} {\bibfnamefont {J.}~\bibnamefont
  {Lucietti}},\ }\href {\doibase 10.12942/lrr-2013-8} {\bibfield  {journal}
  {\bibinfo  {journal} {Living Reviews in Relativity}\ }\textbf {\bibinfo
  {volume} {16}},\ \bibinfo {pages} {8} (\bibinfo {year} {2013})}\BibitemShut
  {NoStop}%
\bibitem [{\citenamefont {Bertotti}(1959)}]{Bertotti59}%
  \BibitemOpen
  \bibfield  {author} {\bibinfo {author} {\bibfnamefont {B.}~\bibnamefont
  {Bertotti}},\ }\href {\doibase 10.1103/PhysRev.116.1331} {\bibfield
  {journal} {\bibinfo  {journal} {Phys. Rev.}\ }\textbf {\bibinfo {volume}
  {116}},\ \bibinfo {pages} {1331} (\bibinfo {year} {1959})}\BibitemShut
  {NoStop}%
\bibitem [{\citenamefont {I.}(1959)}]{Robinson59}%
  \BibitemOpen
  \bibfield  {author} {\bibinfo {author} {\bibfnamefont {R.}~\bibnamefont
  {I.}},\ }\href@noop {} {\bibfield  {journal} {\bibinfo  {journal} {Bull.
  Acad. Polon. Sci.}\ }\textbf {\bibinfo {volume} {7}},\ \bibinfo {pages} {351}
  (\bibinfo {year} {1959})}\BibitemShut {NoStop}%
\bibitem [{\citenamefont {Leaver}(1990)}]{Leaver90}%
  \BibitemOpen
  \bibfield  {author} {\bibinfo {author} {\bibfnamefont {E.~W.}\ \bibnamefont
  {Leaver}},\ }\href {\doibase 10.1103/PhysRevD.41.2986} {\bibfield  {journal}
  {\bibinfo  {journal} {Phys. Rev. D}\ }\textbf {\bibinfo {volume} {41}},\
  \bibinfo {pages} {2986} (\bibinfo {year} {1990})}\BibitemShut {NoStop}%
\bibitem [{\citenamefont {Zenginoglu}(2008)}]{Zenginoglu:2007jw}%
  \BibitemOpen
  \bibfield  {author} {\bibinfo {author} {\bibfnamefont {A.}~\bibnamefont
  {Zenginoglu}},\ }\href {\doibase 10.1088/0264-9381/25/14/145002} {\bibfield
  {journal} {\bibinfo  {journal} {Class. Quant. Grav.}\ }\textbf {\bibinfo
  {volume} {25}},\ \bibinfo {pages} {145002} (\bibinfo {year}
  {2008})}\BibitemShut {NoStop}%
\bibitem [{\citenamefont {Zenginoglu}(2011{\natexlab{a}})}]{Zenginoglu:2011jz}%
  \BibitemOpen
  \bibfield  {author} {\bibinfo {author} {\bibfnamefont {A.}~\bibnamefont
  {Zenginoglu}},\ }\href {\doibase 10.1103/PhysRevD.83.127502} {\bibfield
  {journal} {\bibinfo  {journal} {Phys. Rev.}\ }\textbf {\bibinfo {volume}
  {D83}},\ \bibinfo {pages} {127502} (\bibinfo {year}
  {2011}{\natexlab{a}})}\BibitemShut {NoStop}%
\bibitem [{\citenamefont {Brill}\ \emph {et~al.}(1980)\citenamefont {Brill},
  \citenamefont {Cavallo},\ and\ \citenamefont {Isenberg}}]{brill:2789}%
  \BibitemOpen
  \bibfield  {author} {\bibinfo {author} {\bibfnamefont {D.~R.}\ \bibnamefont
  {Brill}}, \bibinfo {author} {\bibfnamefont {J.~M.}\ \bibnamefont {Cavallo}},
  \ and\ \bibinfo {author} {\bibfnamefont {J.~A.}\ \bibnamefont {Isenberg}},\
  }\href {\doibase 10.1063/1.524400} {\bibfield  {journal} {\bibinfo  {journal}
  {Journal of Mathematical Physics}\ }\textbf {\bibinfo {volume} {21}},\
  \bibinfo {pages} {2789} (\bibinfo {year} {1980})}\BibitemShut {NoStop}%
\bibitem [{\citenamefont {Malec}\ and\ \citenamefont
  {O'Murchadha}(2009)}]{Malec:2009hg}%
  \BibitemOpen
  \bibfield  {author} {\bibinfo {author} {\bibfnamefont {E.}~\bibnamefont
  {Malec}}\ and\ \bibinfo {author} {\bibfnamefont {N.}~\bibnamefont
  {O'Murchadha}},\ }\href {\doibase 10.1103/PhysRevD.80.024017} {\bibfield
  {journal} {\bibinfo  {journal} {Phys. Rev.}\ }\textbf {\bibinfo {volume}
  {D80}},\ \bibinfo {pages} {024017} (\bibinfo {year} {2009})}\BibitemShut
  {NoStop}%
\bibitem [{\citenamefont {Tuite}\ and\ \citenamefont
  {Murchadha}(2013)}]{Tuite:2013hza}%
  \BibitemOpen
  \bibfield  {author} {\bibinfo {author} {\bibfnamefont {P.}~\bibnamefont
  {Tuite}}\ and\ \bibinfo {author} {\bibfnamefont {N.~O.}\ \bibnamefont
  {Murchadha}},\ }\href@noop {} {\enquote {\bibinfo {title} {{Constant Mean
  Curvature Slices of the Reissner-Nordstr\"{o}m Spacetime}},}\ } (\bibinfo
  {year} {2013}),\ \Eprint {http://arxiv.org/abs/1307.4657} {arXiv:1307.4657
  [gr-qc]} \BibitemShut {NoStop}%
\bibitem [{\citenamefont {Schinkel}\ \emph
  {et~al.}(2014{\natexlab{a}})\citenamefont {Schinkel}, \citenamefont
  {Panosso~Macedo},\ and\ \citenamefont {Ansorg}}]{Schinkel:2013tka}%
  \BibitemOpen
  \bibfield  {author} {\bibinfo {author} {\bibfnamefont {D.}~\bibnamefont
  {Schinkel}}, \bibinfo {author} {\bibfnamefont {R.}~\bibnamefont
  {Panosso~Macedo}}, \ and\ \bibinfo {author} {\bibfnamefont {M.}~\bibnamefont
  {Ansorg}},\ }\href {\doibase 10.1088/0264-9381/31/7/075017} {\bibfield
  {journal} {\bibinfo  {journal} {Class. Quant. Grav.}\ }\textbf {\bibinfo
  {volume} {31}},\ \bibinfo {pages} {075017} (\bibinfo {year}
  {2014}{\natexlab{a}})}\BibitemShut {NoStop}%
\bibitem [{\citenamefont {Schinkel}\ \emph
  {et~al.}(2014{\natexlab{b}})\citenamefont {Schinkel}, \citenamefont
  {Ansorg},\ and\ \citenamefont {Panosso~Macedo}}]{Schinkel:2013zm}%
  \BibitemOpen
  \bibfield  {author} {\bibinfo {author} {\bibfnamefont {D.}~\bibnamefont
  {Schinkel}}, \bibinfo {author} {\bibfnamefont {M.}~\bibnamefont {Ansorg}}, \
  and\ \bibinfo {author} {\bibfnamefont {R.}~\bibnamefont {Panosso~Macedo}},\
  }\href {\doibase 10.1088/0264-9381/31/16/165001} {\bibfield  {journal}
  {\bibinfo  {journal} {Class. Quant. Grav.}\ }\textbf {\bibinfo {volume}
  {31}},\ \bibinfo {pages} {165001} (\bibinfo {year}
  {2014}{\natexlab{b}})}\BibitemShut {NoStop}%
\bibitem [{\citenamefont {Zenginoglu}(2011{\natexlab{b}})}]{Zenginoglu:2010cq}%
  \BibitemOpen
  \bibfield  {author} {\bibinfo {author} {\bibfnamefont {A.}~\bibnamefont
  {Zenginoglu}},\ }\href {\doibase 10.1016/j.jcp.2010.12.016} {\bibfield
  {journal} {\bibinfo  {journal} {J. Comput. Phys.}\ }\textbf {\bibinfo
  {volume} {230}},\ \bibinfo {pages} {2286} (\bibinfo {year}
  {2011}{\natexlab{b}})}\BibitemShut {NoStop}%
\bibitem [{\citenamefont {Hubner}(1999)}]{Hubner:1999th}%
  \BibitemOpen
  \bibfield  {author} {\bibinfo {author} {\bibfnamefont {P.}~\bibnamefont
  {Hubner}},\ }\href {\doibase 10.1088/0264-9381/16/9/302} {\bibfield
  {journal} {\bibinfo  {journal} {Class. Quant. Grav.}\ }\textbf {\bibinfo
  {volume} {16}},\ \bibinfo {pages} {2823} (\bibinfo {year}
  {1999})}\BibitemShut {NoStop}%
\bibitem [{\citenamefont {Frauendiener}\ and\ \citenamefont
  {Hein}(2002)}]{Frauendiener2002}%
  \BibitemOpen
  \bibfield  {author} {\bibinfo {author} {\bibfnamefont {J.}~\bibnamefont
  {Frauendiener}}\ and\ \bibinfo {author} {\bibfnamefont {M.}~\bibnamefont
  {Hein}},\ }\href {\doibase 10.1103/PhysRevD.66.124004} {\bibfield  {journal}
  {\bibinfo  {journal} {Phys. Rev. D}\ }\textbf {\bibinfo {volume} {66}},\
  \bibinfo {pages} {124004} (\bibinfo {year} {2002})}\BibitemShut {NoStop}%
\bibitem [{\citenamefont {Bardeen}\ \emph {et~al.}(2011)\citenamefont
  {Bardeen}, \citenamefont {Sarbach},\ and\ \citenamefont
  {Buchman}}]{Bardeen:2011ip}%
  \BibitemOpen
  \bibfield  {author} {\bibinfo {author} {\bibfnamefont {J.~M.}\ \bibnamefont
  {Bardeen}}, \bibinfo {author} {\bibfnamefont {O.}~\bibnamefont {Sarbach}}, \
  and\ \bibinfo {author} {\bibfnamefont {L.~T.}\ \bibnamefont {Buchman}},\
  }\href {\doibase 10.1103/PhysRevD.83.104045} {\bibfield  {journal} {\bibinfo
  {journal} {Phys. Rev.}\ }\textbf {\bibinfo {volume} {D83}},\ \bibinfo {pages}
  {104045} (\bibinfo {year} {2011})}\BibitemShut {NoStop}%
\bibitem [{\citenamefont {Rinne}(2010)}]{Rinne:2009qx}%
  \BibitemOpen
  \bibfield  {author} {\bibinfo {author} {\bibfnamefont {O.}~\bibnamefont
  {Rinne}},\ }\href {\doibase 10.1088/0264-9381/27/3/035014} {\bibfield
  {journal} {\bibinfo  {journal} {Class. Quant. Grav.}\ }\textbf {\bibinfo
  {volume} {27}},\ \bibinfo {pages} {035014} (\bibinfo {year}
  {2010})}\BibitemShut {NoStop}%
\bibitem [{\citenamefont {Rinne}\ and\ \citenamefont
  {Moncrief}(2013)}]{Rinne:2013qc}%
  \BibitemOpen
  \bibfield  {author} {\bibinfo {author} {\bibfnamefont {O.}~\bibnamefont
  {Rinne}}\ and\ \bibinfo {author} {\bibfnamefont {V.}~\bibnamefont
  {Moncrief}},\ }\href {\doibase 10.1088/0264-9381/30/9/095009} {\bibfield
  {journal} {\bibinfo  {journal} {Class. Quant. Grav.}\ }\textbf {\bibinfo
  {volume} {30}},\ \bibinfo {pages} {095009} (\bibinfo {year}
  {2013})}\BibitemShut {NoStop}%
\bibitem [{\citenamefont {Vañó-Viñuales}\ \emph {et~al.}(2015)\citenamefont
  {Vañó-Viñuales}, \citenamefont {Husa},\ and\ \citenamefont
  {Hilditch}}]{Vano-Vinuales:2014koa}%
  \BibitemOpen
  \bibfield  {author} {\bibinfo {author} {\bibfnamefont {A.}~\bibnamefont
  {Vañó-Viñuales}}, \bibinfo {author} {\bibfnamefont {S.}~\bibnamefont
  {Husa}}, \ and\ \bibinfo {author} {\bibfnamefont {D.}~\bibnamefont
  {Hilditch}},\ }\href {\doibase 10.1088/0264-9381/32/17/175010} {\bibfield
  {journal} {\bibinfo  {journal} {Class. Quant. Grav.}\ }\textbf {\bibinfo
  {volume} {32}},\ \bibinfo {pages} {175010} (\bibinfo {year}
  {2015})}\BibitemShut {NoStop}%
\bibitem [{\citenamefont {Morales}\ and\ \citenamefont
  {Sarbach}(2017)}]{Morales:2016rgt}%
  \BibitemOpen
  \bibfield  {author} {\bibinfo {author} {\bibfnamefont {M.~D.}\ \bibnamefont
  {Morales}}\ and\ \bibinfo {author} {\bibfnamefont {O.}~\bibnamefont
  {Sarbach}},\ }\href {\doibase 10.1103/PhysRevD.95.044001} {\bibfield
  {journal} {\bibinfo  {journal} {Phys. Rev.}\ }\textbf {\bibinfo {volume}
  {D95}},\ \bibinfo {pages} {044001} (\bibinfo {year} {2017})}\BibitemShut
  {NoStop}%
\bibitem [{\citenamefont {Hilditch}\ \emph {et~al.}(2018)\citenamefont
  {Hilditch}, \citenamefont {Harms}, \citenamefont {Bugner}, \citenamefont
  {Rüter},\ and\ \citenamefont {Brügmann}}]{Hilditch:2016xzh}%
  \BibitemOpen
  \bibfield  {author} {\bibinfo {author} {\bibfnamefont {D.}~\bibnamefont
  {Hilditch}}, \bibinfo {author} {\bibfnamefont {E.}~\bibnamefont {Harms}},
  \bibinfo {author} {\bibfnamefont {M.}~\bibnamefont {Bugner}}, \bibinfo
  {author} {\bibfnamefont {H.}~\bibnamefont {Rüter}}, \ and\ \bibinfo {author}
  {\bibfnamefont {B.}~\bibnamefont {Brügmann}},\ }\href {\doibase
  10.1088/1361-6382/aaa4ac} {\bibfield  {journal} {\bibinfo  {journal} {Class.
  Quant. Grav.}\ }\textbf {\bibinfo {volume} {35}},\ \bibinfo {pages} {055003}
  (\bibinfo {year} {2018})}\BibitemShut {NoStop}%
\bibitem [{\citenamefont {Vañó-Viñuales}\ and\ \citenamefont
  {Husa}(2018)}]{Vano-Vinuales:2017qij}%
  \BibitemOpen
  \bibfield  {author} {\bibinfo {author} {\bibfnamefont {A.}~\bibnamefont
  {Vañó-Viñuales}}\ and\ \bibinfo {author} {\bibfnamefont {S.}~\bibnamefont
  {Husa}},\ }\href {\doibase 10.1088/1361-6382/aaa4e2} {\bibfield  {journal}
  {\bibinfo  {journal} {Class. Quant. Grav.}\ }\textbf {\bibinfo {volume}
  {35}},\ \bibinfo {pages} {045014} (\bibinfo {year} {2018})}\BibitemShut
  {NoStop}%
\bibitem [{\citenamefont {Panosso~Macedo}\ and\ \citenamefont
  {Ansorg}(2014)}]{Macedo:2014bfa}%
  \BibitemOpen
  \bibfield  {author} {\bibinfo {author} {\bibfnamefont {R.}~\bibnamefont
  {Panosso~Macedo}}\ and\ \bibinfo {author} {\bibfnamefont {M.}~\bibnamefont
  {Ansorg}},\ }\href {\doibase 10.1016/j.jcp.2014.07.040} {\bibfield  {journal}
  {\bibinfo  {journal} {J. Comput. Phys.}\ }\textbf {\bibinfo {volume} {276}},\
  \bibinfo {pages} {357} (\bibinfo {year} {2014})}\BibitemShut {NoStop}%
\bibitem [{\citenamefont {Zenginoglu}\ \emph {et~al.}(2009)\citenamefont
  {Zenginoglu}, \citenamefont {Nunez},\ and\ \citenamefont
  {Husa}}]{Zenginoglu:2008uc}%
  \BibitemOpen
  \bibfield  {author} {\bibinfo {author} {\bibfnamefont {A.}~\bibnamefont
  {Zenginoglu}}, \bibinfo {author} {\bibfnamefont {D.}~\bibnamefont {Nunez}}, \
  and\ \bibinfo {author} {\bibfnamefont {S.}~\bibnamefont {Husa}},\ }\href
  {\doibase 10.1088/0264-9381/26/3/035009} {\bibfield  {journal} {\bibinfo
  {journal} {Class. Quant. Grav.}\ }\textbf {\bibinfo {volume} {26}},\ \bibinfo
  {pages} {035009} (\bibinfo {year} {2009})}\BibitemShut {NoStop}%
\bibitem [{\citenamefont {Zenginoglu}(2010)}]{Zenginoglu:2009ey}%
  \BibitemOpen
  \bibfield  {author} {\bibinfo {author} {\bibfnamefont {A.}~\bibnamefont
  {Zenginoglu}},\ }\href {\doibase 10.1088/0264-9381/27/4/045015} {\bibfield
  {journal} {\bibinfo  {journal} {Class. Quant. Grav.}\ }\textbf {\bibinfo
  {volume} {27}},\ \bibinfo {pages} {045015} (\bibinfo {year}
  {2010})}\BibitemShut {NoStop}%
\bibitem [{\citenamefont {Racz}\ and\ \citenamefont
  {Toth}(2011)}]{Racz:2011qu}%
  \BibitemOpen
  \bibfield  {author} {\bibinfo {author} {\bibfnamefont {I.}~\bibnamefont
  {Racz}}\ and\ \bibinfo {author} {\bibfnamefont {G.~Z.}\ \bibnamefont
  {Toth}},\ }\href {\doibase 10.1088/0264-9381/28/19/195003} {\bibfield
  {journal} {\bibinfo  {journal} {Class. Quant. Grav.}\ }\textbf {\bibinfo
  {volume} {28}},\ \bibinfo {pages} {195003} (\bibinfo {year}
  {2011})}\BibitemShut {NoStop}%
\bibitem [{\citenamefont {Jasiulek}(2012)}]{Jasiulek:2011ce}%
  \BibitemOpen
  \bibfield  {author} {\bibinfo {author} {\bibfnamefont {M.}~\bibnamefont
  {Jasiulek}},\ }\href {\doibase 10.1088/0264-9381/29/1/015008} {\bibfield
  {journal} {\bibinfo  {journal} {Class. Quant. Grav.}\ }\textbf {\bibinfo
  {volume} {29}},\ \bibinfo {pages} {015008} (\bibinfo {year}
  {2012})}\BibitemShut {NoStop}%
\bibitem [{\citenamefont {Harms}\ \emph {et~al.}(2013)\citenamefont {Harms},
  \citenamefont {Bernuzzi},\ and\ \citenamefont {Br{\"u}gmann}}]{Harms:2013ib}%
  \BibitemOpen
  \bibfield  {author} {\bibinfo {author} {\bibfnamefont {E.}~\bibnamefont
  {Harms}}, \bibinfo {author} {\bibfnamefont {S.}~\bibnamefont {Bernuzzi}}, \
  and\ \bibinfo {author} {\bibfnamefont {B.}~\bibnamefont {Br{\"u}gmann}},\
  }\href {\doibase 10.1088/0264-9381/30/11/115013} {\bibfield  {journal}
  {\bibinfo  {journal} {Class. Quant. Grav.}\ }\textbf {\bibinfo {volume}
  {30}},\ \bibinfo {pages} {115013} (\bibinfo {year} {2013})}\BibitemShut
  {NoStop}%
\bibitem [{\citenamefont {Wald}(1984)}]{Wald84}%
  \BibitemOpen
  \bibfield  {author} {\bibinfo {author} {\bibfnamefont {R.~M.}\ \bibnamefont
  {Wald}},\ }\href@noop {} {\emph {\bibinfo {title} {General relativity}}},\
  Wald84\ (\bibinfo  {publisher} {The University of Chicago Press},\ \bibinfo
  {address} {Chicago},\ \bibinfo {year} {1984})\BibitemShut {NoStop}%
\bibitem [{\citenamefont {Leaver}(1985)}]{Leaver85}%
  \BibitemOpen
  \bibfield  {author} {\bibinfo {author} {\bibfnamefont {E.}~\bibnamefont
  {Leaver}},\ }\href@noop {} {\bibfield  {journal} {\bibinfo  {journal} {Proc.
  R. Soc. London, Ser. A}\ }\textbf {\bibinfo {volume} {402}},\ \bibinfo
  {pages} {285} (\bibinfo {year} {1985})}\BibitemShut {NoStop}%
\bibitem [{\citenamefont {Leaver}(1986)}]{Leaver86c}%
  \BibitemOpen
  \bibfield  {author} {\bibinfo {author} {\bibfnamefont {E.~W.}\ \bibnamefont
  {Leaver}},\ }\href@noop {} {\bibfield  {journal} {\bibinfo  {journal} {Phys.
  Rev. D}\ }\textbf {\bibinfo {volume} {34}},\ \bibinfo {pages} {384} (\bibinfo
  {year} {1986})}\BibitemShut {NoStop}%
\bibitem [{\citenamefont {{Bachelot}}\ and\ \citenamefont
  {{Motet-Bachelot}}(1993)}]{Bachelot1993}%
  \BibitemOpen
  \bibfield  {author} {\bibinfo {author} {\bibfnamefont {A.}~\bibnamefont
  {{Bachelot}}}\ and\ \bibinfo {author} {\bibfnamefont {A.}~\bibnamefont
  {{Motet-Bachelot}}},\ }\href@noop {} {\bibfield  {journal} {\bibinfo
  {journal} {Ann.~Inst.~Henri Poincar{\'e}, Phys.~Th{\'e}or., Vol.~59, No.~1,
  p.~3 - 68}\ }\textbf {\bibinfo {volume} {59}},\ \bibinfo {pages} {3}
  (\bibinfo {year} {1993})}\BibitemShut {NoStop}%
\bibitem [{\citenamefont {Batic}\ \emph {et~al.}(2018)\citenamefont {Batic},
  \citenamefont {Nowakowski},\ and\ \citenamefont {Redway}}]{Batic:2018nxk}%
  \BibitemOpen
  \bibfield  {author} {\bibinfo {author} {\bibfnamefont {D.}~\bibnamefont
  {Batic}}, \bibinfo {author} {\bibfnamefont {M.}~\bibnamefont {Nowakowski}}, \
  and\ \bibinfo {author} {\bibfnamefont {K.}~\bibnamefont {Redway}},\ }\href
  {\doibase 10.1103/PhysRevD.98.024017} {\bibfield  {journal} {\bibinfo
  {journal} {Phys. Rev.}\ }\textbf {\bibinfo {volume} {D98}},\ \bibinfo {pages}
  {024017} (\bibinfo {year} {2018})}\BibitemShut {NoStop}%
\bibitem [{\citenamefont {Panosso~Macedo}(2018)}]{Macedo:2018gvw}%
  \BibitemOpen
  \bibfield  {author} {\bibinfo {author} {\bibfnamefont {R.}~\bibnamefont
  {Panosso~Macedo}},\ }\href@noop {} {\enquote {\bibinfo {title} {{Comment:
  Some exact quasinormal frequencies of a massless scalar field in
  Schwarzschild spacetime}},}\ } (\bibinfo {year} {2018}),\ \Eprint
  {http://arxiv.org/abs/1807.05940} {arXiv:1807.05940 [gr-qc]} \BibitemShut
  {NoStop}%
\bibitem [{\citenamefont {Panosso~Macedo}(tion)}]{Macedo2019}%
  \BibitemOpen
  \bibfield  {author} {\bibinfo {author} {\bibfnamefont {R.}~\bibnamefont
  {Panosso~Macedo}},\ }\href@noop {} {\enquote {\bibinfo {title} {{Late time
  decay of higher-dimensional black holes along future null infinity}},}\ }
  (\bibinfo {year} {In preparation})\BibitemShut {NoStop}%
\bibitem [{\citenamefont {Moncrief}(1974{\natexlab{a}})}]{Moncrief74a}%
  \BibitemOpen
  \bibfield  {author} {\bibinfo {author} {\bibfnamefont {V.}~\bibnamefont
  {Moncrief}},\ }\href {\doibase 10.1103/PhysRevD.9.2707} {\bibfield  {journal}
  {\bibinfo  {journal} {Phys. Rev. D}\ }\textbf {\bibinfo {volume} {9}},\
  \bibinfo {pages} {2707} (\bibinfo {year} {1974}{\natexlab{a}})}\BibitemShut
  {NoStop}%
\bibitem [{\citenamefont {Moncrief}(1974{\natexlab{b}})}]{Moncrief74b}%
  \BibitemOpen
  \bibfield  {author} {\bibinfo {author} {\bibfnamefont {V.}~\bibnamefont
  {Moncrief}},\ }\href {\doibase 10.1103/PhysRevD.10.1057} {\bibfield
  {journal} {\bibinfo  {journal} {Phys. Rev. D}\ }\textbf {\bibinfo {volume}
  {10}},\ \bibinfo {pages} {1057} (\bibinfo {year}
  {1974}{\natexlab{b}})}\BibitemShut {NoStop}%
\bibitem [{\citenamefont {Moncrief}(1975)}]{Moncrief75}%
  \BibitemOpen
  \bibfield  {author} {\bibinfo {author} {\bibfnamefont {V.}~\bibnamefont
  {Moncrief}},\ }\href {\doibase 10.1103/PhysRevD.12.1526} {\bibfield
  {journal} {\bibinfo  {journal} {Phys. Rev. D}\ }\textbf {\bibinfo {volume}
  {12}},\ \bibinfo {pages} {1526} (\bibinfo {year} {1975})}\BibitemShut
  {NoStop}%
\bibitem [{\citenamefont {Zerilli}(1974)}]{Zerilli74}%
  \BibitemOpen
  \bibfield  {author} {\bibinfo {author} {\bibfnamefont {F.~J.}\ \bibnamefont
  {Zerilli}},\ }\href {\doibase 10.1103/PhysRevD.9.860} {\bibfield  {journal}
  {\bibinfo  {journal} {Phys. Rev. D}\ }\textbf {\bibinfo {volume} {9}},\
  \bibinfo {pages} {860} (\bibinfo {year} {1974})}\BibitemShut {NoStop}%
\bibitem [{\citenamefont {Onozawa}\ \emph {et~al.}(1996)\citenamefont
  {Onozawa}, \citenamefont {Mishima}, \citenamefont {Okamura},\ and\
  \citenamefont {Ishihara}}]{Onozawa:1995vu}%
  \BibitemOpen
  \bibfield  {author} {\bibinfo {author} {\bibfnamefont {H.}~\bibnamefont
  {Onozawa}}, \bibinfo {author} {\bibfnamefont {T.}~\bibnamefont {Mishima}},
  \bibinfo {author} {\bibfnamefont {T.}~\bibnamefont {Okamura}}, \ and\
  \bibinfo {author} {\bibfnamefont {H.}~\bibnamefont {Ishihara}},\ }\href
  {\doibase 10.1103/PhysRevD.53.7033} {\bibfield  {journal} {\bibinfo
  {journal} {Phys. Rev.}\ }\textbf {\bibinfo {volume} {D53}},\ \bibinfo {pages}
  {7033} (\bibinfo {year} {1996})}\BibitemShut {NoStop}%
\bibitem [{\citenamefont {Kroon}(2016)}]{Kroon:2016ink}%
  \BibitemOpen
  \bibfield  {author} {\bibinfo {author} {\bibfnamefont {J.~A.~V.}\
  \bibnamefont {Kroon}},\ }\href
  {http://www.cambridge.org/de/academic/subjects/physics/cosmology-relativity-and-gravitation/conformal-methods-general-relativity?format=HB}
  {\emph {\bibinfo {title} {{Conformal Methods in General Relativity}}}}\
  (\bibinfo  {publisher} {Cambridge University Press},\ \bibinfo {address}
  {Cambridge},\ \bibinfo {year} {2016})\BibitemShut {NoStop}%
\bibitem [{\citenamefont {Richartz}(2016)}]{Richartz:2015saa}%
  \BibitemOpen
  \bibfield  {author} {\bibinfo {author} {\bibfnamefont {M.}~\bibnamefont
  {Richartz}},\ }\href {\doibase 10.1103/PhysRevD.93.064062} {\bibfield
  {journal} {\bibinfo  {journal} {Phys. Rev.}\ }\textbf {\bibinfo {volume}
  {D93}},\ \bibinfo {pages} {064062} (\bibinfo {year} {2016})}\BibitemShut
  {NoStop}%
\bibitem [{\citenamefont {Jaramillo}\ \emph
  {et~al.}(2012{\natexlab{a}})\citenamefont {Jaramillo}, \citenamefont
  {Panosso~Macedo}, \citenamefont {Moesta},\ and\ \citenamefont
  {Rezzolla}}]{Jaramillo:2011re}%
  \BibitemOpen
  \bibfield  {author} {\bibinfo {author} {\bibfnamefont {J.~L.}\ \bibnamefont
  {Jaramillo}}, \bibinfo {author} {\bibfnamefont {R.}~\bibnamefont
  {Panosso~Macedo}}, \bibinfo {author} {\bibfnamefont {P.}~\bibnamefont
  {Moesta}}, \ and\ \bibinfo {author} {\bibfnamefont {L.}~\bibnamefont
  {Rezzolla}},\ }\href {\doibase 10.1103/PhysRevD.85.084030} {\bibfield
  {journal} {\bibinfo  {journal} {Phys. Rev.}\ }\textbf {\bibinfo {volume}
  {D85}},\ \bibinfo {pages} {084030} (\bibinfo {year}
  {2012}{\natexlab{a}})}\BibitemShut {NoStop}%
\bibitem [{\citenamefont {Jaramillo}\ \emph
  {et~al.}(2012{\natexlab{b}})\citenamefont {Jaramillo}, \citenamefont
  {Macedo}, \citenamefont {Moesta},\ and\ \citenamefont
  {Rezzolla}}]{Jaramillo:2011rf}%
  \BibitemOpen
  \bibfield  {author} {\bibinfo {author} {\bibfnamefont {J.~L.}\ \bibnamefont
  {Jaramillo}}, \bibinfo {author} {\bibfnamefont {R.~P.}\ \bibnamefont
  {Macedo}}, \bibinfo {author} {\bibfnamefont {P.}~\bibnamefont {Moesta}}, \
  and\ \bibinfo {author} {\bibfnamefont {L.}~\bibnamefont {Rezzolla}},\ }\href
  {\doibase 10.1103/PhysRevD.85.084031} {\bibfield  {journal} {\bibinfo
  {journal} {Phys. Rev.}\ }\textbf {\bibinfo {volume} {D85}},\ \bibinfo {pages}
  {084031} (\bibinfo {year} {2012}{\natexlab{b}})}\BibitemShut {NoStop}%
\bibitem [{\citenamefont {Jaramillo}\ \emph {et~al.}(2011)\citenamefont
  {Jaramillo}, \citenamefont {Macedo}, \citenamefont {Moesta},\ and\
  \citenamefont {Rezzolla}}]{Jaramillo:2012rr}%
  \BibitemOpen
  \bibfield  {author} {\bibinfo {author} {\bibfnamefont {J.~L.}\ \bibnamefont
  {Jaramillo}}, \bibinfo {author} {\bibfnamefont {R.~P.}\ \bibnamefont
  {Macedo}}, \bibinfo {author} {\bibfnamefont {P.}~\bibnamefont {Moesta}}, \
  and\ \bibinfo {author} {\bibfnamefont {L.}~\bibnamefont {Rezzolla}},\
  }\bibfield  {booktitle} {\emph {\bibinfo {booktitle} {{Proceedings, Spanish
  Relativity Meeting : Towards new paradigms. (ERE 2011): Madrid, Spain, August
  29-September 2, 2011}}},\ }\href {\doibase 10.1063/1.4734411} {\bibfield
  {journal} {\bibinfo  {journal} {AIP Conf. Proc.}\ }\textbf {\bibinfo {volume}
  {1458}},\ \bibinfo {pages} {158} (\bibinfo {year} {2011})}\BibitemShut
  {NoStop}%
\bibitem [{\citenamefont {Gupta}\ \emph {et~al.}(2018)\citenamefont {Gupta},
  \citenamefont {Krishnan}, \citenamefont {Nielsen},\ and\ \citenamefont
  {Schnetter}}]{Gupta:2018znn}%
  \BibitemOpen
  \bibfield  {author} {\bibinfo {author} {\bibfnamefont {A.}~\bibnamefont
  {Gupta}}, \bibinfo {author} {\bibfnamefont {B.}~\bibnamefont {Krishnan}},
  \bibinfo {author} {\bibfnamefont {A.}~\bibnamefont {Nielsen}}, \ and\
  \bibinfo {author} {\bibfnamefont {E.}~\bibnamefont {Schnetter}},\ }\href
  {\doibase 10.1103/PhysRevD.97.084028} {\bibfield  {journal} {\bibinfo
  {journal} {Phys. Rev.}\ }\textbf {\bibinfo {volume} {D97}},\ \bibinfo {pages}
  {084028} (\bibinfo {year} {2018})}\BibitemShut {NoStop}%
\bibitem [{\citenamefont {Mitsou}(2011)}]{Mitsou:2010jv}%
  \BibitemOpen
  \bibfield  {author} {\bibinfo {author} {\bibfnamefont {E.}~\bibnamefont
  {Mitsou}},\ }\href {\doibase 10.1103/PhysRevD.83.044039} {\bibfield
  {journal} {\bibinfo  {journal} {Phys. Rev.}\ }\textbf {\bibinfo {volume}
  {D83}},\ \bibinfo {pages} {044039} (\bibinfo {year} {2011})}\BibitemShut
  {NoStop}%
\bibitem [{\citenamefont {Bernuzzi}\ \emph
  {et~al.}(2011{\natexlab{a}})\citenamefont {Bernuzzi}, \citenamefont {Nagar},\
  and\ \citenamefont {Zenginoglu}}]{Bernuzzi:2010xj}%
  \BibitemOpen
  \bibfield  {author} {\bibinfo {author} {\bibfnamefont {S.}~\bibnamefont
  {Bernuzzi}}, \bibinfo {author} {\bibfnamefont {A.}~\bibnamefont {Nagar}}, \
  and\ \bibinfo {author} {\bibfnamefont {A.}~\bibnamefont {Zenginoglu}},\
  }\href {\doibase 10.1103/PhysRevD.83.064010} {\bibfield  {journal} {\bibinfo
  {journal} {Phys. Rev.}\ }\textbf {\bibinfo {volume} {D83}},\ \bibinfo {pages}
  {064010} (\bibinfo {year} {2011}{\natexlab{a}})}\BibitemShut {NoStop}%
\bibitem [{\citenamefont {Zenginoglu}\ and\ \citenamefont
  {Khanna}(2011)}]{Zenginoglu:2011zz}%
  \BibitemOpen
  \bibfield  {author} {\bibinfo {author} {\bibfnamefont {A.}~\bibnamefont
  {Zenginoglu}}\ and\ \bibinfo {author} {\bibfnamefont {G.}~\bibnamefont
  {Khanna}},\ }\href {\doibase 10.1103/PhysRevX.1.021017} {\bibfield  {journal}
  {\bibinfo  {journal} {Phys. Rev.}\ }\textbf {\bibinfo {volume} {X1}},\
  \bibinfo {pages} {021017} (\bibinfo {year} {2011})}\BibitemShut {NoStop}%
\bibitem [{\citenamefont {Bernuzzi}\ \emph
  {et~al.}(2011{\natexlab{b}})\citenamefont {Bernuzzi}, \citenamefont {Nagar},\
  and\ \citenamefont {Zenginoglu}}]{Bernuzzi:2011aj}%
  \BibitemOpen
  \bibfield  {author} {\bibinfo {author} {\bibfnamefont {S.}~\bibnamefont
  {Bernuzzi}}, \bibinfo {author} {\bibfnamefont {A.}~\bibnamefont {Nagar}}, \
  and\ \bibinfo {author} {\bibfnamefont {A.}~\bibnamefont {Zenginoglu}},\
  }\href {\doibase 10.1103/PhysRevD.84.084026} {\bibfield  {journal} {\bibinfo
  {journal} {Phys. Rev.}\ }\textbf {\bibinfo {volume} {D84}},\ \bibinfo {pages}
  {084026} (\bibinfo {year} {2011}{\natexlab{b}})}\BibitemShut {NoStop}%
\bibitem [{\citenamefont {Harms}\ \emph {et~al.}(2014)\citenamefont {Harms},
  \citenamefont {Bernuzzi}, \citenamefont {Nagar},\ and\ \citenamefont
  {Zenginoglu}}]{Harms:2014dqa}%
  \BibitemOpen
  \bibfield  {author} {\bibinfo {author} {\bibfnamefont {E.}~\bibnamefont
  {Harms}}, \bibinfo {author} {\bibfnamefont {S.}~\bibnamefont {Bernuzzi}},
  \bibinfo {author} {\bibfnamefont {A.}~\bibnamefont {Nagar}}, \ and\ \bibinfo
  {author} {\bibfnamefont {A.}~\bibnamefont {Zenginoglu}},\ }\href {\doibase
  10.1088/0264-9381/31/24/245004} {\bibfield  {journal} {\bibinfo  {journal}
  {Class. Quant. Grav.}\ }\textbf {\bibinfo {volume} {31}},\ \bibinfo {pages}
  {245004} (\bibinfo {year} {2014})}\BibitemShut {NoStop}%
\bibitem [{\citenamefont {Harms}\ \emph {et~al.}(2016)\citenamefont {Harms},
  \citenamefont {Lukes-Gerakopoulos}, \citenamefont {Bernuzzi},\ and\
  \citenamefont {Nagar}}]{Harms:2015ixa}%
  \BibitemOpen
  \bibfield  {author} {\bibinfo {author} {\bibfnamefont {E.}~\bibnamefont
  {Harms}}, \bibinfo {author} {\bibfnamefont {G.}~\bibnamefont
  {Lukes-Gerakopoulos}}, \bibinfo {author} {\bibfnamefont {S.}~\bibnamefont
  {Bernuzzi}}, \ and\ \bibinfo {author} {\bibfnamefont {A.}~\bibnamefont
  {Nagar}},\ }\href {\doibase 10.1103/PhysRevD.93.044015} {\bibfield  {journal}
  {\bibinfo  {journal} {Phys. Rev.}\ }\textbf {\bibinfo {volume} {D93}},\
  \bibinfo {pages} {044015} (\bibinfo {year} {2016})}\BibitemShut {NoStop}%
\bibitem [{\citenamefont {Thornburg}\ and\ \citenamefont
  {Wardell}(2017)}]{Thornburg:2016msc}%
  \BibitemOpen
  \bibfield  {author} {\bibinfo {author} {\bibfnamefont {J.}~\bibnamefont
  {Thornburg}}\ and\ \bibinfo {author} {\bibfnamefont {B.}~\bibnamefont
  {Wardell}},\ }\href {\doibase 10.1103/PhysRevD.95.084043} {\bibfield
  {journal} {\bibinfo  {journal} {Phys. Rev.}\ }\textbf {\bibinfo {volume}
  {D95}},\ \bibinfo {pages} {084043} (\bibinfo {year} {2017})}\BibitemShut
  {NoStop}%
\bibitem [{\citenamefont {Cardoso}\ \emph {et~al.}(2003)\citenamefont
  {Cardoso}, \citenamefont {Yoshida}, \citenamefont {Dias},\ and\ \citenamefont
  {Lemos}}]{Cardoso:2003jf}%
  \BibitemOpen
  \bibfield  {author} {\bibinfo {author} {\bibfnamefont {V.}~\bibnamefont
  {Cardoso}}, \bibinfo {author} {\bibfnamefont {S.}~\bibnamefont {Yoshida}},
  \bibinfo {author} {\bibfnamefont {O.~J.~C.}\ \bibnamefont {Dias}}, \ and\
  \bibinfo {author} {\bibfnamefont {J.~P.~S.}\ \bibnamefont {Lemos}},\ }\href
  {\doibase 10.1103/PhysRevD.68.061503} {\bibfield  {journal} {\bibinfo
  {journal} {Phys. Rev.}\ }\textbf {\bibinfo {volume} {D68}},\ \bibinfo {pages}
  {061503} (\bibinfo {year} {2003})}\BibitemShut {NoStop}%
\bibitem [{\citenamefont {Cardoso}\ and\ \citenamefont
  {Pani}(2003)}]{Cardoso:2017njb}%
  \BibitemOpen
  \bibfield  {author} {\bibinfo {author} {\bibfnamefont {V.}~\bibnamefont
  {Cardoso}}\ and\ \bibinfo {author} {\bibfnamefont {P.}~\bibnamefont {Pani}},\
  }\href {\doibase 10.1038/s41550-017-0225-y} {\bibfield  {journal} {\bibinfo
  {journal} {Nature Astronomy}\ }\textbf {\bibinfo {volume} {1}},\ \bibinfo
  {pages} {586} (\bibinfo {year} {2003})}\BibitemShut {NoStop}%
\bibitem [{\citenamefont {Sj\"ostrand}(user)}]{Sjost18}%
  \BibitemOpen
  \bibfield  {author} {\bibinfo {author} {\bibfnamefont {J.}~\bibnamefont
  {Sj\"ostrand}},\ }\href@noop {} {\enquote {\bibinfo {title} {Non-self-adjoint
  differential operators, spectral asymptotics and random perturbations},}\ }
  (\bibinfo {year} {to appear in {\em Pseudodifferential Operators, Theory and
  Applications}, Birkh\"auser})\BibitemShut {NoStop}%
\bibitem [{\citenamefont {Ching}\ \emph {et~al.}(1998)\citenamefont {Ching},
  \citenamefont {Leung}, \citenamefont {Maassen van~den Brink}, \citenamefont
  {Suen}, \citenamefont {Tong},\ and\ \citenamefont {Young}}]{ChiLeuMaa98}%
  \BibitemOpen
  \bibfield  {author} {\bibinfo {author} {\bibfnamefont {E.~S.~C.}\
  \bibnamefont {Ching}}, \bibinfo {author} {\bibfnamefont {P.~T.}\ \bibnamefont
  {Leung}}, \bibinfo {author} {\bibfnamefont {A.}~\bibnamefont {Maassen van~den
  Brink}}, \bibinfo {author} {\bibfnamefont {W.~M.}\ \bibnamefont {Suen}},
  \bibinfo {author} {\bibfnamefont {S.~S.}\ \bibnamefont {Tong}}, \ and\
  \bibinfo {author} {\bibfnamefont {K.}~\bibnamefont {Young}},\ }\href
  {\doibase 10.1103/RevModPhys.70.1545} {\bibfield  {journal} {\bibinfo
  {journal} {Rev. Mod. Phys.}\ }\textbf {\bibinfo {volume} {70}},\ \bibinfo
  {pages} {1545} (\bibinfo {year} {1998})}\BibitemShut {NoStop}%
\bibitem [{\citenamefont {Beyer}(1999)}]{Beyer:1998nu}%
  \BibitemOpen
  \bibfield  {author} {\bibinfo {author} {\bibfnamefont {H.~R.}\ \bibnamefont
  {Beyer}},\ }\href {\doibase 10.1007/s002200050651} {\bibfield  {journal}
  {\bibinfo  {journal} {Commun. Math. Phys.}\ }\textbf {\bibinfo {volume}
  {204}},\ \bibinfo {pages} {397} (\bibinfo {year} {1999})}\BibitemShut
  {NoStop}%
\bibitem [{\citenamefont {Lax}\ and\ \citenamefont
  {Phillips}(1989)}]{LaxPhi89}%
  \BibitemOpen
  \bibfield  {author} {\bibinfo {author} {\bibfnamefont {P.~D.}\ \bibnamefont
  {Lax}}\ and\ \bibinfo {author} {\bibfnamefont {R.~S.}\ \bibnamefont
  {Phillips}},\ }\href@noop {} {\emph {\bibinfo {title} {Scattering theory}}},\
  \bibinfo {edition} {second edition}\ ed.,\ \bibinfo {series} {Pure and
  Applied Mathematics}, Vol.~\bibinfo {volume} {26}\ (\bibinfo  {publisher}
  {Academic Press},\ \bibinfo {address} {Boston},\ \bibinfo {year}
  {1989})\BibitemShut {NoStop}%
\bibitem [{\citenamefont {Vainberg}(1973)}]{Vainb73}%
  \BibitemOpen
  \bibfield  {author} {\bibinfo {author} {\bibfnamefont {B.~R.}\ \bibnamefont
  {Vainberg}},\ }\href@noop {} {\bibfield  {journal} {\bibinfo  {journal} {Mat.
  Sb. (N.S.)}\ }\textbf {\bibinfo {volume} {92}},\ \bibinfo {pages} {224}
  (\bibinfo {year} {1973})}\BibitemShut {NoStop}%
\end{thebibliography}%

\end{document}